\documentclass[12pt]{article}
\usepackage{myaasmacros}
\usepackage{graphicx}
\usepackage{natbib}
\usepackage{bm}
\usepackage{array}
\usepackage{wrapfig}
\usepackage[pdftex,
        colorlinks=true,
        urlcolor=blue,		
        filecolor=blue,		
        citecolor=blue,		
        linkcolor=blue,		
        pdftitle={The Local Dark Matter Density},
        pdfauthor={J. I. Read},
        pdfsubject={A review of current efforts to measure the mean density of dark matter near the Sun.},
        pdfkeywords={},
        pdfproducer={pdfLaTeX},
        pagebackref,
        pdfpagemode=None,
        bookmarksopen=true]{hyperref}
\makeatletter
\newcommand{\thickhline}{%
    \noalign {\ifnum 0=`}\fi \hrule height 1pt
    \futurelet \reserved@a \@xhline
}
\newcolumntype{"}{@{\hskip\tabcolsep\vrule width 1pt\hskip\tabcolsep}}
\makeatother

\usepackage[paper=a4paper,left=25mm,width=160mm,top=25mm,bottom=25mm]{geometry} 
\usepackage[labelfont=bf]{caption}

\begin{document}

\def\ltsima{$\; \buildrel < \over \sim \;$}
\def\simlt{\lower.5ex\hbox{\ltsima}}   
\def\gtsima{$\; \buildrel > \over \sim \;$}
\def\simgt{\lower.5ex\hbox{\gtsima}}
\newcommand\bcite[1]{\citeauthor{#1} \citeyear{#1}}

\def\gprior{{\tt gprior}}
\def\cpfrior{{\tt cprior}}
\def\bprior{{\tt bprior}}
\def\lbprior{{\tt lbprior}}

\def\vztwo{\overline{v_z^2}}
\def\vztwoi{\overline{v_{z,i}^2}}
\def\rhodisc{\rho_\mathrm{disc}(z)}
\def\rhodmext{\rho_\mathrm{dm,ext}}
\def\rhodm{\rho_\mathrm{dm}}
\def\rhodmlab{\tilde{\rho}_\mathrm{dm}}
\def\rhoeff{\rho_\mathrm{dm}^\mathrm{eff}}
\def\nuobs{\nu_\mathrm{obs}(z)}
\def\msun{{\rm M}_\odot}
\def\brhodmext{{\bm \rho_\mathrm{dm,ext}}}
\def\brhodm{{\bm \rho_\mathrm{dm}}}

\def\Msunpcth{\,\mathrm{M}\,_\odot\,\mathrm{pc}^{-3}}
\def\Msunpctw{\,\mathrm{M}\,_\odot\,\mathrm{pc}^{-2}}
\def\GeVcmth{\,\mathrm{GeV}\,\mathrm{cm}^{-3}}

\newcommand{\HI}{{H{\footnotesize I}}}

\def\newblock{\hskip .11em plus .33em minus .07em}

\title{The Local Dark Matter Density}

\author{J. I. Read$^{1}$\thanks{E-mail: justin.inglis.read@gmail.com}\\
$^1${\small Department of Physics, University of Surrey, Guildford, GU2 7XH, Surrey, UK}\\
}

\date{}

\maketitle

\begin{abstract}
\noindent
I review current efforts to measure the mean density of dark matter near the Sun. This encodes valuable dynamical information about our Galaxy and is also of great importance for `direct detection' dark matter experiments. I discuss theoretical expectations in our current cosmology; the theory behind mass modelling of the Galaxy; and I show how combining local and global measures probes the shape of the Milky Way dark matter halo and the possible presence of a `dark disc'. I stress the strengths and weaknesses of different methodologies and highlight the continuing need for detailed tests on mock data -- particularly in the light of recently discovered evidence for disequilibria in the Milky Way disc. I collate the latest measurements of $\rhodm$ and show that, once the baryonic surface density contribution $\Sigma_b$ is normalised across different groups, there is remarkably good agreement. Compiling data from the literature, I estimate $\Sigma_b = 54.2 \pm 4.9 \Msunpctw$, where the dominant source of uncertainty is in the \HI\ gas contribution. Assuming this contribution from the baryons, I highlight several recent measurements of $\rhodm$ in order of increasing data complexity and prior, and, correspondingly, decreasing formal error bars (see Table \ref{tab:measurements}). Comparing these measurements with spherical extrapolations from the Milky Way's rotation curve, I show that the Milky Way is consistent with having a spherical dark matter halo at $R_0 \sim 8$\,kpc. The very latest measures of $\rhodm$ based on $\sim 10,000$ stars from the Sloan Digital Sky Survey appear to favour little halo flattening at $R_0$, suggesting that the Galaxy has a rather weak dark matter disc (see Figure \ref{fig:darkdisc}), with a correspondingly quiescent merger history. I caution, however, that this result hinges on there being no large systematics that remain to be uncovered in the SDSS data, and on the local baryonic surface density being $\Sigma_b \sim 55 \Msunpctw$. 

I conclude by discussing how the new Gaia satellite will be transformative. We will obtain much tighter constraints on both $\Sigma_b$ and $\rhodm$ by having accurate 6D phase space data for millions of stars near the Sun. These data will drive us towards fully three dimensional models of our Galactic potential, moving us into the realm of precision measurements of $\rhodm$. 
\end{abstract}

\tableofcontents 

\section{Introduction}\label{sec:introduction}

The local dark matter density ($\rhodm$) is an average over a small volume, typically a few hundred parsecs$^{\rm \footnotemark}$, around the Sun.\footnotetext{1 parsec = 3.26 light years = 3.086$\times 10^{16}$\,m.} It is of great interest for two main reasons. Firstly, it encodes valuable information about the local shape of the Milky Way's dark matter halo$^{\rm \footnotemark}$ near the disc plane. This provides interesting constraints on galaxy formation models and cosmology \citep[e.g.][]{1994ApJ...431..617D,2001ApJ...551..294I,2004ApJ...611L..73K,2007MNRAS.378...55M,2007arXiv0707.0737D,2012MNRAS.424L..16L}; on the merger history of our Galaxy \citep[e.g.][]{1989AJ.....98.1554L,2008MNRAS.389.1041R,2009MNRAS.397...44R}; and on alternative gravity theories \citep[e.g.][]{2001MNRAS.326.1261M,2004MNRAS.347.1055K,2005MNRAS.361..971R,2007MNRAS.379..597N}. Secondly, $\rhodm$ is important for direct detection experiments that hope to find evidence for a dark matter particle in the laboratory. The expected recoil rate (per unit mass, nuclear recoil energy $E$, and time) in such experiments is given by \citep[e.g.][]{1996APh.....6...87L}: 

\begin{equation}
\frac{dR}{dE} = \frac{\rhodmlab \sigma_W |F(E)|^2}{2 m_W \mu^2} \int_{v>\sqrt{m_N E/2\mu^2}}^{v_\mathrm{max}} \frac{f({\bf v},t)}{v}d^3 {\bf v} 
\label{eqn:recoilrate} 
\end{equation} 
where $\sigma_W$ and $m_W$ are the interaction cross section and mass of the dark matter particle (that we would like to measure); $|F(E)|$ is a nuclear form factor that depends on the choice of detector material; $m_N$ is the mass of the target nucleus; $\mu$ is the reduced mass of the dark matter-nucleus system; $v = |{\bf v}|$ is the speed of the dark matter particles; $f({\bf v},t)$ is the velocity distribution function; $v_\mathrm{max} = 533^{+54}_{-41}$\,km/s (at 90\% confidence) is the Galactic escape speed \citep{2014A&A...562A..91P}; and $\rhodmlab$ is the dark matter density within the detector. 
 
It is clear from equation \ref{eqn:recoilrate} that the ratio $\sigma_W/m_W$ trivially degenerates with $\rhodmlab$. Thus, to measure the nature of dark matter from such experiments (in the event of a signal), we must have an independent measure of $\rhodmlab$. This can be obtained by extrapolating from $\rhodm$ to the lab, accounting for potential fine-grained structure \citep{2008arXiv0801.3269K,2008MNRAS.385..236V,2009MNRAS.394..641Z,2009PhRvD..79j3531P,2011MNRAS.418.2648F}; I discuss this in \S\ref{sec:cosmotheory}. We also need to know the velocity distribution function of dark matter particles passing through the detector: $f({\bf v},t)$. In the limit of small numbers of detected dark matter particles, this must be estimated from numerical simulations (\S\ref{sec:cosmotheory}). However, for several thousand detections across a wide range of recoil energy, it can be measured directly \citep{Peter_2011}.

\footnotetext{I use the standard terminology `halo' to mean a gravitationally bound collection of dark matter particles. I also define here `subhalo' to mean a bound halo orbiting within a larger halo.}

\begin{figure}[t]
\begin{center}
\includegraphics[width = 0.65\textwidth]{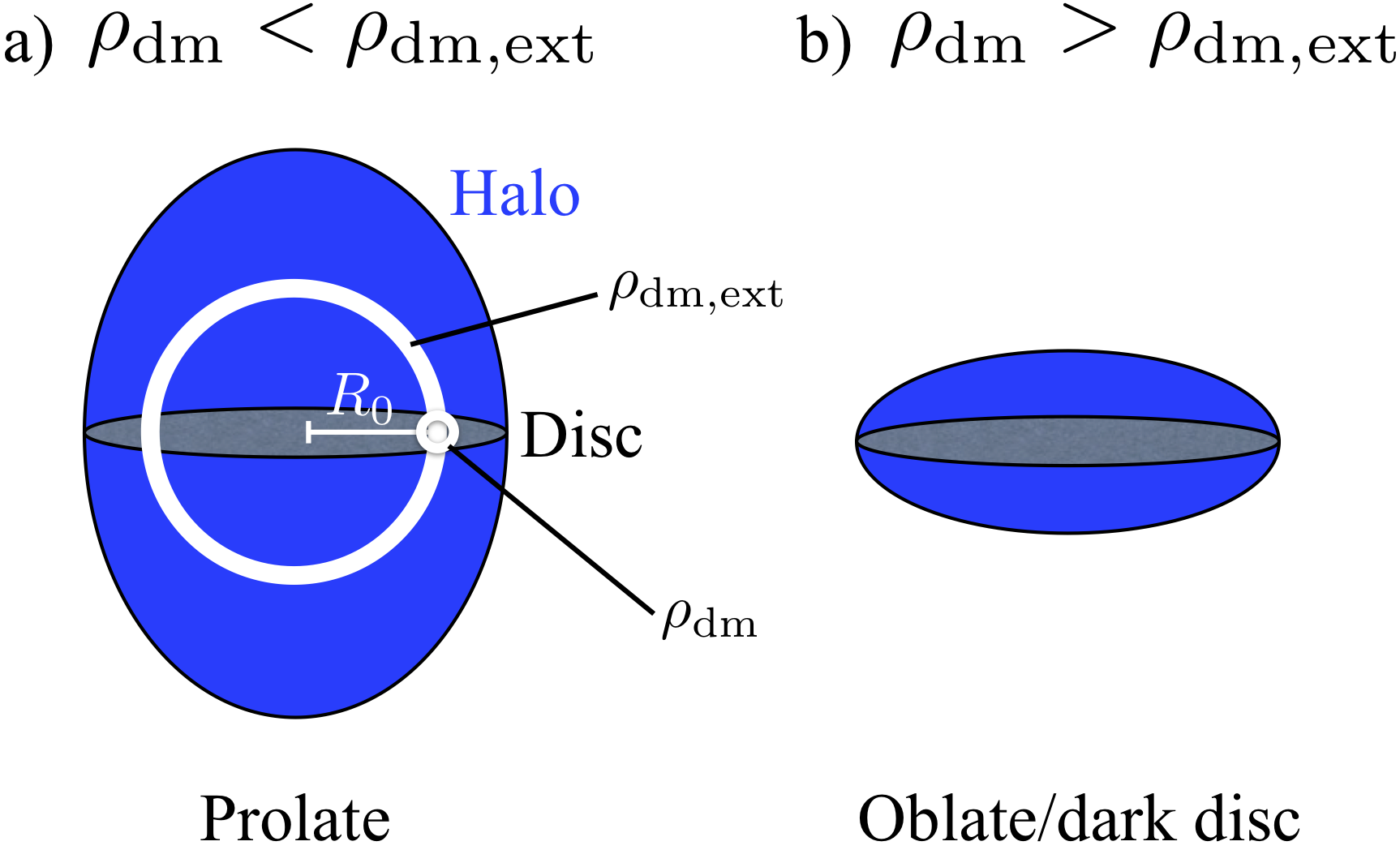}
\end{center}
\vspace{-5mm}
\caption{\small A schematic representation of local versus global measures of the dark matter density. The Milky Way disc is marked in grey; the dark matter halo in blue. Local measures -- $\rhodm$ -- are an average over a small volume, typically a few hundred parsecs around the Sun. Global measures -- $\rhodmext$ -- are extrapolated from larger scales and rely on assumptions about the shape of the Milky Way dark matter halo. (Here I define $\rhodmext$ such that the halo is assumed to be spherical.) Such probes are complementary. If $\rhodm < \rhodmext$, this implies a stretched or prolate dark matter halo (situation {\it a}, left). Conversely, if $\rhodm > \rhodmext$, this implies a squashed halo, or the presence of additional dark matter near the Milky Way disc (situation {\it b}, right). This latter is expected if our Galaxy has a `dark disc' (see \S\ref{sec:cosmotheory}).}
\label{fig:combinedprobes} 
\end{figure}

There are two main approaches to measuring $\rhodm$. Local measures use the vertical kinematics of stars near the Sun -- called `tracers' \citep[e.g.][]{1922ApJ....55..302K,oort_force_1932,1960BAN....15....1H,oort_note_1960,bahcall_self-consistent_1984,bahcall_k_1984,1987AA...180...94B,1989MNRAS.239..571K,1989MNRAS.239..605K,1989MNRAS.239..651K,1991ApJ...367L...9K,bahcall_local_1992,creze_distribution_1998,holmberg_local_2000,2003A&A...399..531S,2004MNRAS.352..440H,2006AA...446..933B,2012MNRAS.425.1445G,2012ApJ...746..181S,2012ApJ...751...30M,2012ApJ...756...89B,2013ApJ...772..108Z}. Global measures extrapolate $\rhodm$ from the rotation curve\footnote{Actually, many modern studies use the local surface density of matter as a constraint on their models, typically taking the value from \citet{1991ApJ...367L...9K}. However, \citet{1991ApJ...367L...9K} use a prior from the rotation curve that assumes a spherical halo (see \S\ref{sec:theory}, \S\ref{sec:tests} and \S\ref{sec:measurements}). For this reason, I still consider  global models that include a prior from \citet{1991ApJ...367L...9K} as `spherical halo' models that measure $\rhodmext$.} \citep[e.g.][]{1998MNRAS.294..429D,1989ApJ...342..272F,1992AJ....103.1552M,sofue_unified_2008,weber_determination_2010,2010JCAP...08..004C,2011MNRAS.414.2446M}. More recently, there have been attempts to bridge these two scales by modelling the phase space distribution of stars over larger volumes around the Solar neighbourhood \citep{2013arXiv1309.0809B}. The global measures often result in very small error bars (\bcite{2010JCAP...08..004C}; though see \bcite{2010AA...523A..83S} and \bcite{2011JCAP...11..029I}). However, these small errors hinge on strong assumptions about the Galactic halo shape -- particularly near the disc plane \citep{weber_determination_2010}. By contrast, local measures rely on fewer assumptions, but have correspondingly larger errors \citep[e.g.][]{2012MNRAS.425.1445G,2012ApJ...746..181S,2013ApJ...772..108Z}. To avoid confusion, I will refer to results from global estimates that assume a spherically symmetric dark matter halo as an `extrapolated' dark matter density, denoted $\rhodmext$, while I will refer to local measures as $\rhodm$. Combining measures of $\rhodm$ and $\rhodmext$, we can probe the local shape of the Milky Way halo. If $\rhodm < \rhodmext$, then the dark matter halo at the Solar position $R_0 \sim 8$\,kpc is likely prolate (stretched) along a direction perpendicular to the disc plane. If $\rhodm > \rhodmext$, this could imply an oblate (squashed) halo, or a local dark matter disc (see Figure \ref{fig:combinedprobes}). I discuss the theoretical implications of these different scenarios in \S\ref{sec:cosmotheory}. 

Measurements of $\rhodm$ have a long history dating back to \citet{1922ApJ....55..302K} who was one of the first to coin the term {\it ``dark matter"}. Using the measured vertical velocity of stars near the Sun, he compared the sum of their masses to the vertical gravitational force required to keep them in equilibrium, finding that:

\begin{quote}
{\it ``As matters stand it appears at once that this} [{\it dark matter}\hspace{0.4mm}] {\it mass cannot be excessive."}
\end{quote}
However, this early pioneering work treated the stars as a collisional gas, whereas stars are really a {\it collisionless fluid} that obeys similar but different equations of motion. This was corrected the same year by \citet{1922MNRAS..82..122J}, who laid down the basic theory for mass modelling of stellar systems that I outline in \S\ref{sec:theory}. The technique was later refined and applied to improved data by \citet{oort_force_1932}, \citet{1960BAN....15....1H}, \citet{oort_note_1960}, and \citet{bahcall_k_1984,bahcall_self-consistent_1984}. However, there were several problems with these early works: i) their measurements relied on poorly calibrated `photometric' estimates of the distances (\S\ref{sec:tracers}); (ii) stars were chosen that were sometimes too young to be dynamically well mixed in the disc (see \S\ref{sec:theory}); (iii) populations were often assumed to be `isothermal' with the vertical velocity dispersion constant with height \citep[typically a poor approximation: ][]{1989MNRAS.239..651K,2011MNRAS.416.2318G}; and (iv) it was often unclear if the stars for which photometric density distributions could be estimated were the same stars for which the velocity distribution was measured \citep[][and see \S\ref{sec:theory}]{1989MNRAS.239..651K}. A key series of papers by \citet{1989MNRAS.239..571K,1989MNRAS.239..605K,1989MNRAS.239..651K,1991ApJ...367L...9K}, improved on this by collecting an unprecedented amount of data, and compiling a {\it volume complete} sample of K-dwarf stars (that are particularly good for measuring photometric distances; \S\ref{sec:tracers}) towards the South Galactic Pole. A quarter of these had radial velocity measurements.

A further key improvement came with the {\it Hipparcos} satellite that launched in August 1989, providing positions and proper motions for $\sim 100,000$ stars within $\sim 100$\,pc of the Sun \citep{vanLeeuwen2007}. It was a boon for the field, since prior to this only radial Doppler velocities and photometric distances were available. Several new measurements of $\rhodm$ using these new data followed \citep{creze_distribution_1998,holmberg_local_2000,2003A&A...399..531S,2004MNRAS.352..440H,2006AA...446..933B}. 

Most recently, there have been a series of new measurements coming from new Galactic surveys -- the Sloan Digital Sky Survey (SDSS; \citealt{2012ApJ...746..181S,2013ApJ...772..108Z}), and the RAdial Velocity Experiment (RAVE; \citealt{2008MNRAS.391..793S}). These same surveys have recently found evidence for vertical density waves in the Milky Way disc \citep{2012ApJ...750L..41W,2013arXiv1302.2468W,2013arXiv1309.2300Y}, perhaps caused by the recent Sagittarius dwarf merger \citep{2013MNRAS.429..159G}. This is something that may prove both a blessing and a curse for attempts to measure $\rhodm$; I discuss this further in \S\ref{sec:disequilibria}.

All of the above measurements use stellar kinematics to probe the total Galactic potential near the Sun. To extract the local dark matter density from this, we must assume some weak field theory of gravity (to link the potential to the matter density; see \S\ref{sec:cosmomodel} and \S\ref{sec:theory}), and we must subtract off the contribution from visible matter (i.e. stars, gas, stellar remnants etc.). I call this from here on the {\it baryonic} matter density $\rho_b$. Estimates of this have also evolved with time, from an early estimate of $\rho_b \sim 0.038\Msunpcth$$^{\rm \footnotemark}$ \citep{oort_force_1932} to the more modern value $\rho_b = 0.0914 \pm 0.009\Msunpcth$ \citep{2006MNRAS.372.1149F}. I discuss the latest constraints on $\rho_b$ in \S\ref{sec:massmodel}. 

\begin{figure}
\begin{center}
\includegraphics[width = 0.99\textwidth]{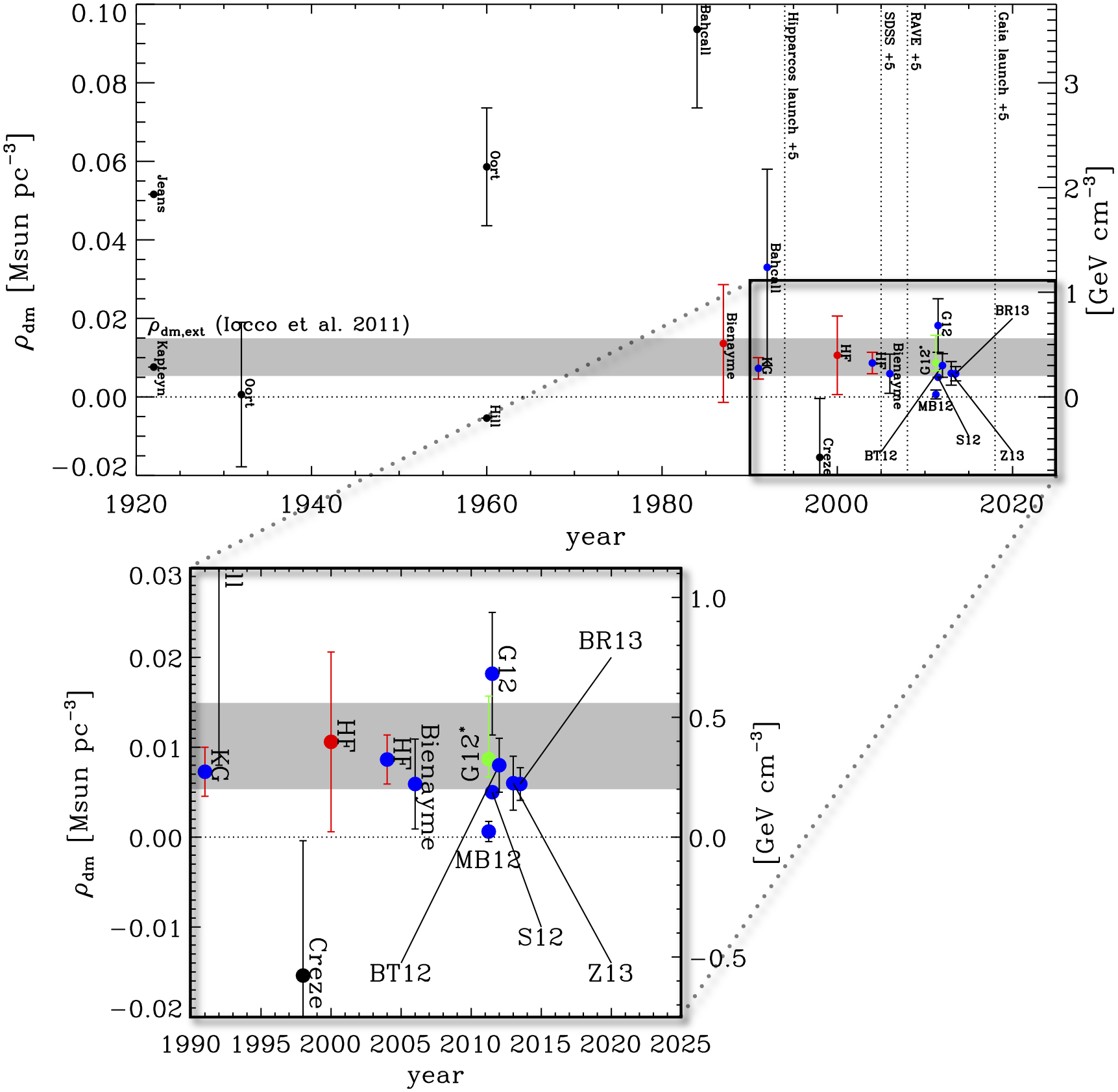}
\end{center}
\vspace{-8mm}
\caption{\small A century of measurements of $\rhodm$. In all cases, I assume the same matter density and surface density of $\rho_b = 0.0914\Msunpcth$ and $\Sigma_b = 55\Msunpctw$ \citep{2006MNRAS.372.1149F}. Values derived from a surface density rather than a volume density have a blue filled circle; red data points indicate the use of a `rotation curve' prior (see \S\ref{sec:rotprior}). The green data point is derived from \citet{2012MNRAS.425.1445G} assuming a stronger prior on $\Sigma_b = 55 \pm 1 \Msunpctw$ (see \S\ref{sec:measurements}). All error bars represent either $1\sigma$ uncertainties or 68\% confidence intervals. Overlaid are: $\rhodmext$ extrapolated from the rotation curve assuming spherical symmetry (grey band); the launch dates plus 5 years for the Hipparcos and Gaia astrometric satellite missions; and the start date plus 5 years of the SDSS and RAVE surveys. Where no error bar was calculated for a given measurement, there is simply a horizontal line through that data point. All data and references (including definitions of abbreviations) are given in Table \ref{tab:measurements}.}
\label{fig:rhodmhistory} 
\end{figure}

\footnotetext{Particle physicists may be more used to seeing these mass densities in units of $\GeVcmth$; a useful conversion is: $0.008\Msunpcth = 0.3\GeVcmth$. I mark all densities also in $\GeVcmth$ along the right $y$-axis of Figure \ref{fig:rhodmhistory}, and in Table \ref{tab:measurements}.}

In addition to the above improvements in data, there has been a concerted push to better understand the model systematics that go into the measurement of $\rhodm$. Early work by \citet{1989ApJ...344..217S} and \citet{1989MNRAS.239..571K} explored the effects of un-modelled coupling between radial and vertical star motions (see \S\ref{sec:theory}), while tests on simple mock data drawn from an analytic Galactic model have been useful in determining the effect of errors due to measurement uncertainties and poor sampling (\citealt{1991ApJ...367L...9K}; \citealt{2013A&A...555A.105I}; and see \S\ref{sec:tests}). But a full test of methods on dynamically realistic $N$-body mock data has only come recently with \citet{2011MNRAS.416.2318G}. This has exposed some rather surprising model biases that I discuss further in \S\ref{sec:theory} and \S\ref{sec:tests}. Finally, new methods to combat such systematics are being developed \citep[e.g.][]{2011MNRAS.416.2318G,2013MNRAS.433.1411M} resulting in further new measurements of $\rhodm$ \citep{2012MNRAS.425.1445G}. I discuss these techniques in \S\ref{sec:theory} and compare and contrast the latest measurements in \S\ref{sec:measurements}. 

A summary of measurements of $\rhodm$ from Kapteyn through to the present day is given in Figure \ref{fig:rhodmhistory}, where I mark also the latest limits on $\rhodmext$ from the rotation curve assuming a spherically symmetric dark matter halo (grey band\footnote{This is taken from \citealt{2011JCAP...11..029I}, but is consistent with other recent measures \citep{1998MNRAS.294..429D,1989ApJ...342..272F,1992AJ....103.1552M,sofue_unified_2008,2010AA...523A..83S,weber_determination_2010,2010JCAP...08..004C}.}); all data and references are given in Table \ref{tab:measurements}. I discuss this Figure in detail along with the latest constraints on $\rhodm$ in \S\ref{sec:measurements}.

With the successful launch of the Gaia satellite, measurements of $\rhodm$ are set to enter a golden age \citep[e.g.][]{2001A&A...369..339P,2005MNRAS.359.1306W}. There are significant challenges to be overcome \citep{2013A&ARv..21...61R,2013arXiv1309.2794B}, but as has happened post-Hipparcos, Gaia will likely drive another step-wise improvement in the error bars on $\rhodm$. I discuss this in \S\ref{sec:gaia}. 

This article is organised as follows. In \S\ref{sec:cosmotheory}, I discuss theoretical expectations for $\rhodm$ and its laboratory extrapolation $\rhodmlab$ in our current cosmology. In \S\ref{sec:theory}, I present the key theory behind both local and global measures of the local dark matter density, with a particular focus on moment methods. In \S\ref{sec:tests}, I present tests of different methods on simple 1D mock data, determining what quality and type of data best constrain $\rhodm$. In \S\ref{sec:measurements}, I discuss historical measures of $\rhodm$ and summarise the latest measurements from different groups. I compare and contrast the advantages and disadvantages of different methods and data, and I assess where the key uncertainties remain. In \S\ref{sec:gaia}, I discuss how the Gaia satellite will transform our measurements of $\rhodm$. Finally, in \S\ref{sec:conclusions}, I present my conclusions. 

\section{Theoretical expectations for $\rhodm$ and its laboratory extrapolation $\rhodmlab$}\label{sec:cosmotheory}

\begin{figure}
\begin{center}
\includegraphics[width = 0.95\textwidth]{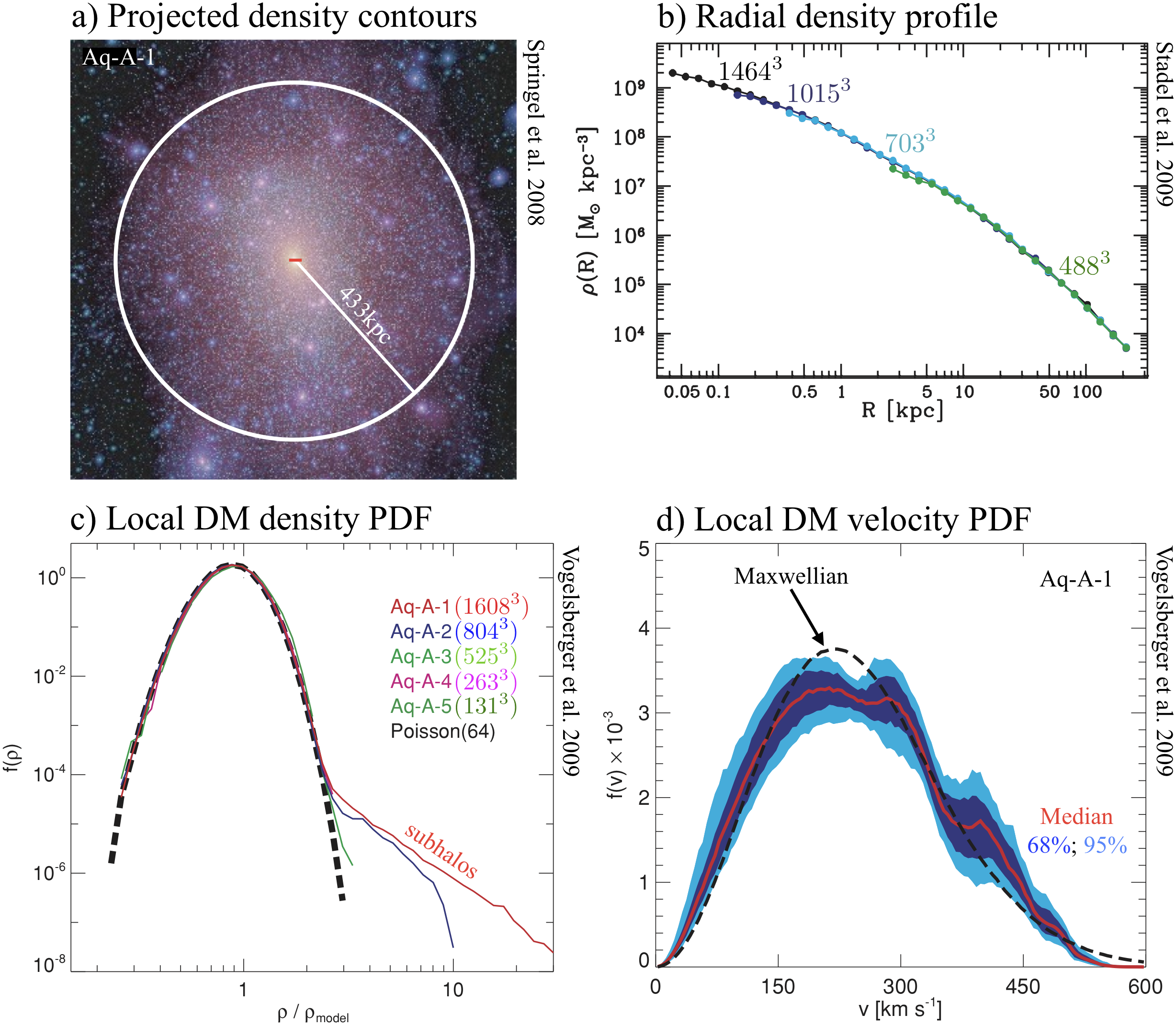}
\end{center}
\vspace{-8mm}
\caption{\small Key predictions from dark-matter-only (DMO) cosmological simulations. {\bf a)} Projected density contours of the {\it Aquarius} Aq-A-1 DMO cosmological simulation of a halo of Milky Way mass ($M_{200} \sim 10^{12}$\,M$_\odot$), run with 4.2 billion dark matter super-particles \citep{2008MNRAS.391.1685S}. The size of the Galactic disc out to the Sun position $R_0 = 8$\,kpc (not modelled in this simulation) is marked by the red horizontal line. {\bf b)} The spherically averaged dark matter density profile from the {\it GHALO} suite of Milky Way mass halo simulations \citep{2009MNRAS.398L..21S}. Four different resolutions (super-particle numbers) are marked, showing excellent numerical convergence. {\bf c)} The dark matter density Probability Distribution Function (PDF) in the {\it Aquarius} suite, calculated using a kernel average (64 smoothing neighbours) at each super-particle, normalised to a power law model fit over a thick ellipsoidal shell at 6-12\,kpc from the halo centre \citep{2009MNRAS.395..797V}. Simulations Aq-A-1 through Aq-A-5 (of decreasing numerical resolution, as marked) are over-plotted; only Aq-A-1 and Aq-A-2 resolve the high density tail due to subhalos. The black dashed line shows the intrinsic scatter due to Poisson noise in the density estimator. {\bf d)} The dark matter velocity PDF averaged over 2\,kpc boxes at 7-9\,kpc from the halo centre of Aq-A-1.}
\label{fig:cdmsims} 
\end{figure}

Before discussing mass modelling theory and the latest results, it is worth a short digression to describe our theoretical expectations for $\rhodm$ (averaged over a few hundred parsecs), and its extrapolation to the dark matter density in the laboratory $\rhodmlab$. 

\subsection{The cosmological model}\label{sec:cosmomodel}

Throughout this review, I will assume the `standard' $\Lambda$ Cold Dark Matter model, or $\Lambda$CDM, where the $\Lambda$ refers to `dark energy' -- an apparent acceleration of the Universe at the present time. This is supported by a wealth of observational data. The cosmic microwave background radiation \citep{1992ApJ...396L..13W,2013arXiv1303.5076P}; galaxy clustering \citep{2002ApJ...581...20C}; baryon acoustic oscillations \citep{2013JCAP...04..026S}; and Type Ia SNe standard candles \citep{1998AJ....116.1009R,1999ApJ...517..565P} all point towards a cosmological model where the energy density of the Universe comprises just 5\% in baryons ($\Omega_b$); 27\% in dark matter ($\Omega_{\rm dm}$); and 68\% in dark energy ($\Omega_\Lambda$). This is further supported by evidence for dark matter within galaxies and clusters from stellar/galaxy kinematics \citep[e.g.][]{1937ApJ....86..217Z,1984ApJ...278...81V,KleynaEtal2001,2012ApJ...745...92A}; stellar/gaseous rotation curves \citep[e.g.][]{1959BAN....14..323V,Freeman1970,1977A&A....57..373B,1979A&A....79..281B,1980ApJ...238..471R,1985ApJ...295..305V}; and gravitational lensing \citep[e.g.][]{1979Natur.279..381W,2006ApJ...648L.109C}.

The $\Lambda$CDM model has two unknown elements: dark energy and dark matter. The former appears to be consistent with a `cosmological constant' that could result from vacuum energy, though this is far from established \citep[e.g.][]{2013arXiv1303.5076P}. The latter, we are better able to pin down. While it remains unclear exactly what dark matter is, it does appear to move as a collisionless non-relativistic fluid at least at the present time \citep{2006ApJ...648L.109C}. Alternative gravity theories like MOND \citep{1983ApJ...270..365M} and its relativistic extension TeVeS \citep{2004PhRvD..70h3509B} face a host of observational challenges\footnotemark\ \citep[e.g.][]{2006ApJ...648L.109C,2008MNRAS.389..250N,2011ApJ...738..186I,2011IJMPD..20.2749D}. By contrast, dark matter as a collisionless fluid appears to give an excellent match to the growth of large scale structure in the Universe \citep[e.g.][]{2008PhRvL.100d1304V}. On smaller scales, there have been many claims of problems with $\Lambda$CDM, most notably the {\it missing satellites} and {\it cusp-core} problems. The former is a large discrepancy between the predicted and observed number of satellite galaxies around the Milky Way and M31 \citep{1999ApJ...522...82K,1999ApJ...524L..19M}; the latter is a discrepancy between predicted `cuspy' dark matter density at the centres of dwarf galaxies ($\rho = \rho_0 \left[r/r_0\right]^{-1}$) and observed constant density cores ($\rho = \rho_0$; \citealt{1994ApJ...427L...1F,1994Natur.370..629M}). Both problems have stood the test of time, with dark matter cores now being reported even within tiny dwarf spheroidals (dSphs) orbiting the Milky Way \citep[e.g.][]{Goerdt:2006rw,2011ApJ...742...20W,2012MNRAS.426..601C}. These small scale problems may be telling us something exciting about the nature of dark matter \citep[e.g.][]{2001ApJ...556...93B,2013MNRAS.430...81R} or inflation physics \citep[e.g.][]{2002PhRvD..66d3003Z}. However, on scales below $\sim 1$\,Mpc `baryon physics' (radiative cooling, star formation and feedback from stellar winds, supernovae and active galactic nuclei) become important. These difficult-to-model processes could physically reshape the dark matter at the centres of galaxies, solving the cusp-core problem without the need to resort to exotic cosmology \citep{2005MNRAS.356..107R,MashchenkoEtAl2006,2012MNRAS.421.3464P,2013MNRAS.429.3068T}. Such cored dwarfs are then much more easily tidally disrupted by the Milky Way (MW), plausibly solving the missing satellites problem too \citep{2006MNRAS.tmp..153R,2012ApJ...761...71Z}. I discuss this in more detail in \S\ref{sec:baryons}. 

While dark matter is most likely some sort of collisionless fluid, it is not clear what it is made up of. Microlensing constraints from the Milky Way bulge and the nearby Large and Small Magellanic Clouds put an upper bound of the mass of `compact object' dark matter of $M_{\rm dm} < 10^{-7}$\,M$_\odot$ \citep[e.g.][]{2007A&A...469..387T}. While no smoking gun, this and the other results above point towards dark matter being comprised of some new yet-to-be discovered weakly interacting particle that lies beyond the standard model of particle physics \citep[e.g.][]{1996PhR...267..195J,2009ARNPS..59..191B}.  The precise nature of this particle, however, remains elusive. It could be quite massive ($\sim 10 - 1000$\,GeV), as predicted by some supersymmetric extensions to the standard model \citep[e.g.][]{1996PhR...267..195J}. This would make it non-relativistic at all times, so-called `Cold Dark Matter' (CDM). However, other popular models like axions or sterile neutrinos predict a lighter particle ($\sim 1 - 50$\,keV) that would be relativistic for a time in the early Universe \citep[e.g.][]{2009ARNPS..59..191B}, so-called `Warm Dark Matter' (WDM). I focus on CDM in this review as it remains better-studied than WDM (see also the discussion in \S\ref{sec:dmosims}), but note that WDM remains an exciting proposition that deserves to be more fully explored.

\footnotetext{Note that one of the key pieces of evidence in favour of a collisionless fluid dark matter is the so-called `bullet cluster' \citep{2006ApJ...648L.109C}. Due to a recent merger between two galaxy clusters, this system has a large offset between the weak lensing mass peaks (that correlate well with the galactic light) and the bulk of the visible mass that is in the form of hot X-ray emitting gas. This is hard to reproduce in alternative gravity (AG) models, despite some heroic attempts to do so \citep{2006MNRAS.371..138A}. Some proponents of AG, while side-stepping the thorny issue of the bullet cluster, have pointed out that other cluster collision systems appear to produce rather different results from the bullet cluster. The problem poster-child is Abel 520 which was reported to have a `dark core' that correlates well with the X-ray emission but not the galactic light -- the exact opposite of the bullet cluster \citep{2007ApJ...668..806M}. However, lensing is known to suffer from degeneracies that can masquerade as phantom mass peaks or rings \citep{2008MNRAS.386..307L}. It is likely that the dark core in Abel 520 is one of these examples, disappearing with improved data and models (\citealt{2012ApJ...758..128C}, but see \citealt{2014ApJ...783...78J}).}

\subsection{Dark matter only (DMO) simulations}\label{sec:dmosims}

In $\Lambda$CDM, structure grows via the hierarchical accretion of smaller sub-structures \citep{1978MNRAS.183..341W}. The process is highly non-linear, requiring numerical $N$-body simulations to integrate the equations of motion \citep{1991ApJ...378..496D,1996ApJ...462..563N,2009MNRAS.398L..21S,2008MNRAS.391.1685S,2011EPJP..126...55D}. Such simulations solve Newtonian gravity between $N$ `super-particles' on the background of an expanding Freedmann-Lema\^itre-Robertson-Walker metric. The super-particles have mass typically $\simgt 10^3$\,M$_\odot$ each and represent large smoothed patches of the collisionless dark matter fluid; they should not be confused with dark matter particles that are likely $>10^{60}$ of magnitude smaller in mass.  This ``Newtonian approximation" is extremely good \citep{2013arXiv1308.6524A}, and certainly more than adequate for calculating the phase space distribution function of dark matter in the Galaxy. 

A detailed discussion of cosmological $N$-body simulations is beyond the scope of this present work (see e.g. \citealt{2011EPJP..126...55D} and \citealt{2012PDU.....1...50K} for reviews). Here, I simply note that the results from these simulations -- at least for non-relativistic cold dark matter -- numerically converge on a well-defined asymptotic solution as the number of super-particles is increased \citep[][and see Figure \ref{fig:cdmsims}, panel b]{2007arXiv0706.1270H,2013arXiv1308.2669K}. In this sense, the results from these `Dark Matter Only', or DMO simulations as I will call them from now on, are robust. That said, problems still remain for simulations where there is a strong suppression in the small scale power spectrum, as in WDM simulations \citep{2001ApJ...556...93B,2001ApJ...559..516A,2007astro.ph..2575W,2013MNRAS.tmp.1799H}. There, discreteness noise due to anisotropic force errors leads to the growth of spurious numerical substructures. A full solution to the problem remains elusive, though recent work shows promise. \citet{2013MNRAS.tmp.1799H} suggest a radical break from the standard $N$-body method by numerically modelling the folding of the dark matter phase sheet in phase space. Unlike standard $N$-body methods, they explicitly calculate the phase space distribution of the sheet by interpolating between particles. They then integrate over velocity to obtain the dark matter density field. The method shows great promise but becomes computationally expensive in regions where the sheet becomes highly foliated -- i.e. at the centres of forming halos. \citet{2014MNRAS.439..300L} propose a much less computationally expensive post-processing algorithm to prune spurious structures from standard $N$-body simulations. However, this can only remove {\it surviving} spurious structure, leading to the worry that already merged spurious halos may remain problematic.

\subsubsection{Key predictions from DMO simulations}\label{sec:dmopredictions}

In this section, I summarise the key predictions, relevant for this review, from $\Lambda$CDM DMO simulations of Milky Way-mass Galactic halos. These are collated in Figure \ref{fig:cdmsims}.

\paragraph{The spherically averaged radial density profile}
A first key prediction from DMO simulations is the spherically averaged radial density profile of dark matter halos. This is well-fit (at the $\sim 10\%$ level; \citealt{2006AJ....132.2685M,2009MNRAS.398L..21S}) by a split power law that goes as roughly $r^{-1}$ in the centre and $r^{-3}$ at the edge \citep{1991ApJ...378..496D,1996ApJ...462..563N}, the `NFW' profile (see Figure \ref{fig:cdmsims}b): 

\begin{equation} 
\rho = \rho_0 \left(\frac{r}{r_s}\right)^{-1}\left(1+\frac{r}{r_s}\right)^{-2}
\label{eqn:nfwprofile}
\end{equation} 
where $r_s$ is a radial scale length; and $\rho_0$ is a density normalisation. These are usually defined in terms of a `concentration parameter' $c= r_{200} / r_s$; a `virial radius' $r_{200}$; and a `virial mass' $M_{200}$:  

\begin{equation} 
\rho_0 = \frac{200}{3} \frac{c^3}{\ln(1+c)-\frac{c}{1+c}} \rho_{\rm crit};
\end{equation}
where $\rho_{\rm crit} = 128.2$\,M$_\odot$\,kpc$^{-3}$ is the critical density of the Universe at redshift $z=0$;  

\begin{equation}
r_{200} = \left(\frac{3M_{200}}{4 \pi 200 \rho_{\rm crit}}\right)^{1/3}
\end{equation}
is the `virial radius' at which the mean enclosed density is 200 times $\rho_{\rm crit}$; and $M_{200}$ is the `virial mass' -- the mass enclosed within $r_{200}$.

The NFW profile appears to be `universal' in the sense that it gives a good fit to the full range of halo masses probed to date, from dwarf galaxy subhalos to giant galaxy cluster halos \citep{1996ApJ...462..563N,2008MNRAS.391.1685S,2009MNRAS.398L..21S}, though the physical reason for this universality remains to be fully understood \citep[e.g.][]{2006ApJ...653...43M,2013MNRAS.430..121P}. 

Although there is significant scatter in $r_s$ at a given $M_{200}$, there is a correlation between the two \citep{1996ApJ...462..563N,2001ApJ...555..240B,2007MNRAS.378...55M}. At redshift $z=0$, \citet{2007MNRAS.378...55M} find:

\begin{equation} 
\log c = 1.02 [\pm 0.015]-0.109 [\pm 0.005] \left(\log \left[\frac{M_{200}}{1.47\,{\rm M}_\odot}\right] -12 \right)
\end{equation}
with an intrinsic scatter about this mean relation of $\sigma_{\log c} = 0.14 \pm 0.013$. Thus, for Milky Way mass halos ($M_{200} \sim 10^{12}$\,M$_\odot$; \citealt{1999MNRAS.310..645W,2002ApJ...573..597K,2011MNRAS.414.2446M,2013arXiv1309.4293P}), we have $r_{200} = 210$\,kpc and $r_s = 19^{+7.5}_{-5.4}$\,kpc at 68\% confidence.

\paragraph{The shape of dark matter halos}
DMO simulations also make predictions for the shape of dark matter halos, which are found to be triaxial \citep[][and see Figure \ref{fig:cdmsims}a]{1991ApJ...378..496D,1992ApJ...399..405W,1996ApJ...462..563N,2002ApJ...574..538J}. Consistent with earlier work, \citet{2007MNRAS.378...55M} find a mean shape parameter $\langle q \rangle = (b+c)/2a \sim 0.8$ when averaged over the whole halo, where $a > b > c$ are the long, intermediate and short axes of the figure. This corresponds to a typically {\it prolate} (egg-shaped) halo. Like the halo concentration parameter, $\langle {q} \rangle$ shows significant scatter at a given halo mass, slightly decreasing with halo mass \citep{2007MNRAS.378...55M}. When not averaged over the whole halo, the shape parameter $q$ is also a function of ellipsoidal radius \citep{2002ApJ...574..538J}. An understanding of the expected distribution of halo shapes is important for $\rhodm$ when we try to extrapolate its value from larger scales (see Figure \ref{fig:combinedprobes}), and when studying the expected scatter in $\rhodm$ at a given Galactocentric radius. I discuss this latter, next. 

\paragraph{The local dark matter density}\label{sec:dmorhodm}
Defining the `Solar neighbourhood' as a small volume around the Sun, we can use the above simulations to theoretically estimate $\rhodm$ for halos of Milky Way mass. The first and simplest analysis is to average $\rhodm$ in a spherical shell at the `Solar circle', $R_0 \sim 8$\,kpc. \citet{2009MNRAS.394..641Z} perform this exercise for the high resolution {\it `VL-II'} DMO simulation of a Milky Way mass halo, finding $\langle \rhodm \rangle_s = 0.01056$\,M$_\odot$\,pc$^{-3}$, which is remarkably close to that measured for the real Milky Way (see Figure \ref{fig:rhodmhistory}).

We can go further, however, and use the DMO simulations to estimate the expected {\it scatter} in $\rhodm$. This is encoded in the dark matter density Probability Density Function (PDF). \citet{2009MNRAS.395..797V} calculate this at $\sim 8$\,kpc from the halo centre for the {\it Aquarius} suite of high resolution DMO simulations (Figure \ref{fig:cdmsims}c). They use a `smoothed particle' kernel weighted density estimate calculated at the position of each super-particle in a thick ellipsoidal shell at 6-12\,kpc from the halo centre. This is then normalised to a power law model fit to this same ellipsoidal shell. With this analysis, the resultant scatter in $\rhodm$ is remarkably small -- consistent with the Poisson noise in the density estimator (black dashed line; Figure \ref{fig:cdmsims}c). (In other words, the scatter in $\rhodm$ is so small that they are unable to measure it above the intrinsic super-particle noise in the simulation.) However, this small scatter relies on the analysis being performed over ellipsoidal shells. \citet{2009MNRAS.394..641Z} perform a similar exercise for the {\it VL-II} simulation \citep{2007ApJ...667..859D}, but averaging $\rhodm$ over spherical volumes of radius $500$\,pc and normalising to $\langle \rhodm \rangle_s$. With this `spherical' analysis, they find a scatter in $\rhodm$ of up to a factor of $2-3$ within the 68\% confidence interval of their density PDF. When averaging instead along just one axis of the triaxial halo figure, they find a small scatter similar to that reported in \citet{2009MNRAS.395..797V}. Thus, the scatter in $\rhodm$ reported by \citet{2009MNRAS.394..641Z} is entirely due to systematic differences in $\rhodm$ along the long, intermediate and short axis of the triaxial halo. If the Milky Way halo is triaxial and we allow the disc to be aligned along any of the principle axes, then such scatter should be considered as part of our theoretical uncertainty on $\rhodm$. In practice, however, we cannot align discs arbitrarily within triaxial halos. Discs are unstable if aligned perpendicular to the intermediate axis of the figure \citep{1979ApJ...233..872H,1981MNRAS.196..455B,2013MNRAS.434.2971D}. More importantly, baryons -- stars and gas -- that are not included in the DMO models, likely alter the expected halo shape, making halos much rounder and reducing the expected scatter in $\rhodm$. I discuss this in \S\ref{sec:baryons}. 

The two highest resolution {\it Aquarius} simulations -- Aq-A-1 and Aq-A-2 -- are able to resolve the high density tail in the PDF due to subhalos at 8\,kpc (Figure \ref{fig:cdmsims}c, blue and red lines). While subhalos can significantly boost $\rhodm$, the likelihood of this happening is very small (see \S\ref{sec:rhodmlab}).

Finally, it is straightforward to show from these DMO simulations that, even up to $\sim 1$\,kpc above the disc of the Milky Way, we expect $\rhodm$ to be roughly constant when averaged over small `Solar neighbourhood' volumes \citep{2011MNRAS.416.2318G}. This will provide a valuable simplification when trying to derive $\rhodm$ from real data, as we shall see in \S\ref{sec:theory}.

\paragraph{The local velocity distribution function of dark matter} 
We can also use DMO simulations to predict the local velocity distribution function of dark matter in the Milky Way. This is important for direct detection experiments as I already discussed in \S\ref{sec:introduction}. The latest simulations are consistent with being close to Maxwellian, but not quite \citep[][and see Figure \ref{fig:cdmsims}d]{2009MNRAS.394..641Z,2009MNRAS.395..797V}. The ``not-quite" is important, particularly at the high velocity tail end of the distribution. This is boosted in the simulations with respect to a pure Maxwellian profile, where the highest velocity particles come from recently accreted structure that is not fully phase-mixed (so-called `debris flows' \citealt{2012PhRvD..86f3505K,2012PDU.....1..155L}). These structures are a super-position of many tidal streams that intersect the Solar neighbourhood volume; they are particularly important for direct detection experiments that are sensitive to light or inelastic dark matter, or those with directional sensitivity \citep{2012PhRvD..86f3505K}. Even more pronounced effects occur if an undisrupted but significant stream penetrates the Sun position \citep{2001PhRvD..64h3516S}. This is statistically unlikely, but -- at least for the more massive streams -- can be observationally tested by hunting for the visible stream-stars that would accompany such a `dark stream' \citep{2005PhRvD..71d3516F}. Lower mass satellite streams are potentially more problematic. These could also alter the local velocity PDF while being completely devoid of stars and essentially undetectable. I discuss these in \S\ref{sec:rhodmlab}, below. 

An example velocity PDF averaged over 2\,kpc boxes at 7-9\,kpc from the halo centre of the Aq-A-1 {\it Aquarius} simulation is shown in Figure \ref{fig:cdmsims}d. Notice that, while the distribution is reasonably Maxwellian, there are prominent bumps and wiggles of larger magnitude than the box-to-box scatter. These depend on the particular formation history of a given dark matter halo (see Figure 4 from \citealt{2009MNRAS.395..797V}). As pointed out by those authors, if we enter an era where dark matter particles are routinely detected, then we could actually measure such bumps and wiggles for our own Galaxy. Since these encode information about our Galactic accretion history, we could conceive of unravelling our past via detailed modelling of the dark matter velocity PDF. I caution, however, that such bumps and wiggles may be at least partially erased by baryonic processes during Galaxy formation (\S\ref{sec:baryons}); this remains to be explored.

\subsubsection{Extrapolating from $\rhodm$ to $\rhodmlab$}\label{sec:rhodmlab}

Even with over a billion super-particles, the spatial resolution of the {\it Aquarius} Aq-A-1 DMO simulation is $\sim 20$\,pc \citep{2008MNRAS.391.1685S}. While this is sufficient to model $\rhodm$ on the scales that we can hope to measure it in our Galaxy, it is many orders of magnitude away from $\rhodmlab$. Thus, we must {\it extrapolate} from $\rhodm$ to obtain $\rhodmlab$. The key concerns here are:

\begin{enumerate} 
\item unresolved substructrues, tidal debris/streams, and caustics \citep{2001PhRvD..64h3516S,2005PhRvD..71d3516F,2008arXiv0801.3269K,2009MNRAS.400.2174V,2011MNRAS.413.1419V,2011MNRAS.418.2648F}; and
\item the effect of the Solar system on the dark matter phase space distribution function \citep[e.g.][]{2009PhRvD..79j3531P}. 
\end{enumerate}

\paragraph{Streams \& Caustics} \citet{2011MNRAS.413.1419V} use a novel `sub-grid' stream model applied to the {\it Aquarius} simulation suite to show that unresolved streams are unlikely to significantly affect the smoothed results found in high resolution cosmological simulations (see also \citealt{2011MNRAS.418.2648F}). This is because of the sheer number of criss-crossing streams ($\sim 10^{14}$) that co-add to make the distribution very smooth. The result is rather fortunate. Massive streams that could affect the velocity PDF are rare and in any case detectable because of their accompanying stars; lower mass streams that may be undetectable due to a lack of accompanying stars are common and, as a result, co-add to make the velocity PDF smooth. Caustics (regions of extremely high density caused by foliations of the dark matter phase sheet) appear to be similarly unimportant \citep{2009MNRAS.400.2174V}.

\paragraph{Unresolved substructure} \citet{2008arXiv0801.3269K} discuss the possibility that we lie within a small dark matter subhalo, significantly increasing $\rhodmlab$ with respect to $\rhodm$. While this can occur, the probability that we lie on top of such a subhalo is quite small. \citet{2008arXiv0801.3269K} derive a density PDF for $\rhodmlab$ that has a peak at $\rhodmlab < \rhodm$, with a power law tail to high density caused by subhalos. The peak is {\it lower} than $\rhodm$ because of mass conservation. If we move more dark matter into substructures, then the tail to high density is boosted because there are more dense substructures, but the peak of the density PDF is shifted to lower density because there is less mass in the remaining smooth component. Since we are most likely to lie at or near the peak of the distribution, substructure halos have the effect, statistically, of {\it reducing} $\rhodmlab$ with respect to $\rhodm$. \citet{2008arXiv0801.3269K} extrapolate mass functions from $N$-body simulations down to the free streaming scale. Assuming a total mass fraction in substructure of 10\%, the peak of the density PDF for $\rhodmlab$ is only very slightly shifted to $\sim 0.9\rhodm$, while the probability that $\rhodmlab$ is larger than $\rhodm$ is very small. 

\paragraph{Solar system capture \& scattering} Finally, the effect of scattering within the Solar system is also likely to be small \citep{2009PhRvD..79j3531P}, once both the capture {\it and ejection} of dark matter particles is taken into account \citep{2010arXiv1004.5258E}.\\

\noindent
In conclusion, current state-of-the-art DMO simulations that achieve a spatial resolution of $\sim 20$\,pc appear to be adequate for making predictions for both $\rhodm$ and $\rhodmlab$, under the assumption that baryons do not significantly alter the dark matter distribution. However, this assumption is most likely a poor one, as I discuss next.

\subsection{The effect of baryons}\label{sec:baryons}

While the DMO simulations are well understood, when including `baryonic' matter (stars and gas) the simulations become significantly more complex \citep[e.g.][]{2008arXiv0801.3845M}. At present, the state-of-the art still leaves important physics below the resolution limit -- so-called `sub-grid' physics -- leading to large discrepancies between groups \citep{2008arXiv0801.3845M,2012MNRAS.423.1726S}. However, this situation is set to improve rapidly as both software algorithms and hardware improve \citep[e.g.][]{2011EPJP..126...55D}. Recent simulations have now passed a critical resolution threshold of $\sim 100$\,pc that allows the most massive star forming regions to be resolved \citep{2011ApJ...742...76G,2011MNRAS.410.1391A,2013arXiv1311.2073H}, as well as beginning to resolve the scale height of the Milky Way thin disc ($\sim 200$\,pc) for the first time. The most massive star forming regions are where the majority of massive stars explode as supernovae, returning heat and metals to the inter-stellar medium (ISM). This stellar {\it feedback} appears to be critical in forming galaxies that match the observed properties of real galaxies in the Universe \citep[e.g.][]{2008arXiv0801.3845M,2011ApJ...742...76G,2011MNRAS.410.1391A,2013arXiv1311.2073H}, though at present rather strong feedback -- where a significant fraction of the available SNe energy couples very efficiently to the surrounding gas -- appears to be required \citep[e.g.][]{2008Sci...319..174M,GovernatoEtAl2010,2011ApJ...742...76G,2013MNRAS.429.3068T}. Such feedback is not yet problematic given our uncertainties in how feedback operates \citep[e.g.][]{2013ApJ...770...25A}, but more work needs to be done on modelling the small scale physics and its coupling to larger scales to determine whether or not feedback can really regulate the growth of galaxies, or whether we are missing some important ingredient in our cosmological model.  

\subsubsection{Qualitative predictions}

While we are currently unable to make strong {\it predictions} when including baryonic processes, we can still study the expected changes to the DMO predictions in a more qualitative manner using the latest simulations. I discuss the key results from these here. 

\paragraph{Most of the local mass near the Sun is in baryons, not dark matter}
The first important point to realise is that although we expect (and indeed observe) a significant amount of dark matter in our galaxy, the amount of dark matter expected in the vicinity of the Sun is actually rather small. This is because gas is a dissipative fluid. Unlike dark matter, gas can condense to form a rotationally supported disc that dominates the local gravitational potential. We can estimate the approximate about of dark matter expected in the vicinity of the Sun from the rotation curve assuming spherical symmetry. The enclosed mass at the Solar position $R_0 \sim 8$\,kpc is given by:

\begin{equation} 
M_{\rm dm}(R_0) \sim \frac{v_c^2 R_0}{G} - M_d
\end{equation} 
where $G$ is Newton's gravitational constant; $v_c \sim 220$\,km/s is the local circular speed \citep{2012ApJ...759..131B,2012MNRAS.427..274S,2013arXiv1307.6073G}; and $M_d \sim 6\times 10^{10}$\,M$_\odot$ is the mass of the Milky Way stellar disc \citep[e.g.][]{BinneyTremaine2008}. This gives $M_{\rm dm}(R_0) \sim 3\times10^{10}$\,M$_\odot$. Thus, {\it about half of the mass of the Milky Way interior to $R_0$ is actually in baryons rather than dark matter} (see e.g. \citealt{2002ApJ...573..597K} for a more detailed analysis that arrives at the same conclusion). As we approach the disc plane, this becomes even more extreme. The scale height of the Milky Way thin disc is $z_0 \sim 200$\,pc, with most of the disc mass lying within $\sim 500$\,pc \citep[e.g.][]{BinneyTremaine2008}. Thus, assuming a halo like that simulated in \citet{2008MNRAS.391.1685S} normalised to the Milky Way rotation curve, dark matter comprises just $\sim$ one tenth of the mass in the Solar neighbourhood volume ($8 < R_0 < 9$\,kpc; $|z| < 500$\,pc).

The above makes hunting for the gravitational effect of dark matter near the Sun rather like looking for the proverbial needle in the haystack. This is one motivation for using extrapolations from larger scales where the dark matter dominates the potential. We are left in the end with a trade-off. We can average over large volumes over which we will see significant dark matter, but be necessarily less `local', or we can average over a very small volume near the Sun, but be significantly more sensitive to our assumed baryonic mass model. I discuss this further in \S\ref{sec:theory}. 

\begin{figure}[t]
\begin{center}
\hspace{-4mm}
\includegraphics[width = 0.99\textwidth]{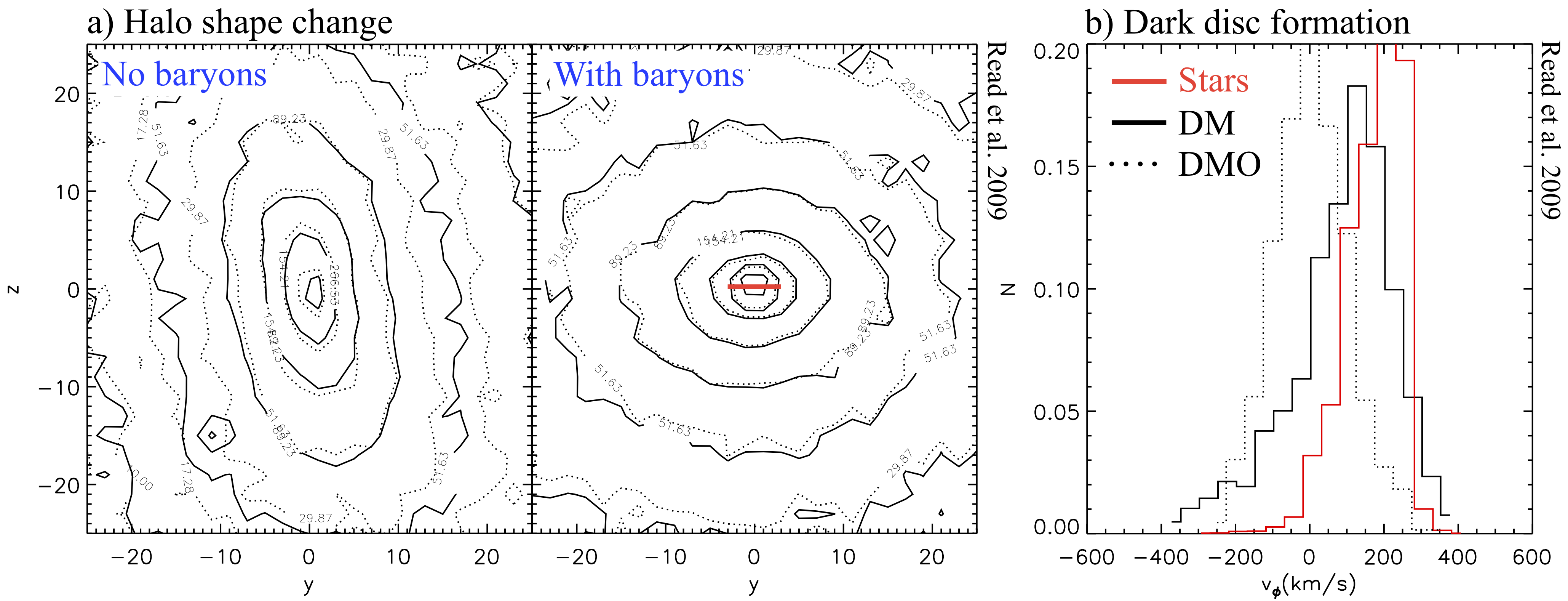}
\end{center}
\vspace{-8mm}
\caption{\small Including baryons in the cosmological simulations alters the predictions for $\rhodm$. {\bf a)} Adding dissipative baryonic matter causes the dark matter halo become oblate and aligned with the disc (red horizontal line; \citealt{2009MNRAS.397...44R}). {\bf b)} The presence of a massive disc at high redshift biases the accretion of satellites causing their tidal debris -- both stars and dark matter -- to settle into a rotating disc. This plot shows the distribution function of rotational velocity in the disc plane $v_\phi$ for the LPM simulation (Table \ref{tab:darkdisc}; \citealt{2009MNRAS.397...44R}). Without baryons (DMO; dotted), the dark matter distribution is well-fit by a single Gaussian. Including baryons (DM; black), it is skewed towards the rotating stellar disc (red); it is well-fit by a double Gaussian. This is a particularly extreme example since the LPM simulation had a massive near-planar merger at redshift $z\sim 1$.}
\label{fig:baryonsims} 
\end{figure}

\paragraph{Cusp-core transformations and halo shape change}\label{sec:barcuspcoreshape}
As gas collects and dominates the central potential of galaxies, it can cause a physical rearrangement of the dark matter distribution (simply through the gravitational interaction). Dark matter can contract in response to gas condensation \citep{1980ApJ...242.1232Y,1986ApJ...301...27B}, or even expand if energetic supernovae, or active galactic nuclei eject a significant amount of mass \citep{1996MNRAS.283L..72N}. This latter process needs to repeat multiple times for the effects to be significant \citep{2005MNRAS.356..107R,MashchenkoEtAl2006,2012MNRAS.421.3464P,2013MNRAS.429.3068T,2014Natur.506..171P}. But if it does act, it will gradually transform dark matter cusps, predicted by DMO simulations (\S\ref{sec:dmosims}), into constant density dark matter cores. Such cores have been observed in dwarf galaxies for over two decades now \citep[e.g.][]{1994Natur.370..629M,1994ApJ...427L...1F}, lending support to such an idea. Further observational evidence has come more recently. If such cusp-core transformations occur, then the star formation history of dwarf galaxies should be bursty with a duty cycle of $\sim 250$\,Myrs, while their stars should be similarly heated leading to -- at least in the older stellar populations -- a significant vertical dispersion. Both of these predictions are consistent, and perhaps even favoured, by the latest data \citep{2013MNRAS.429.3068T}. Such processes may even be important for galaxies as massive as the Milky Way \citep{2010MNRAS.407....2D,2012ApJ...744L...9M}. 

Gas condensation also alters the shape of dark matter halos making them oblate and aligned with the disc, at least within $\sim 10$ disc scale lengths \citep[][and see Figure \ref{fig:baryonsims}a]{1991ApJ...377..365K,1994ApJ...431..617D,2007arXiv0707.0737D,2009MNRAS.397...44R}. This has three important effects on $\rhodm$. Firstly, it makes assumptions of spherical symmetry for our Galaxy not unreasonable, despite the expectation in a DMO Universe that halos are triaxial \citep[][and see \S\ref{sec:dmopredictions}]{1991ApJ...378..496D,1992ApJ...399..405W,1996ApJ...462..563N,2002ApJ...574..538J}. This means that spherical extrapolations from the rotation curve $\rhodmext$ could give a reasonable estimate of $\rhodm$ (see Figure \ref{fig:combinedprobes}). Secondly, a more spherical halo significantly reduces the expected scatter in $\rhodm$ at the Solar neighbourhood \citep[][and see discussion in \S\ref{sec:dmorhodm}]{2010PhRvD..82b3531P}. Thirdly, oblate halos enhance $\rhodm$.  We can think of this enhancement as coming from a contraction of the dark matter halo due to the addition of a massive stellar disc. \citet{2012ApJ...756...89B} use a back-of-the-envelope calculation to argue that for the Milky Way, this enhancement should be about $\sim 30$\%. This matches recently published numerical results remarkably well \citep{2010PhRvD..82b3531P,2013arXiv1308.1703K}. 

\paragraph{The formation of a `dark disc'}\label{sec:darkdisctheory}
Finally, if a star/gas disc is already in place at high redshift then it will bias the further accretion of subhalos towards the disc plane. This is a result of momentum exchange due to gravitational scattering between the satellite and the disc stars: `dynamical friction' \citep[e.g.][]{BinneyTremaine2008}. The frictional force goes as: 

\begin{equation} 
M \dot{\bf v} = C \frac{\rho M^2}{v^3}{\bf v} 
\end{equation} 
where $M$ is the mass of the satellite; $\dot{\bf v}$ is the deceleration due to dynamical friction; $\rho$ is the background density (i.e. stars, gas, dark matter etc.); $C$ is some constant of proportionality; and $v = |{\bf v}|$ is the velocity of the satellite relative to the background\footnote{Apart from especially resonant situations, the above formula that owes to \citet{Chandrasekhar1943} works remarkably well \citep{2006astro.ph..6636R}.}. 

Assuming a disc density (see \S\ref{sec:theory}):

\begin{equation}
\rho = \rho_0 \exp(-|z|/z_0)
\end{equation}
and assuming that the satellite travels on a straight line at velocity $v$ through the disc, then its change in velocity over a single passage is given by: 

\begin{equation}
\Delta v =  |\dot{\bf v}| \Delta t =  |\dot{\bf v}| \frac{2z_0}{v} \simeq \frac{2 C \rho_0 z_0 M}{v^3}
\label{eqn:discfriction}
\end{equation}
This frictional force acts to drag the most massive satellites down towards the disc plane, leading to an accreted disc that contains both stars and dark matter \citep[][and see Figure \ref{fig:baryonsims}b]{1989AJ.....98.1554L,2008MNRAS.389.1041R,2009MNRAS.397...44R,2009ApJ...703.2275P,2010JCAP...02..012L,2013arXiv1308.1703K}. 

There are three important points to note from equation \ref{eqn:discfriction}. Firstly, the force depends on the satellite mass $M$ and so will only be important for the most massive mergers \citep{2008MNRAS.389.1041R}. Secondly, the force is approximately proportional to the product of the disc scale height and central density: $2 \rho_0 z_0$, which is nothing more than the disc surface density: 

\begin{equation}
\Sigma = 2 \int_0^\infty \rho(z)dz = 2\rho_0 z_0
\end{equation}
Thus, even if simulations do not properly resolve $z_0$ (most cosmological simulations do not), they can still largely capture the disc-plane dragging process correctly so long as they correctly capture $\Sigma$ \citep{2009MNRAS.397...44R}. 

Finally, notice that the friction force goes as $1/v^2$, where $v = |{\bf v}| \simeq |{\bf v_{\rm sat}} - {\bf  v_{\rm disc}}|$ is the {\it difference} in velocity between the satellite and the background. Thus, the friction is significantly enhanced for satellites that co-rotate with the disc. For this reason, we expect the accreted disc stars and dark matter to largely co-rotate \citep{2008MNRAS.389.1041R,2009MNRAS.397...44R}. Retrograde accreted material must also be present, but it is most likely to be sub-dominant to the prograde material, particularly as we approach the Solar neighbourhood.

\begin{table}
\begin{center}
\resizebox{0.99\textwidth}{!}{%
\begin{tabular}{llcccc}
\thickhline
\multicolumn{2}{l}{\bf Description} & $\rho_{\rm dd}/\rhodm$ & $\sigma_{\rm dd}$(km/s) & $v_{\rm rot,dd}$(km/s) & $(\rhodm - \rhodmext)/\rhodmext$\\
\thickhline
\multicolumn{1}{l |}{Q} & Quiescent (MW1) & 0.23 & 50 &  54 & 0.175 \\
\multicolumn{1}{l |}{LM} & Late Mergers (h204) & 1.1 & 76 & 144 & 0.35 \\ 
\multicolumn{1}{l |}{LPM} & Large ($\sim$1:1) Planar Merger (h258) & 1.65 & 88 & 140 & 0.47\\
\thickhline
\end{tabular}
}
\end{center}
\vspace{-5mm}
\caption{\small Dark disc properties for three numerical simulations of Milky Way mass galaxies taken from \citet{2009MNRAS.397...44R}. The original simulation labels are given in brackets; I use the more descriptive labels Q, LM and LPM in this review. Each galaxy had a rather different merger history: Q was rather Quiescent with no massive mergers since redshift $z=2$; LM had two Large Mergers at $z < 0.5$; and LPM had a Large near-Planar Merger at $z \sim 1$ ($\sim 8$\,Gyr ago). The columns show: a description of the simulation; the dark disc to smooth halo density ratio averaged over $|z| < 2.1$\,kpc; $7 < R < 8$\,kpc; the vertical velocity dispersion of the dark disc; the rotational velocity of the dark disc; and the ratio of the local to extrapolated dark matter density evaluated at $7 < R < 8$\,kpc (see text for details).}
\label{tab:darkdisc}
\end{table}

\citet{2008MNRAS.389.1041R} and \citet{2009MNRAS.397...44R} estimate that the dark disc should contribute $\sim 0.25 - 1.5$ times $\rhodm$ from the non-rotating smooth halo in our current cosmology, depending on the (rather uncertain) merger history and mass of our Galaxy. The dark disc also changes the velocity PDF of dark matter particles, producing a distribution that is better-fit by a double rather than single Gaussian, with interesting implications for both direct and indirect dark matter particle searches \citep{2009PhLB..674..250B,2009ApJ...696..920B}. Table \ref{tab:darkdisc} summarises the range of dark disc properties found by \citet{2009MNRAS.397...44R} for three Milky Way mass galaxies with rather different merger histories, as marked. The most quiescent galaxy Q has a rather puny dark disc that contributes just $\sim 20$\% to $\rhodm$, while the LPM simulation has a massive $\sim$1:1 near-planar merger that produces a dark disc that dominates $\rhodm$. This latter simulation introduces an alternate dark disc formation mechanism: if the mass ratio of the merger is small enough, then a gas rich merger can {\it define} the resultant disc plane, leading to a very significant dark disc \citep{2009MNRAS.397...44R}. Such a scenario is not immediately implausible for the Milky Way. The LPM merger occurred at redshift $z \sim 1$ which corresponds to $\sim 8$\,Gyr ago in our current cosmology. This is about the age separation of the Milky Way thin and thick discs (if there are indeed such distinct entities \citealt{2012ApJ...751..131B}). Thus, any stellar heating induced by the merger could be hidden entirely in the thick disc stars -- perhaps even explaining the origin of the thick disc. 

While the ratio $\rho_{\rm dd}/\rhodm$ is of great interest for direct dark matter detection experiments (\S\ref{sec:introduction}), it is difficult to measure directly. Much more accessible is a comparison of the local to extrapolated dark matter density (c.f. Figure \ref{fig:combinedprobes}):

\begin{equation} 
\zeta = (\rhodm - \rhodmext)/\rhodmext
\end{equation} 
For $\zeta > 0$, we have a flattened halo near the disc plane and/or a dark disc, while $\zeta < 0$ implies a prolate halo. To calculate $\zeta$ from the simulation data, I average $\rhodm$ over $|z| < 0.5$\,kpc and $7 < R < 8$\,kpc; and calculate $\rhodmext$ from the cumulative enclosed dark matter mass assuming spherical symmetry: 

\begin{equation} 
\rhodmext = \frac{M_{\rm dm}(R_2) - M_{\rm dm}(R_1)}{4\pi \overline{R}^2 \Delta R}
\end{equation}
where $R_2 = 8$\,kpc; $R_1 = 7$\,kpc; $\overline{R} = 7.5$\,kpc; and $\Delta R = R_2 - R_1$. I compare the values of $\zeta$ for the Q, LM and LPM simulations to real data for the Milky Way in \S\ref{sec:shapeddcomp}.

The above findings for dark discs have largely been confirmed by more recent works \citep{2009ApJ...703.2275P,2010JCAP...02..012L,2013arXiv1308.1703K}; however, there is some significant debate about how quiescent the merger history of our Galaxy was. The {\it Eris} simulation explored by \citet{2013arXiv1308.1703K}, for example, has a particularly quiescent merger history as compared to typical dark matter halos of similar mass. \citet{2009ApJ...703.2275P} argue that this must be so, as otherwise mergers would dynamically over-heat the Milky Way thick stellar disc. However, such heating is reduced if mergers are of lower inclination and orbital eccentricity (exactly the same mergers that give rise to significant dark discs; \citealt{2008MNRAS.389.1041R}); or if gas -- not present in the \citet{2009ApJ...703.2275P} models -- is included \citep{2010MNRAS.403.1009M}.

Turning the above around, however, if it can be demonstrated that the Milky Way has a rather puny dark disc, then the implication is that its merger history must have indeed been rather quiescent. I discuss the possibility of empirically constraining the dark disc -- and therefore the merger history of our Galaxy -- next.  

\subsubsection{Hunting for the Milky Way's dark disc}\label{sec:darkdischunting}

One approach to constrain the Milky Way's dark disc is to hunt for the stars that must have been accreted along with it. These should show distinct chemistry and kinematics from the in-situ Milky Way population leading to the hope that they can be detected \citep{2008MNRAS.389.1041R}. A second possibility for detecting the `dark disc' is via its dynamical influence -- i.e. its contribution to $\rhodm$ \citep{2008MNRAS.389.1041R,2012MNRAS.425.1445G}. The expected scale height of the dark disc is large ($2-3$\,kpc) and so, unless measurements probe high up above the Galactic disc, the approximation that $\rhodm$ is constant over the Solar neighbourhood remains reasonable even when considering the dark disc \citep{2008MNRAS.389.1041R}. This means, however, that the `dark disc' will likely degenerate with the flattened oblate halo that is expected due to adiabatic contraction of the dark matter halo (see \S\ref{sec:barcuspcoreshape}). By combining measures of $\rhodm$ and $\rhodmext$ extrapolated from the rotation curve (Figure \ref{fig:combinedprobes}) with chemo-dynamic Galactic `archaeology' in the Milky Way, we can hope to break this degeneracy. There are several interesting scenarios: 

\begin{enumerate} 
\item $\rhodm < \rhodmext$. In this case, there is no dark disc and the dark matter halo is likely {\it prolate}. This would have interesting implications for $\Lambda$CDM cosmology and/or galaxy formation theories as such a scenario is not expected. It would also essentially rule out weak field alternative gravity theories that require the gravitational potential to share symmetry properties with the disc \citep{2004ApJ...610L..97H,2005MNRAS.361..971R}.

\item $\rhodm \simeq \rhodmext$. In this case the dark matter halo is near-spherical and there is no significant `dark disc'. This implies a rather quiescent merger history for the Milky Way \citep{2008MNRAS.389.1041R,2009ApJ...703.2275P,2013arXiv1308.1703K}.

\item $\rhodm > \rhodmext$. This implies either an oblate/squashed dark matter halo and/or a `dark disc'. The degeneracy between these two scenarios can also be broken with improved data:

\begin{enumerate}
\item An oblate halo will additionally show flattening far from the disc plane. There may already be hints of such a flattening in the tidal debris of satellites orbiting around the Milky Way \citep[e.g.][]{2012MNRAS.424L..16L} and in the kinematics of distant Milky Way `halo stars' \citep[e.g.][]{2012arXiv1209.2708L}. Neither probe is conclusive at present. However, relatively small improvements in data promise significantly improved constraints \citep{2013arXiv1308.2235L}.
\item A `dark disc' can be found via the stars that are accreted with it. These should show distinct chemistry and kinematics from the underlying in-situ disc population.
\end{enumerate}  
\end{enumerate} 

We explore which of the above scenarios, given current data, is most likely for the Milky Way in \S\ref{sec:measurements}. 

\subsubsection{Towards ab-initio simulations including baryonic physics} 

Ideally, we would like to be able to make robust quantitative predictions from numerical simulations that model both the dark matter and baryonic fluids. While this remains a significant challenge, resolving the correct spatial locations of the most massive star forming regions within galaxies ($\sim 100$\,pc) is a key milestone that we have recently passed \citep{2011ApJ...742...76G,2011MNRAS.410.1391A}. For this reason, we can expect that the next generation of galaxy formation simulations will be significantly more predictive \citep[e.g.][]{2013arXiv1308.2669K}.

\section{Mass modelling theory}\label{sec:theory}

In this section, I briefly review the theory behind calculating the gravitational potential from an equilibrium distribution of `tracer' stars moving in that potential. I focus mainly on stellar tracers in this review, discussing gas briefly in \S\ref{sec:gas}.

A population of tracer stars obeys the collisionless Boltzmann equation: 

\begin{equation}
\frac{df}{dt} = \frac{\partial f}{\partial t} + \nabla_x f\cdot{\bf v} - \nabla_v f\cdot \nabla_x \Phi = 0
\label{eqn:colboltz}
\end{equation}
where $f({\bf x},{\bf v})$ is the stellar distribution function; ${\bf x}$ and ${\bf v}$ are the positions and velocities, respectively; and $\Phi$ is the gravitational potential. 

Assuming Newtonian weak field gravity, the force $\nabla_x \Phi$ is related to the total mass density $\rho$ (stars, gas, dark matter etc.) through Poisson's equation:

\begin{equation}
\nabla_x \cdot \nabla_x \Phi = \nabla_x^2 \Phi = 4\pi G \rho
\label{eqn:poisson}
\end{equation}

If the system is in dynamic equilibrium (steady state), then we may neglect the partial time derivative of $f$ in equation \ref{eqn:colboltz}. This may not be a good approximation for the Milky Way if it has been recently bombarded by a satellite, or if the chosen tracers are not dynamically `well mixed' in the disc. I discuss the choice of tracer stars in \S\ref{sec:tracers}; and recent evidences for disequilibria in the Milky Way disc in \S\ref{sec:disequilibria}.

Assuming equilibrium tracers for now, we drop the $\partial f/\partial t$ term. With this assumption -- and armed with a measurement of the phase space distribution function $f$ of our tracers -- in principle, we can directly measure the gravitational force $\nabla_x \Phi$ by solving equation \ref{eqn:colboltz}. In practice, however, this is hard because $f$ is six-dimensional (even a million stars gives only 10 sample points per dimension) and we need to estimate the (noisy) partial derivatives of $f$. There are several solutions to this problem, each with advantages and disadvantages. I detail these, next. 

\subsection{Distribution function modelling} 

In distribution function modelling, we write down some parameterised (but possibly rather general) functional form for $f({\bf x}, {\bf v})$. With a particular form in mind, the derivatives may be calculated either analytically or numerically without noise being an issue. Furthermore, since $f$ -- appropriately normalised -- is really just a probability density distribution, we can directly calculate the likelihood of the data given the model: 

\begin{equation} 
\mathcal{L} = \prod_i \frac{f({\bf x}_i, {\bf v}_i)}{\int d^3 {\bf x} d^3 {\bf v} f({\bf x}, {\bf v})}
\label{eqn:likelihood}
\end{equation}
where the product is over all stars $i$ with phase space position $[{\bf x}_i,{\bf v}_i]$, while the integral is over the full distribution function. A useful trick is to take the logarithm of equation \ref{eqn:likelihood} that transforms the product into a more computationally manageable sum. 

The advantages of such an approach are: i) we can directly model discrete data; and ii) we maximise the information content in the data by using the full shape information in the distribution function. The key disadvantage is that we must assume some form for $f$ up-front. If our choice(s) for $f$ do not include the correct solution, then we will obtain biased results no matter what quality or abundance of data are available (I give an example of this in \S\ref{sec:tests}). Furthermore, it can often be very difficult to work out when this is happening. 

One way to combat the above is to make $f$ as general as possible. There are several approaches that may be considered as variants of one-another. I briefly describe these next, before shifting to moment methods (\S\ref{sec:jeans}) that are the main focus of this review. 

\subsubsection{`Schwarzschild' or orbit modelling}\label{sec:schwarzschild} 

In Schwarzschild modelling, we model the distribution function as a linear combination of many stellar {\it orbits} \citep{Schwarzschild1979}. Starting with some assumed gravitational potential $\Phi$, we build an {\it orbit library}: a large collection of representative orbits within this potential. This is usually comprised of regular orbits, though chaotic orbits can also be modelled as a constant additive phase space contribution \citep[e.g.][]{BinneyTremaine2008,1996MNRAS.283..149Z}. The observed distribution of stars is then fit using a weighted sum over these orbits \citep[for recent examples, see:][]{2008MNRAS.385..614V,2008MNRAS.385..647V,2013MNRAS.434.3174V}.

Schwarzschild modelling has the advantage that the distribution function is directly constrained by the data in an essentially parameter free way (once the potential is prescribed). The disadvantages are mostly due to the computational cost of exploring a wide range of models. For discrete data, we require a large number of orbits to properly span the phase space (error-free data formally require infinitely many orbits; \citealt{2013arXiv1303.6099M}); while for each trial potential, we must begin over building the orbit-library from scratch. \citet{2013MNRAS.433.1411M} have argued recently that the intrinsic noise in the method owing to the finite number of orbits within the library could be a major barrier for exquisite data, unless the data are binned (for a discussion of the perils and pitfalls of binning data, see \S\ref{sec:jeans}). Furthermore, moving to libraries with an enormous number of orbits can lead to the danger of over-fitting noise in the data.

\subsubsection{Made to Measure (M2M)}\label{sec:mtwom}

The made to measure (M2M) method was first proposed by \citet{1996MNRAS.282..223S}. At heart, it is really an $N$-body method. However, it is different from typical $N$-body techniques in that each star has a constantly evolving orbit weight that pushes the simulated $N$-body system towards the real data. The idea is to maximise a {\it merit function} \citep{2009MNRAS.395.1079D}:

\begin{equation}
Q = \mu S - \frac{1}{2} C
\end{equation}
where $C$ is some {\it constraint function} that measures the goodness of fit; $\mu$ is a Lagrange multiplier; and $S$ is some {\it penalty function} that forces us towards a single optimal solution; more on this shortly. The functions $C$ and $S$ are a matter of choice, but typically $C$ is a $\chi^2$-like measure: 

\begin{equation} 
C = \sum_j^n \left(\frac{Y_j - y_j}{\sigma_j}\right)^2
\end{equation} 
where $Y_j$ are the data values with uncertainties $\sigma_j$; and $y_j = \sum_i w_i K_j ({\bf x}_i, {\bf v}_i)$ are moments of the model weighted by a smoothing kernel $K_j$ and some individual weights $w_i$ (typically, a time averaged weight is used to avoid oscillating solutions; \citealt{2009MNRAS.395.1079D}); and $S$ is a pseudo-entropy: 

\begin{equation}
S = -\sum_i \hat{w}_i \log\left(\frac{\hat{w}_i}{p_i}\right)
\end{equation} 
where $\hat{w}_i = w_i / \sum_j w_j$ are normalised weights, and $p_i$ are {\it priors} on these weights.

The basic idea is then to solve the motion of the particles as a usual $N$ body problem: 

\begin{equation} 
\ddot{\bf x}_i = \nabla_x \Phi
\end{equation} 
where the potential $\Phi$ and accelerations $\nabla_x \Phi$ are calculated using standard numerical techniques \citep[e.g.][]{2011EPJP..126...55D}, while evolving the weights $w_i$ with time to maximise $Q$:

\begin{equation}
\dot{w}_i = \epsilon w_i \frac{\partial Q}{\partial w_i} 
\end{equation}
where $\epsilon$ is a normalisation parameter.

Modern implementations of the M2M method include: \citet{2004ApJ...601L.155B}, \citet{2007MNRAS.376...71D}, \citet{2009MNRAS.392..904R}, \citet{2009MNRAS.395.1079D}, \citet{2010MNRAS.405..301L} and \citet{2013MNRAS.430.1928H}. Each of these authors have extended and adapted the above classic methodology mainly to cope with the problem of orbit weight convergence. 

The key advantage of M2M is that it naturally avoids assumptions about the form or shape of the gravitational potential, or the distribution function. Unlike the Schwarzschild method, the potential is fit simultaneously along with the orbit weights. However, it shares many of the same issues as Schwarzschild modelling. Searching through many models can be slow since M2M converges only on one `best' solution; there may be others that are equally good \citep{2009MNRAS.395.1079D}. There is a danger that solutions will not converge \citep{2009MNRAS.395.1079D} and, as with Schwarzschild, there is a danger of over-fitting noise in the data \citep{2007MNRAS.376...71D}. However, most of these issues will continue to improve with time as software and hardware algorithms improve \citep[e.g.][]{2011EPJP..126...55D}. Indeed, this is what has driven a sudden interest in the method -- largely untouched since \citet{1996MNRAS.282..223S} -- over the past few years.

\subsubsection{Action modelling}\label{sec:torus}

The Jeans theorem states that for regular orbits -- and assuming a steady state galaxy -- the distribution function may be written in terms of isolating integrals \citep[e.g.][]{BinneyTremaine2008}. A particularly useful choice of canonical coordinates for the isolating integrals are the {\it Action-Angle} variables \citep[e.g.][]{BinneyTremaine2008,2013arXiv1309.2794B}. These have the useful property that the actions ${\bf J}$ are conserved along each orbit, while the angles ${\bm \theta}$ increase linearly with time. From Hamilton's equations, we have: 

\begin{equation}
\dot{\bf J} = \frac{\partial H}{\partial {\bm \theta}} = 0 \,\,\,\, ; \,\,\,\, \dot{\bm \theta} = \frac{\partial H}{\partial {\bf J}} = {\bf \Omega}({\bf J}) = {\rm const.} 
\end{equation} 
$\Rightarrow$
\begin{equation} 
{\bf J} = {\rm const.} \,\,\,\, ; \,\,\,\, \theta_i = \theta_{0,i} + \Omega_i t
\end{equation}
where $H$ is the Hamiltonian. 

In one dimension, the constant action and linearly increasing angle maps out a circle in phase space. In two dimensions, this becomes a {\it torus}; while in three dimensions, it is a 3-torus (recall that a circle is a 1-torus). 

By the Jeans theorem, we can write the distribution function solely in terms of these actions: $f \equiv f({\bf J})$. Thus, once the orbital actions for a set of stars are known, the full distribution function is immediately known. This is a key strength of action modelling\footnotemark. Like other methods, however, it also has some disadvantages. Firstly, the map from the observables $[{\bf x}, {\bf v}]$ to the Actions $[{\bf J},{\bm \theta}]$ and visa-versa is non-trivial. Simple solutions are known for separable {\it St\"ackel} potentials \citep{stackel,1985MNRAS.216..273D}, but more general potentials require a numerical solution. One potential approach is {\it torus modelling}, where orbital tori in a general Galactic potential are fit by warping known tori from a simple toy potential \citep{1994MNRAS.268.1033K,2012MNRAS.426..128S,2013arXiv1309.2794B}. A full solution for general potentials has not yet been presented, but may be achievable as an extension of existing techniques \citep{2013arXiv1309.2794B}. Secondly, only regular orbits can be modelled in this way. \citet{2013arXiv1309.2794B} cast this as an advantage in that it allows us to study the departure from regularity in a controlled manner. Perturbation theory about the best-fitting regular model, for example, has already proven to be able to recover the behaviour of irregular orbits in the case of a planar logarithmic potential \citep{1994MNRAS.268.1041K}.

\footnotetext{Action modelling is also very promising for studies of tidal debris, since the locus of debris material in action space is rather simple, while in configuration space it can be rather complex \citep[e.g.][]{2011MNRAS.413.1852E,2013MNRAS.433.1813S,2013arXiv1308.2235L}.}

\citet{2012MNRAS.426.1324B} have recently introduced a useful approximation for calculating actions in potentials that are close to St\"ackel form. This was applied to fit a simple parameterised distribution function to Solar neighbourhood data in \citet{2012MNRAS.426.1328B}, illustrating the power of such an approach. The axisymmetric distribution function is assumed to take a `quasi-isothermal' form: 

\begin{equation}
f(J_r,J_z,L_z) = \frac{\Omega \Sigma \epsilon}{2 \pi^2 \sigma_r^2 \sigma_z^2 \kappa}\left[1 + \tanh(L_z/L_0)\right]e^{-\kappa J_z / \sigma_r^2} e^{-\epsilon J_z / \sigma_z^2}
\label{eqn:simpledistbinney}
\end{equation} 
where $J_r, J_z$ are the radial and vertical actions, respectively; $L_z$ is the specific angular momentum of orbits within the disc plane; and $\Omega(L_z), \kappa(L_z)$ and $\epsilon(L_z)$ are the circular, radial and vertical epicyclic frequencies set by the gravitational potential. Under the epicycle approximation of near-circular orbits, these are given by \citep[e.g.][]{BinneyTremaine2008}:

\begin{equation}
\Omega^2 = \frac{L_z^2}{R^4} \,\,;\,\, \kappa^2 = \left(R\frac{d\Omega^2}{dR} + 4\Omega^2\right)_{R_c,0} \,\,;\,\, \epsilon^2 = \left(\frac{\partial^2\Phi}{\partial z^2}\right)_{R_c,0}
\end{equation} 
We must then further specify a form for the disc surface density $\Sigma$, the functions $\sigma_r(L_z)$ and $\sigma_z(L_z)$, and the gravitational potential $\Phi$. Some simple choices for these (exponentials for $\Sigma, \sigma_r$ and $\sigma_z$; and a \citet{DehnenBinney1998} model for the potential) are adopted in \citet{2012MNRAS.426.1328B}. The `St\"ackel action' approximation is then required in order to map the observables $[{\bf x}, {\bf v}]$ onto the actions ${\bf J}$ that appear in equation \ref{eqn:simpledistbinney} for a given potential $\Phi(R,z)$ \citep{2012MNRAS.426.1324B}. 

This same model has also been used recently by \citet{2013arXiv1309.0809B} to measure the surface density of the Milky Way disc over a range of radii ($4.5 < R < 9$\,kpc), for the first time. I discuss these measurements in \S\ref{sec:measurements}. 

\subsection{Moment methods: the Jeans equations}\label{sec:jeans}

A completely different approach to distribution function modelling is to take instead {\it moments} of equation \ref{eqn:colboltz}. Casting the steady state collisionless Boltzmann equation (equation \ref{eqn:colboltz} without the $\partial f/\partial t$ term) in cylindrical polar coordinates $[R,\phi,z]$, we have \citep[e.g.][]{BinneyTremaine2008}: 

\begin{equation} 
v_R \frac{\partial f}{\partial R} + \frac{v_\phi}{R}\frac{\partial f}{\partial \phi} + v_z \frac{\partial f}{\partial z} - \left(\frac{\partial \Phi}{\partial R} - \frac{v_\phi^2}{R}\right)\frac{\partial f}{\partial v_R} - \frac{1}{R}\left(v_Rv_\phi + \frac{\partial \Phi}{\partial \phi}\right)\frac{\partial f}{\partial v_\phi} - \frac{\partial \Phi}{\partial z}\frac{\partial f}{\partial v_z} = 0 
\end{equation} 
Multiplying through $v_R$, $v_\phi$ or $v_z$ and integrating over all velocities derives the three Jeans equations \citep{1922MNRAS..82..122J,BinneyTremaine2008}: 

\begin{equation} 
\frac{\partial(\nu \sigma_R^2)}{\partial R} + \frac{\partial (\nu \sigma_{Rz})}{\partial z} + \nu\left(\frac{\sigma_R^2 - \sigma_\theta^2}{R} + \frac{\partial \Phi}{\partial R}\right) = 0 \hspace{21mm} R-\mathrm{Jeans}
\label{eqn:Rjeans}
\end{equation} 
\begin{equation} 
\frac{1}{R^2}\frac{\partial(R^2\nu\sigma_{R\phi})}{\partial R} + \frac{\partial (\nu\sigma_{\phi z})}{\partial z} = 0 \hspace{49mm} \phi-\mathrm{Jeans}
\label{eqn:phijeans}
\end{equation} 
\begin{equation} 
\frac{1}{R}\frac{\partial\left(R\nu \sigma_{Rz}\right)}{\partial R} + \frac{\partial }{\partial z}\left(\nu\sigma_z^2\right) + \nu\frac{\partial \Phi}{\partial z} = 0 \hspace{37mm} z-\mathrm{Jeans}
\label{eqn:zjeans}
\end{equation}
where:
\begin{equation} 
\nu = \int d^3 {\bf v} f({\bf x},{\bf v})
\end{equation} 
is the density of the tracer stars, which is the zeroth moment of the distribution function;

\begin{equation}
\langle v\rangle_i = \frac{1}{\nu}\int d^3 {\bf v} v_i f({\bf x},{\bf v})
\end{equation}
is the mean velocity (with $i = R,\phi,z$), which is the first moment of the distribution function; and 

\begin{equation}
\sigma_{ij} = \frac{1}{\nu}\int d^3 {\bf v} (v_i - \langle v\rangle_i) (v_j - \langle v\rangle_j) f({\bf x},{\bf v})
\end{equation}
is the velocity dispersion tensor, which is a second velocity moment of the distribution function. (Note that $\nu$ should not be confused with the total matter density $\rho$ that appears in the Poisson equation (equation \ref{eqn:poisson}). The equality $\nu = \rho$ is only valid if the tracer stars comprise all of the gravitating mass.)

In principle, we may continue in the same vein adding ever higher order moment equations (for example, multiplying through by $v_R^2$ and integrating). This begins to constrain the shape of $f$ at each point through its moments. (A Gaussian is fully defined by its first and second moments and thus the above equations are sufficient. However, more complex distributions will have non-trivial third, fourth and higher moments.) This is potentially valuable but highlights a key problem: such a set of moment equations has no closure relation \citep[e.g.][]{BinneyTremaine2008}. Some distribution functions can be pathological, requiring a infinite set of moment equations\footnotemark. Even then, such a set of moments may not correspond to a unique distribution function (the log-normal distribution is a simple example; e.g. \citealt{2012PhRvL.108g1301C}).

\footnotetext{One way to see this is to consider the Fourier transform of some function $f(x)$: $\mathcal{F}(k) = \int_{-\infty}^{\infty} e^{-2\pi i k x}f(x)dx$. Taking the derivative at $k=0$, we obtain: $\left. \frac{d\mathcal{F}}{dk}\right|_{k=0} \equiv \mathcal{F}^1(0) = -2\pi i \int_{-\infty}^{\infty} x f(x) dx$, which is nothing more than a first moment of $f(x)$. Thus, the moments of $f$ give us the Taylor expansion coefficients for $\mathcal{F}(k) = \sum_{n=0} \frac{\mathcal{F}^n(0)}{n!} k^n$ and thereby fully define the functions $\mathcal{F}(k)$ and $f(x)$. The trouble is that there is no guarantee that the Taylor expansion of $\mathcal{F}$ will converge.}

The key advantages of Jeans methods are: i) they are extremely fast as compared to other methods, allowing large parameter spaces to be explored; and ii) no assumption about the form of $f$ is required since we just constrain its moments. The key disadvantages are that we must bin the data in order to calculate the moments; the shape of the distribution function is not used; the set of moment equations is not closed (see above); and it is possible in some cases that a solution is found for which no actual distribution function exists \citep{2006ApJ...642..752A,BinneyTremaine2008}. Data binning is a particular problem since it averages information away, while it must be performed in `model' rather than `data' space which can make it tricky to properly account for observational uncertainties. I discuss this further in \S\ref{sec:degeneracies}.

\subsection{The 1D approximation}\label{sec:onedimapprox}

Given current data, solving all three Jeans equations (\ref{eqn:Rjeans}, \ref{eqn:phijeans} and \ref{eqn:zjeans}) is neither practical nor possible (though this is beginning to change; see \S\ref{sec:measurements}). For this reason, simplifying assumptions are a necessity. Fortunately, for measurements close the Solar neighbourhood, we can approximately reduce the dimensionality of the problem to just motion in the $z$ direction. 

Consider the Jeans equation perpendicular to the disc: 

\begin{equation} 
\underbrace{\frac{1}{R}\frac{\partial\left(R\nu \sigma_{Rz}\right)}{\partial R}}_{{\rm tilt\,\,term\,\,} \mathcal{T}} + \frac{\partial}{\partial z}\left(\nu\sigma_z^2\right) + \nu\frac{\partial \Phi}{\partial z} = 0 \hspace{37mm} z-\mathrm{Jeans}
\label{eqn:zjeansrepeat}
\end{equation} 
In this equation, the radial and vertical motions couple only through the `tilt' term $\mathcal{T}$, marked above. Close to the disc plane, we may expand the gravitational potential in a Taylor series about $[R_0,0]$: 

\begin{equation}
\Phi(R_0+\Delta R,\Delta z) \simeq \Phi(R_0,0) + \Delta z \left. \frac{\partial \Phi}{\partial z}\right|_{R_0,0}+ \Delta R \left. \frac{\partial \Phi}{\partial R}\right|_{R_0,0} + O(\Delta^2)
\label{eqn:rhosep}
\end{equation}
that to leading order is separable in $\Delta R$ and $\Delta z$. Therefore, close to the disc plane, the cross term in the velocity ellipsoid must vanish: $\sigma_{Rz} = 0$, and the term $\mathcal{T}$ should be small as compared to the other terms in equation \ref{eqn:zjeansrepeat}. The question remains, however, how close is `close'? This can be estimated by assuming some simple but well-motivated model for the Milky Way disc: 

\begin{equation} 
\nu \simeq \nu_0 \exp(-R/R_0)\exp(-z/z_0)
\label{eqn:nusimp}
\end{equation} 
\begin{equation}
\sigma_z^2 \simeq \sigma_{z,0}^2 \exp(-R/R_1)
\label{eqn:sigzsimp}
\end{equation}
\begin{equation}
\sigma_{Rz} \simeq \sigma_{Rz,0} \exp(-R/R_2) \left(\frac{z}{z_0}\right)^n
\label{eqn:sigRzsimp}
\end{equation}
The vertical and radial exponential dependencies are reasonable given our current knowledge of the Milky Way \citep[e.g.][]{BinneyTremaine2008,2008MNRAS.391..793S,2013A&ARv..21...61R}. The vertical polynomial term for $\sigma_{Rz} \propto z^n$ ensures that $\sigma_{Rz}(R,0) = 0$, while allowing it to rise arbitrarily steeply otherwise.

Putting equations \ref{eqn:nusimp}, \ref{eqn:sigzsimp} and \ref{eqn:sigRzsimp} into equation \ref{eqn:zjeansrepeat} gives: 

\begin{equation} 
\sigma_{Rz}\left[\frac{1}{R} - \frac{1}{R_0} - \frac{1}{R_2}\right] - \sigma_z^2\frac{1}{z_0} + \frac{\partial \Phi}{\partial z} = 0
\end{equation} 
Using $R = R_0 \sim 8$\,kpc; $R_0 \sim R_2 \sim 2$\,kpc; and $z_0 \sim 0.2$\,kpc, we can take the ratio of the first two terms to assess the relative importance of the tilt $\mathcal{T}$ for the Milky Way at the Solar Neighbourhood: 

\begin{equation}
f_\mathcal{T} \sim \frac{7}{40} \frac{\sigma_{Rz}(R_0,z)}{\sigma_z^2(R_0,z)}
\label{eqn:ft}
\end{equation}
Equation \ref{eqn:ft} can be thought of as a percentage error introduced by neglecting $\mathcal{T}$. Current constraints for the Milky Way \citep{2008MNRAS.391..793S} suggest that the tilt angle of the velocity ellipsoid at $\sim 1$\,kpc is:

\begin{equation} 
\tan(2\delta) = \frac{2\sigma_{Rz}^2}{\sigma_z^2 \sigma_R^2} \simeq 2\delta = 14.6 \pm 3.6^\circ
\end{equation} 
Thus, at $|z| \sim 1$\,kpc and using $\sigma_z \sim 20$\,km/s; $\sigma_R \sim 40$\,km/s \citep{2003A&A...398..141S}, we have $f_\mathcal{T}(1\,{\rm kpc}) \sim 0.12$; it will be smaller than this at lower heights. Thus, for $|z| \simlt 1$\,kpc we can reasonably ignore $\mathcal{T}$ at the 10\% level. For larger heights, we will need to measure $\mathcal{T}$ and include it in the analysis. 

From here on, we drop the tilt term $\mathcal{T}$. This gives us a one dimensional equation in $z$:
\begin{equation} 
\frac{\partial}{\partial z}\left(\nu\sigma_z^2\right) + \nu\frac{\partial \Phi}{\partial z} = 0
\label{eqn:zjeansapprox}
\end{equation} 
which has a formal analytic solution: 

\begin{equation}
\frac{\nu}{\nu(0)} = \frac{\sigma_z^2(0)}{\sigma_z^2}\exp\left(-\int_0^z \frac{1}{\sigma_z^2(z')}\frac{\partial \Phi(z')}{\partial z'}dz' \right) 
\label{eqn:zjeansapproxsolve}
\end{equation} 
Finally, we can relate the potential $\Phi$ to the total matter density via Poisson's equation. In cylindrical coordinates (and assuming azimuthal symmetry), this is: 

\begin{eqnarray} 
4\pi G \rho & = & \frac{\partial^2 \Phi}{\partial z^2} + \frac{1}{R}\frac{\partial}{\partial R} \left(R \frac{\partial \Phi}{\partial R}\right)  \nonumber \\ 
& = &  \frac{\partial^2 \Phi}{\partial z^2} + \underbrace{\frac{1}{R}\frac{\partial v_c^2(R,z)}{\partial R}}_{\mathrm{rotation\,\,curve\,\,term}\,\,\mathcal{R}}
\label{eqn:poissoncylin}
\end{eqnarray} 
If the rotation curve term $\mathcal{R}$ is also small, then equation \ref{eqn:poissoncylin} becomes an equation also only in $z$ and our system of equations (equations \ref{eqn:zjeansapprox} and \ref{eqn:poissoncylin}) reduces to 1D motion perpendicular to the disc. We might expect $\mathcal{R}$ to be small given the flatness of the Milky Way rotation curve ($v_c \sim \mathrm{const}$ gives $\mathcal{R}(z=0) \sim 0$). At heights $|z| \simlt 1.5$\,kpc, \citet{1989MNRAS.239..571K} show, for a range of plausible Milky Way potential models, that $\mathcal{R}(z)$ is also small, amounting to a correction of order a few percent. \citet{2012ApJ...756...89B} show that this rises to $\sim 10$\% at $|z| \sim 4$\,kpc, while the error always leads to an {\it underestimate} of $\rhodm$. Thus, for $|z| \simlt 1$\,kpc, we may also safely drop the $\mathcal{R}$ term, leading to a 1D system of equations: the {\it 1D approximation}. 

Armed with our 1D system of equations, we are left with a number of choices in how to solve them. Firstly, we can either simultaneously solve the Jeans and Poisson equations (equations \ref{eqn:zjeansapproxsolve} and \ref{eqn:poissoncylin}), or we can first solve equation \ref{eqn:zjeansapproxsolve} for the vertical force: 

\begin{equation} 
K_z = -\frac{\partial \Phi}{\partial z} 
\end{equation}
and then consider what this means for the mass distribution in the disc \citep{1960BAN....15....1H}. This latter has the advantage that we need not specify a gravitational model until the last possible moment \citep[e.g.][]{2007MNRAS.379..597N}. 

Another choice enters in that we can solve equation \ref{eqn:zjeansapproxsolve} for $\nu(z)$ given some measured or fitted $\sigma_z(z)$, or we can do this the other way round: 

\begin{equation}
\sigma_z(z)^2 = \frac{1}{\nu(z)}\int_0^z \nu(z') K_z(z') dz' + \frac{\sigma_z(0)^2 \nu(0)}{\nu(z)}
\label{eqn:sigmazint}
\end{equation}
which can be advantageous since $\nu(z)$ is often better constrained than $\sigma_z$ \citep[e.g.][]{1989MNRAS.239..571K}.

Finally, we can choose to constrain either the volume density $\rho$ or the {\it surface mass density} $\Sigma$. Neglecting the rotation curve term $\mathcal{R}$, this is given by: 

\begin{equation}
\Sigma(z) = \int_{-z}^z \rho(z')dz' = 2 \int_0^z \frac{1}{4\pi G} \frac{\partial^2 \Phi}{\partial z^2} dz' = \frac{|K_z|}{2\pi G}
\end{equation} 
This has the advantage that is it directly related to the vertical force $K_z$, whereas $\rho$ requires another derivative of the potential. The mean enclosed dark matter density can be calculated from $\Sigma$ as:

\begin{equation} 
\langle \rho \rangle_\mathrm{dm}(z_\mathrm{max}) = \frac{\Sigma_z(z_\mathrm{max}) - \Sigma_b(z_\mathrm{max})}{2z_\mathrm{max}}
\label{eqn:rhodmmean}
\end{equation}
where $\Sigma_b(z_\mathrm{max})$ is the baryonic contribution. 

\subsection{A 1D distribution function method}\label{sec:1Ddist}

If the tilt term is zero rather than just small ($\mathcal{T} = 0$), then we can make a further approximation that the distribution function is fully separable up to $z \sim 1$\,kpc: 

\begin{equation} 
f = f_{R,\phi}(R,v_R,v_\phi) \times f_z(z, v_z)
\label{eqn:sepdistfunc}
\end{equation}
This is a stronger assumption than we have assumed so far as I will discuss in \S\ref{sec:tests}, but it is powerful. Now we can write the vertical density fall-off as an integral over a one-dimensional distribution function in the vertical energy $E_z = \frac{1}{2}v_z^2 + \Phi$ \citep{1989MNRAS.239..571K}:

\begin{equation} 
\nu(z) = \int_{-\infty}^{\infty} dv_z f(z,v_z) = 2\int_\Phi^\infty \frac{f(E_z)}{\sqrt{2\left(E_z - \Phi \right)}}dE_z
\label{eqn:1Ddistfalloff}
\end{equation}
Applying an Abel transformation, we obtain \citep{1989MNRAS.239..571K,BinneyTremaine2008}: 

\begin{equation} 
f(E_z) = -\frac{1}{\pi} \int_{E_z}^\infty \frac{\partial \nu}{\partial \Phi}\frac{1}{\sqrt{2\left(\Phi - E_z\right)}}d\Phi
\end{equation}
which may be {\it directly} compared with discrete data $[z,v_z]$ to obtain a likelihood function: 

\begin{equation} 
\mathcal{L} = \prod_i^N \frac{f(E_{z,i})}{\int_0^\infty f(E_z) dE_z} 
\end{equation}
This is the method derived and used by \citet{1989MNRAS.239..605K}. We call this the `KG' method from here on. 

\citet{1994MNRAS.270..471F}, \citet{2000MNRAS.313..209H} and \citet{2004MNRAS.352..440H} employ a very similar method, but rather than calculating a likelihood from $f(E_z)$, they use the 1D distribution function to calculate the density fall-off of a tracer population moving in a potential $\Phi(z)$. Starting from equation \ref{eqn:1Ddistfalloff}, we define a $z=0$ vertical velocity without loss of generality: 

\begin{equation} 
w = \sqrt{2[E_z - \Phi(0)]} = \sqrt{2E_z} \,\,\,\, ; \,\,\,\, \Phi(0) \equiv 0 
\end{equation} 
Substituting this into equation \ref{eqn:1Ddistfalloff}, we obtain: 
\begin{equation} 
\nu(z) = 2 \int_{\sqrt{2\Phi}}^\infty  \frac{f(w)w dw}{\sqrt{w^2-2\Phi}}
\label{eqn:hfeq}
\end{equation}
which has the advantage that $\nu(z)$ may be calculated using only the vertical velocity distribution function of {\it nearby} stars in the plane, $f(w)$. Comparing this with the observed distribution $\nu_{\rm obs}(z)$, we can hone in on the best-fitting $\Phi(z)$. 

Since both the KG and HF methods assume a separable distribution function (equation \ref{eqn:sepdistfunc}), we focus on the HF method as a proxy for both when confronting 1D methods with mock data in \S\ref{sec:tests}.

\subsection{The mass model}\label{sec:massmodel}

The total matter density $\rho$ is a sum over all baryonic components (stars, gas, stellar remnants etc.) and dark matter. The dark component is likely constant, at least up to $z \sim 1$\,kpc for which the 1D approximation is valid (\S\ref{sec:cosmotheory}). (Recall that for $z < 1$\,kpc this is true even if there is a `dark disc', since this is expected to have a scale height of $\sim 2-3$\,kpc (\S\ref{sec:darkdischunting}). Data probing to $z \simgt 2$\,kpc would be potentially sensitive to the density fall-off of such a dark disc, making it interesting to relax the $\rhodm \sim {\rm const.}$ assumption. For this review, however, where most of the data are for $z \simlt 2$\,kpc and we work typically under the assumption that the tilt is small, I assume $\rhodm = {\rm const.}$)

The baryonic components can be treated as a sum over many isothermals with constant $\sigma_z$ \citep{2006MNRAS.372.1149F}. Isothermals are a convenient decomposition for the disc, since the solution to equation \ref{eqn:zjeansapprox} is then analytic \citep{bahcall_self-consistent_1984}: 

\begin{equation} 
\nu_i = \nu_{0,i} \exp\left(-\frac{\Phi(z)}{\sigma_{z,i}^2}\right)
\end{equation}
(Note that such a decomposition need not refer to physically distinct tracers, though it does in the \citet{2006MNRAS.372.1149F} model. A particular stellar type could be described, for example, by a linear sum over several isothermal components.)

This gives a total mass model: 

\begin{eqnarray} 
\rho & = & \rho_\mathrm{disc} + \rhodm \nonumber \\ 
& = & \sum_i \nu_{0,i} \exp\left(-\frac{\Phi(z)}{\sigma_{z,i}^2}\right) + \rhodm
\label{eqn:surfforce}
\end{eqnarray}
The \citet{2006MNRAS.372.1149F} mass model is described in Table \ref{tab:massmodel}. Integrating the total surface density, we obtain $\Sigma_b = \Sigma_g + \Sigma_* + \Sigma_\bullet = 49.3 \pm 7.5 \Msunpctw$, where the gas contribution is $\Sigma_g = 13.2 \pm 6.6 \Msunpctw$; the stellar contribution is $\Sigma_* = 28.9 \pm 2.9 \Msunpctw$; and stellar remnants/brown dwarfs contribute $\Sigma_\bullet = 7.2 \pm 0.7$. This can be compared with a recent determination of $\Sigma_* = 30 \pm 1 \Msunpctw$ from SDSS\footnote{Note that this error does not include the systematic uncertainty due to the choice of initial stellar mass function (IMF). \citet{2012ApJ...751..131B} estimate that this is small, however, contributing an additional $1 \Msunpctw$ to the error budget.} \citep{2012ApJ...751..131B}.

It is clear that the total error budget is dominated by the gas. Assuming a constant $\rhodm$ up to $\sim 1$\,kpc, the expected dark matter contribution is $\Sigma_{\rm dm} \sim 16 \Msunpctw$ which is only $\sim 2$ times the error on $\Sigma_b$. Thus, the only reason we can hope to measure $\rhodm$ at all is because we expect $\Sigma_b$ and $\Sigma_{\rm dm}$ to have very different vertical dependences, with $\Sigma_b$ largely reaching its asymptote by $z \sim 0.5$\,kpc, and $\Sigma_{\rm dm}$ continuing to grow up to 1\,kpc and beyond (\S\ref{sec:cosmotheory}).

Given the importance of the baryonic mass model, it is worth a moment to understand the origin of the above uncertainties and how we might do better. With the advent of SDSS, the uncertainty in the local stellar surface mass density $\Sigma_*$ is now very small. Combining the \citet{2006MNRAS.372.1149F} constraints for $\Sigma_\bullet$ with the \citet{2012ApJ...751..131B} value for $\Sigma_*$, we obtain a very accurate $\Sigma_* + \Sigma_\bullet = 37.2 \pm 1.2 \Msunpctw$. The major source of error, however, is in the gas surface density $\Sigma_g$, which is primarily \HI\ gas (see Table \ref{tab:massmodel}). The large error on the \HI\ contribution arises because of the difficultly of measuring distance for gas (see \S\ref{sec:gas}). To convert the observations of temperature and velocity as a function of Galactic coordinates on the sky: $T_{\rm gas}(l,b,v)$ to a surface density $\Sigma_g$, we must assume some underlying mass model for the Galaxy  \citep[for a review see][]{2009ARA&A..47...27K}. Using the results from such an analysis independently of measurements of $\rhodm$ immediately creates some inconsistency since the best fit mass model used to derive $\Sigma_g$ may be rather different from the best fit that arises from the measurement of $\rhodm$. I discuss this problem further in \S\ref{sec:gas}. For now, I will side-step this thorny issue and simply discuss the measurements of $\Sigma_g$ available in the literature to date. \citet{holmberg_local_2000} split the \HI\ into hot and cold components that each contribute $\Sigma_{\rm \HI} \sim 4\Msunpctw$ (see Table \ref{tab:massmodel}), whereas \citet{2003ApJ...587..278W} favour $\Sigma_{\rm \HI} \sim 5 \Msunpctw$, and \citet{2008A&A...487..951K} $\Sigma_{\rm \HI} \sim 12 \Msunpctw$. If we take the very latest value to be correct (not necessarily a safe thing to do) and assign an error based on the radial fluctuations in \HI\ reported by \citet{2008A&A...487..951K}, then we obtain $\Sigma_{\rm \HI} = 12 \pm 4\,\Msunpctw$. Including the contribution from warm gas and H$_2$ reported in Table \ref{tab:massmodel}, I derive $\Sigma_b = 54.2 \pm 4.9 \Msunpctw$, where I have assumed a 50\% error on the H$_2$ and warm gas contribution as previously. This formally more accurate $\Sigma_g$ is reported also in Table \ref{tab:massmodel}. I stress, however, that in future we ought to simultaneously fit for $\Sigma_g$ alongside our fit for $\rhodm$.

\begin{table}
\center
\includegraphics[width=0.99\textwidth]{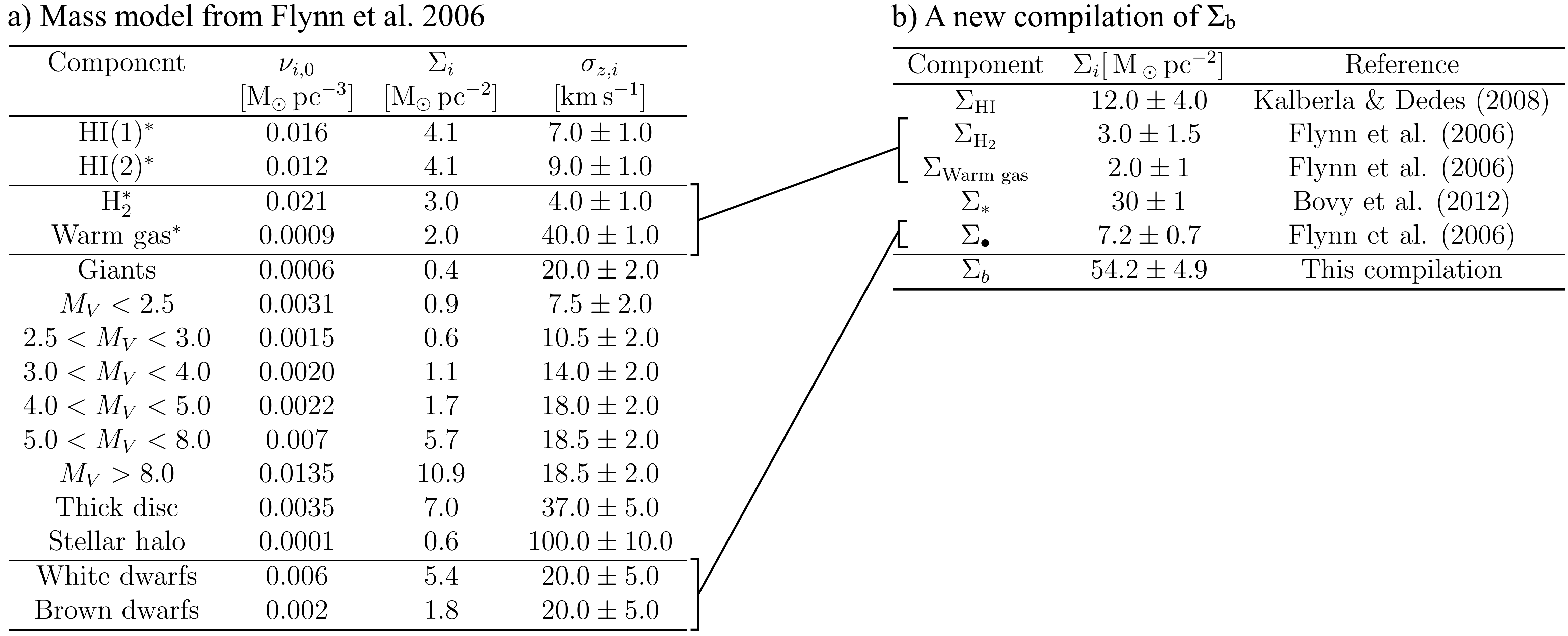}
\caption{\small {\bf a)} The disc mass model from \citet{2006MNRAS.372.1149F}. The columns show: the mass component (stars/gas/stellar remnant); the mass density in the midplane $\rho(0)$; the total column density $\Sigma$; and the vertical velocity dispersion $\sigma_z$. Uncertainties on the densities are of order $\sim 50$\% for all the gas components (indicated with $^*$) and $\sim 10$\% for all the stellar components. {\bf b)} A new compilation of the integrated baryonic surface density $\Sigma_b$ in gas $\Sigma_g = \Sigma_{\rm \HI} + \Sigma_{{\rm H}_2} + \Sigma_{\rm Warm\,\,gas}$; stars $\Sigma_*$; and stellar remnants/brown dwarfs $\Sigma_\bullet$.}
\label{tab:massmodel}
\end{table}

\subsubsection{The rotation curve prior}\label{sec:rotprior}

It is desirable to model the local dark matter density $\rhodm$ independently of the rotation curve if possible for the reasons outlined in \S\ref{sec:introduction}. However, we can still use the rotation curve to put sensible bounds on $\rhodm$. Some authors like \citet{1989MNRAS.239..571K} have applied such priors, while others like \citet{bahcall_self-consistent_1984} have not (see Figure \ref{fig:rhodmhistory}). The precise form of any such prior depends on the choice of mass model. \citet{1989MNRAS.239..571K,1989MNRAS.239..605K,1989MNRAS.239..651K,1991ApJ...367L...9K}, for example, use a series of spherical-halo Galactic mass models that are consistent with the known rotation curve to inform their prior. I describe this prior in more detail and explore its effect in \S\ref{sec:tests}. 

\subsection{The choice of tracer}\label{sec:tracers}

So far, we have assumed the existence of some equilibrium tracer stars with known position and velocity. In the 1D approximation, this means having perfect knowledge of the height and vertical velocity of each star: $[z,v_z]$. If using moment methods, we may then extract from this a density $\nu \equiv \nu(z)$, and a vertical velocity dispersion $\sigma_z \equiv \sigma_z(z)$. However, there are several practical problems that arise when attempting to measure $[z,v_z]$ for real stars in the Milky Way. I briefly discuss these, next. 

\paragraph{Selecting stars in `equilibrium'} Firstly, we require that the tracers are in dynamical equilibrium (steady state) such that we can neglect the partial time derivative of the distribution function (see \S\ref{sec:theory}). For this reason, authors usually avoid young stars since these may not have had time to dynamically mix through the disc \citep[e.g.][]{bahcall_local_1992}. However, there is no guarantee that the disc has not been recently disturbed such that even old stars are currently out of equilibrium; I discuss recent evidence for such disequilibria in the Milky Way in \S\ref{sec:measurements}. 

\paragraph{Selecting stars that reach to high ${\bm z}$} Secondly, we require stars that orbit relatively high up above the disc plane ($z > 0.75$\,kpc) in order to break a degeneracy between the dark and stellar mass in the disc \citep{2011MNRAS.416.2318G}. I discuss this degeneracy further in \S\ref{sec:tests}. 

\begin{figure}[t]
\begin{center}
\includegraphics[width=0.99\textwidth]{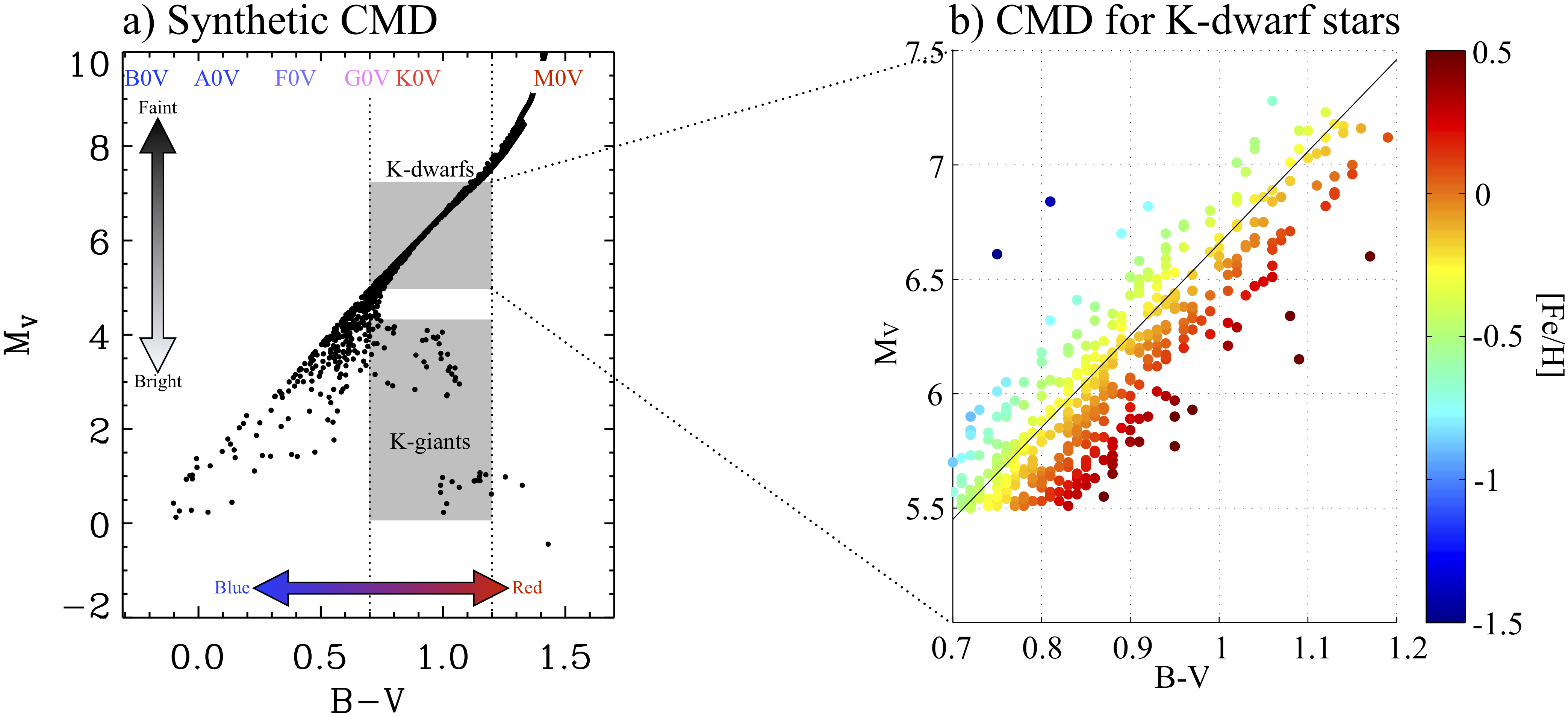}
\vspace{-5mm}
\caption{\small {\bf a)} A synthetic Colour-Magnitude Diagram (CMD) generated using the {\it IAC-star} code \citep{2004AJ....128.1465A}; the stellar {\it spectral type} (B0V, A0V, ... etc.) as a function of colour $B-V$, is marked (see footnote 11 for a definition of colour, magnitude, and spectral type). {\bf b)} A real CMD for 431\,K-dwarf (K0V) stars selected from \citet{2002MNRAS.336..879K} with Hipparcos distances ($z < 100$\,pc; \citealt{2012MNRAS.425.1445G}). These can be used to calibrate a photometric distance at heights $z > 100$\,pc for which Hipparcos distances are not accurate. Notice that the scatter in the relationship between $M_V$ and $B-V$ is largely due to metallicity [Fe/H] (colour contours); it can be significantly reduced if [Fe/H] is known.}
\label{fig:photodist}
\end{center}
\end{figure}

\paragraph{Obtaining a good measure of distance} Thirdly, it is difficult to measure the distance $z$ of a star accurately. In an ideal world, we would use the parallax distance method, since this is the most accurate available \citep[e.g.][]{BinneyTremaine2008}. However, using the Hipparcos satellite, this is currently only possible for bright stars within $\sim 100$\,pc of the Sun \citep{vanLeeuwen2007}. This will change soon with the advent of Gaia (see \S\ref{sec:gaia} and Figure \ref{fig:gaia_accuracy}). In the meantime, we must make do with a {\it photometric distance} estimate. This relies on finding stars of a known luminosity\footnotemark\ $L$ -- so-called `standard candles'. The distance then follows from a flux measurement: 

\footnotetext{For readers not familiar with astronomical nomenclature, it is worth a brief digression in this footnote to explain some common jargon. Astronomers usually use a logarithmic scale for luminosity, called {\it absolute magnitude}, integrated over a range of wavelengths called a {\it waveband}: 

\begin{equation} 
M_V \equiv -2.5 \log_{10}\left(L_V/L_\odot\right) + 4.83 \,\,;\,\, M_B \equiv -2.5 \log_{10}\left(L_B/L_\odot\right) + 5.48
\label{eqn:MV}
\end{equation}
where the $V$ waveband is centred on $\lambda = 550$\,nm; the $B$ waveband is centred on $\lambda = 440$\,nm; and the normalisations are historical. Astronomers also use a similar logarithmic measure of photon flux called {\it apparent magnitude}:

\begin{equation} 
m_V = M_V + 5\log_{10}\left(\frac{d}{10\,{\rm pc}}\right)
\label{eqn:apparentmag}
\end{equation} 
where the normalisation at 10\,pc is historical.

To a very good approximation, stars are black body radiators \citep[e.g.][]{Phillips1999} and are therefore well-described by just three numbers: a colour (that is simply the difference in flux between two wavebands, e.g. $B-V$); a luminosity; and an age. This is why we can use at least some stars as standard candles. Important also, but to a lesser extent is the chemical composition of a star that astronomers call {\it metallicity} (everything heavier than hydrogen is confusingly called a `metal' by astronomers).

Astronomers also often use the {\it spectral type} of a star as a proxy for colour. This is a system of letters, numbers and Roman numerals that goes, in order of blue to red stars: B05, A0V, F0V, G0V, K0V and M0V. The numbers denote finer colour gradation between the letters, and the Roman numeral V denotes a dwarf or `main sequence' star. I mark these spectral types on Figure \ref{fig:photodist}a (see e.g. \citealt{Phillips1999} for further details).
}

\begin{equation} 
d_p^2 = \left[\frac{L_\lambda}{4\pi f_\lambda}\right]
\label{eqn:photodist}
\end{equation} 
where $d_p$ is the photometric distance to the star; and $L_\lambda$ and $f_\lambda$ are the luminosity and flux at a given wavelength $\lambda$. 

To use equation \ref{eqn:photodist} to obtain a photometric distance, we must have some independent measure of $L_\lambda$ for a given stellar type. We can obtain this by calibrating the relationship between $L_\lambda$, colour $B-V$, and metallicity [Fe/H] using nearby stars that have Hipparcos distances (see Figure \ref{fig:photodist}\footnote{Note that the Colour Magnitude Diagram (CMD) in Figure \ref{fig:photodist}a is upside down as compared what is usually plotted \citep[e.g.][]{Phillips1999}. Both choices have a certain logic since large and positive absolute magnitude $M_V$ corresponds to {\it faint} stars.}). In doing this, however, there is a danger of mis-classifying the stellar type in the first place. Notice from Figure \ref{fig:photodist}a that K-giant stars (stars that have evolved off the main sequence; \citealt{Phillips1999}) can masquerade as main sequence K-dwarf stars since they share similar colours. In practice, this is not a major problem because K-giants are so much brighter that K-dwarfs. Beyond about $\sim 200$\,pc, it becomes implausible that a distant K-giant could be mistaken for a nearby faint K-dwarf \citep{1989MNRAS.239..571K}. Thus, a simple distance cut on $z < 200$\,pc is sufficient to weed out K-giant contamination \citep{2012MNRAS.425.1445G}. 

\paragraph{Obtaining good velocities} We also require good velocities $v_z$ for the stars. Radial velocities (along the line of sight) are most accurate since these derive from doppler shifts; however, transverse velocities (so-called {\it proper-motions}) can also be obtained by waiting long enough that the stars move across the sky with respect to a fixed background \citep[e.g.][]{2005MNRAS.359.1306W}. Here, Gaia will also be transformative, obtaining accurate proper motions out $\sim 1$\,kpc even for faint K dwarf stars (see \S\ref{sec:gaia}). Since in the 1D approximation, we require only $v_z$, one useful trick is to look in a direction where $v_z$ can be measured using only Doppler shifts (i.e. where the line of sight points perpendicular to the disc plane); this trick was used by \citet{1989MNRAS.239..571K} to obtain their K-dwarf sample.

\paragraph{The advantage of a `volume complete' sample} If we know that we have observed every single star of a given type up to some height $z_c$, then that sample is said to be {\it volume complete} up to $z_c$. The advantage of using such volume complete samples is that the density $\nu(z)$ simply follows from counting statistics. If, however, we are missing some stars because they become too faint to be reliably detected, or because they are obscured by dust, then we must correct for such incompleteness. Provided we know both the luminosity function of our stars (that can be a function of height), and our selection function, then there is no problem. But this is an area where systematic errors can creep in.

\paragraph{Ensuring consistency} Ideally, we should use the same tracers for $\sigma_z(z)$ that we use for $\nu(z)$. However, in practice, this is often not done as it is much easier to obtain data for $\nu(z)$ (that requires only imaging), than for $\sigma_z(z)$ (that requires spectra and/or proper motions). This leads to an additional source of systematic error \citep{1989MNRAS.239..651K}. The only way to truly avoid this problem is to use a consistent set of tracer stars. 

\paragraph{Modern survey data} Modern surveys RAVE and SDSS have collected velocities each for of order $\sim 10,000$ stars within $\sim 2$\,kpc of the disc plane \citep{2008MNRAS.391..793S,2012ApJ...746..181S,2013ApJ...772..108Z}. Such velocity data are exquisite and have been driving significant improvements in measurements of $\rhodm$ (see Figure \ref{fig:rhodmhistory} and \S\ref{sec:measurements}). However, the challenge with these data is in understanding the survey selection function well enough that $\nu$ for a given tracer may be reliably determined \citep[see e.g. discussion in][]{2012ApJ...746..181S}.\\

\noindent
Given the above list of complications, it is prudent to measure $\rhodm$ both with a very clean sample of stellar tracers, and using the latest survey data that have much improved statistics but for which it is significantly harder to estimate the systematic errors. I take this approach in \S\ref{sec:measurements}. For the `clean' stellar sample, I present results from a recent re-analysis of the K-dwarf data from \citet{1989MNRAS.239..571K}. This uses a new distance calibration that takes advantage of the more modern Hipparcos data \citep[][and see Figure \ref{fig:photodist}]{2012MNRAS.425.1445G}. These data amount to some $\sim 2000$\,K-dwarf stars, a quarter of which have measured $v_z$; they are volume complete up to $1.1$\,kpc above the disc plane. The `less clean' stellar sample comes from SDSS survey data. There, some $\sim 10,000$ stars are available with measured $[z,v_z]$ up to $\sim 2$\,kpc above the disc plane. However, the selection function for these stars is significantly more complex \citep{2012ApJ...746..181S,2013ApJ...772..108Z}. Finally, I review results for a recent study that also uses the SDSS data, but slices the stars into narrow `Mono-Abundance Populations' (MAPs) \citep{2013arXiv1309.0809B}. These MAPs, appear to be well-fit by very simple quasi-isothermal distribution functions (see \S\ref{sec:torus}), allowing for greatly simplified models to be applied to the data. If such assumptions are correct, then even tighter constraints on $\rhodm$ follow. 

\subsection{Errors and model degeneracies}\label{sec:degeneracies}

\paragraph{Degeneracies} If using the mass model described in \ref{sec:massmodel}, then we have over 30 parameters that may degenerate with one another. To cope with this, \citet{2011MNRAS.416.2318G} use a Markov Chain Monte Carlo (MCMC) method to efficiently explore this parameter space \citep[see also][]{2013ApJ...772..108Z}. As we move beyond 1D models, such methods for efficient parameter exploration will become increasingly important; I discuss this briefly in \S\ref{sec:measurements}. The MCMC method is also very useful when folding in observational uncertainties. I discuss this, next. 

\paragraph{Observational Errors} So far, we have assumed perfect data with no observational errors. In general, we will have non-Gaussian probability distribution functions that describe the likelihood of a position,  velocity and tracer membership (star type, metallicity etc.) of a given tracer star. If using a distribution function approach, including these errors is a straightforward (though perhaps computationally expensive) convolution\footnote{The convolution follows from the sum and product probability rules \citep[e.g.][]{sahaprinciples}.}:

\begin{equation} 
\mathcal{L}({\bf a}| {\bf m}) =  \int_{\bf a_0} d {\bf a}_0 g({\bf a} - {\bf a}_0) \mathcal{L}_0({\bf a}_0|{\bf m})
\label{eqn:errorconv}
\end{equation} 
where $\mathcal{L}({\bf a}|{\bf m})$ is the probability of obtaining imperfect data ${\bf a}$ given some model parameters ${\bf m}$; $\mathcal{L}_0({\bf a}_0\mid {\bf m})$ is the probability of obtaining perfect data ${\bf a_0}$ given {\bf m}; and $g({\bf a}|{\bf a_0})$ is the probability of obtaining ${\bf a}$ given ${\bf a_0}$ (i.e. the error probability distribution function).

If not using a distribution function method, the errors can be included in one of two ways. We may include the observational errors, along with the Poisson noise uncertainties, in the calculation of the binned $\nu(z)$ and $\sigma_z(z)$. However, the resultant error PDFs are unlikely to be Gaussian and we should not use the usual $\chi^2$ statistic when comparing these binned data with a given model. Alternatively, we can sample the error PDF to generate many different data sets that are each compared with a given model (this amounts to a Monte Carlo sampling of the convolution integral in equation \ref{eqn:errorconv}). If using an MCMC, this can then be easily included as a `Monte Carlo within a Monte Carlo' \citep{2011MNRAS.416.2318G}. The downside to this approach is that we must generate many more models in our MCMC chain to ensure that both the model parameters and the data uncertainties are properly sampled.

\subsection{Gas as a tracer of the potential}\label{sec:gas} 

In addition to using stars, we may also use gas as a tracer of the Milky Way potential. For gas, the equations are slightly different since gas is a collisional rather than collisionless fluid. Like the stars, gas will obey the Poisson equation, but the collisionless Boltzmann equation (\ref{eqn:colboltz}) is replaced by the equation of hydrostatic equilibrium that balances pressure forces and gravity: 

\begin{equation}
\nabla P_{\rm gas} = -\rho_{\rm gas} \nabla \Phi 
\label{eqn:hydroeq}
\end{equation} 
where $P_{\rm gas}$ and $\rho_{\rm gas}$ are the gas pressure and density, respectively. Equation \ref{eqn:hydroeq} amounts to an assumption of equilibrium for the gas that is potentially much more precarious than the similar assumption of steady state for the stars. This is because typically un-modelled physical process in the interstellar medium, like supernovae, cosmic ray radiation, gas turbulence and magnetic fields, contribute an effective $P_{\rm gas,eff}$ that is not included in equation \ref{eqn:hydroeq} \citep[e.g.][]{2004ARA&A..42..211E}. This could lead to potentially large systematic errors on $\Phi$. \citet{2008ApJ...679.1288L} recently found, for example, that their derived vertical derivative of the \HI\ rotation curve in the Milky Way is too large to be explained by gravity alone.

To solve equation \ref{eqn:hydroeq}, we must also specify an equation of state of the gas that relates pressure to temperature. Usually, a polytrope is assumed: $P_{\rm gas} = A \rho_{\rm gas}^\gamma$, where $A$ is a constant. For an ideal gas, the gas temperature then follows from $P_{\rm gas} = \rho_{\rm gas} k_B T_{\rm gas} / (\mu m_H)$, where $k_B = 1.38 \times 10^{-23}$\,m$^2$\,kg\,s$^{-2}$\,K$^{-1}$ is the Boltzmann constant; $\mu$ is the mean molecular weight; and $m_H$ is the mass of a proton.

Aside from disequilibria and un-modelled physics, a further key complication when using gas is determining the distance. In practice, for the Milky Way we can only measure the temperature as a function of angle on the sky, usually expressed in Galactic coordinates $l,b$; and the line of sight velocity $v$ that follows from the Doppler shift of the \HI\ 21cm line \citep[e.g.][]{1998gaas.book.....B}. To obtain a distance from this, we must model the gas assuming both hydrostatic equilibrium and some background potential for the Milky Way \citep{2003ApJ...588..805K,2007A&A...469..511K}. I discuss the results of such fits to the new Leiden-Argentina-Bonn (LAB) survey data in \S\ref{sec:measurements}. 


\section{Tests using mock data}\label{sec:tests} 

Given the wide array of different methods outlined in \S\ref{sec:theory}, it is helpful to compare and contrast these by applying them to mock data. This allows us to assess systematic errors that occur when model assumptions are violated, and to assess what type and quality of data are most important to improve estimates of $\rhodm$.

\citet{1989ApJ...344..217S} was one of the first to worry about systematic errors in measuring $\rhodm$, focussing on the typically un-modelled tilt-term $\mathcal{T}$; \citet{1989MNRAS.239..571K} estimated the order of magnitude effect of neglecting $\mathcal{T}$ and the rotation curve term $\mathcal{R}$ (\S\ref{sec:onedimapprox}), and the effect of measurement errors; \citet{1989MNRAS.239..651K} discussed the problems that can arise if tracers are inconsistent (see \S\ref{sec:tracers}); and \citet{1991ApJ...367L...9K} and \citet{2013A&A...555A.105I} performed Monte-Carlo simulations of their full analysis pipeline, similar to those that I will perform \S\ref{sec:1dmock}. However, the first detailed investigation using dynamically realistic mocks generated from evolved $N$-body simulations was performed by \citet{2011MNRAS.416.2318G}. I discuss this work in \S\ref{sec:nbodymock}.

\subsection{Simple 1D mock data}\label{sec:1dmock}

It is beyond the scope of this short review to compare and contrast all of the methods outlined in \S\ref{sec:theory}. Instead, in this section I focus on very simple tests of the 1D Jeans method described in \S\ref{sec:onedimapprox}. I will discuss distribution function methods in \S\ref{sec:nbodymock}. As we will see, this is already instructive. All of the mock data tests presented in this paper are available for download from the Gaia Challenge wiki site\footnote{\href{http://astrowiki.ph.surrey.ac.uk/dokuwiki/}{http://astrowiki.ph.surrey.ac.uk/dokuwiki/}.}, where tests of ever increasing sophistication are on-going. 

To set up some simple mock data that are dynamically self-consistent, I use the 1D distribution function approximation outlined in \S\ref{sec:1Ddist}. This assumes that $\mathcal{T} = 0$ at all heights above the disc plane. I will also assume no observational uncertainties. This is essentially {\it ``as good as it gets"} and so such tests should allow us to estimate the absolute minimum uncertainty expected from data sets of a given size.

To make life even easier, I will assume a very simple parameterised form for the tracer density and gravitational potential as in \citet{1989MNRAS.239..571K}$^{\rm \footnotemark}$:

\footnotetext{KG actually use a double exponential for the light profile since this provides a better match to their real data.} 

\begin{equation} 
\nu(z) = \nu_0 \exp(-z/z_0)
\label{eqn:denmock}
\end{equation}
and:
\begin{equation}
\Phi(z) = K\left(\sqrt{z^2 + D^2} - D\right) + Fz^2 
\end{equation}
which gives: 
\begin{equation} 
K_z = -\left[\frac{Kz}{\sqrt{z^2 + D^2}} + 2 F z\right]
\label{eqn:kzkgforce}
\end{equation} 
where $z_0$ is the tracer scale height; $D$ is the disc scale height; and $K$ and $F$ set the vertical force contribution from the disc and dark halo, respectively. I adopt a system of units: kpc, M$_\odot$, km/s.

Assuming Newtonian gravity and that the rotation curve term $\mathcal{R}$ (\S\ref{sec:onedimapprox}) is small, we can relate the vertical force to a surface density (${\rm M}_\odot\,{\rm pc}^{-2}$) via the Poisson equation: 

\begin{equation} 
\Sigma_z(z) \simeq \frac{|K_z|}{2\pi G} 
\label{eqn:surfdenunit}
\end{equation} 
where in the above system of units, $G = 4.299$.

Using these simple analytic forms, we can calculate the distribution function as a function of vertical energy: 

\begin{equation}
f(E_z) = -\frac{1}{\pi}\int_{E_z}^{\infty} \frac{\mathcal{G}(\Phi) d\Phi}{\sqrt{2(\Phi - E_z)}} \,\,\,\, ; \,\,\,\, \mathcal{G} \equiv \frac{1}{z_0}\frac{\nu}{K_z}
\label{fEznum}
\end{equation} 
This needs to be solved numerically which requires us to transform away the infinity in the upper integral limit and the root in the denominator. Using the substitution $\Phi = E_z \sec^2\theta$ gives: 

\begin{equation} 
f(E_z) = \frac{-\sqrt{2 E_z}}{\pi} \int_{0}^{\pi/2} \sec^2\theta \mathcal{G}(\theta,E_z) d\theta
\label{eqn:fEznum}
\end{equation}
To draw a population of $i$ stars, I first draw the positions $z_i$ from equation \ref{eqn:denmock}. Then for each star, assuming values for $[z_0, D, F, K]$, I calculate $f(E_z)$ by numerically integrating equation \ref{eqn:fEznum}. The vertical velocities are then drawn from $f(E_z)$ using an accept/reject method, remembering to normalise $f(E_z)/\max[f(E_z)]$ at each star position $z_i$. 

I set up three mock data sets as described in Table \ref{tab:mockdata}, chosen to be a reasonable match to the Milky Way. The different mocks are designed to explore the effect of sampling and priors (Simple); modelling multiple populations with different scale height simultaneously (Simple2); and having stellar tracers high up above the disc plane (High).

I then attempt to recover the surface mass density $\Sigma_z(z)$ from these mock data. In the spirit of `as good as it gets', I fit exactly the same input mass model to the data (equation \ref{eqn:kzkgforce}). I use the 1D Jeans approximation for this (\S\ref{sec:onedimapprox}), and an MCMC to explore parameter degeneracies (\S\ref{sec:degeneracies}). I run 500,000 models for each MCMC chain and conservatively discard the first half to avoid bias induced by the initial chain parameters. 

\begin{table}
\begin{center}
\begin{tabular}{lccccl}
\thickhline
{\bf Model} & $z_0$ & $K$ & $F$ & $D$ & {\bf Plot} \\
\thickhline
Simple & 0.4 & 1500 & 267.65 & 0.18 & a)-f) \\
Simple2 & 0.9 & 1500 & 267.65 & 0.18 & f) \\
High & 0.65 & 1500 & 267.65 & 0.18  & g), h)\\
\thickhline
\end{tabular}
\end{center}
\vspace{-5mm}
\caption{Mock data parameters. The columns show: mock description; tracer scale height $z_0$; disc and dark matter vertical force parameters $K, F$; disc scale height $D$; and a plot label that indicates which panels in Figure \ref{fig:testmock} use each given mock. (Note that panel f) explores simultaneously fitting two tracers: Simple and Simple2 with different scale heights.) I adopt a system of units: kpc, M$_\odot$, km/s. The disc surface mass density follows from equation \ref{eqn:surfdenunit}: $\Sigma_b = K/(2\pi G) = 55.53 \Msunpctw$. The dark matter density follows similarly: $\Sigma_{\rm dm} = Fz /(2\pi G) \Rightarrow \rhodm = F / (2\pi G 1000) = 0.01 \Msunpcth$.}
\label{tab:mockdata}
\end{table}

\subsubsection{The effect of sampling error and priors} 

\begin{figure}
\begin{center}
\includegraphics[width = 0.99\textwidth]{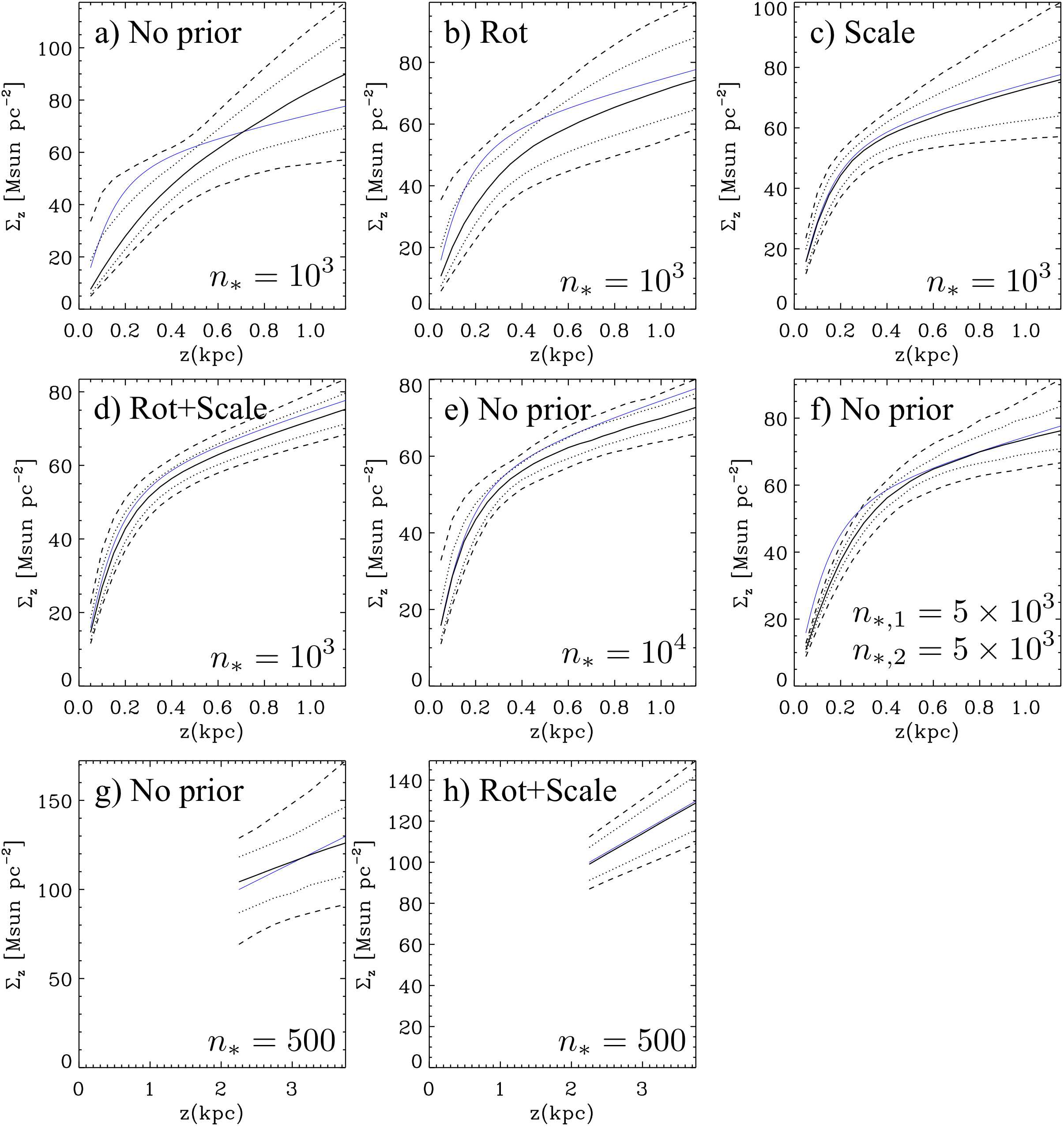}
\end{center}
\vspace{-8mm}
\caption{\small 1D Mock data tests of the recovery of the disc surface mass density $\Sigma_z(z)$. The solid, dotted and dashed lines show the median, 68\% and 95\% confidence intervals for 250,000 models sampled with an MCMC. The blue line shows the input model. The mock data are described in Table \ref{tab:mockdata} and are `as good as it gets' in that I assume no observational errors; zero tilt and rotation curve terms $\mathcal{T}=0$, $\mathcal{R}=0$; and perfect self-consistent tracer stars. Panels a) - d) explore the effect of increasingly strong model priors (as marked) for $n_*=10^3$ tracers from the Simple model (see text for details). Panel e) shows results for $10^4$ tracers; and f) the same split into two populations (Simple and Simple2; see Table \ref{tab:mockdata}) with different scale heights. Finally, panels g) and h) explore results for just 500 tracers high above the disc plane (the `High' mock data set).}
\label{fig:testmock} 
\end{figure}

First, I consider how well we do using just 1000 stars but applying different levels of prior information. The results are shown in Figure \ref{fig:testmock}a-d. I consider two different priors:

\begin{enumerate} 

\item {\bf Rot:} The KG rotation curve prior \citep{1989MNRAS.239..571K,1989MNRAS.239..605K}: 

\begin{equation} 
F = (0.041 - 0.0094 K \pm 0.008) \frac{c}{a} 
\end{equation} 
where $c/a$ describes the the halo flattening perpendicular to the disc. I assume the default choice used by KG $c/a = 1$ which is valid for a spherical dark matter halo.

\item {\bf Scale:} Here, I assume significant knowledge about the baryonic mass distribution such that I can place strong priors on $0.1 < D < 0.25$\,kpc and $50 < \Sigma_b < 60 \Msunpctw$. 

\end{enumerate} 
Without any priors (Figure \ref{fig:testmock}a), the recovery of $\Sigma_z(z)$ is poor. The input model (blue) is recovered within the 95\% confidence interval, but there are significant model degeneracies. Panels b), c) and d) explore the effect of increasing the prior constraints. In panel b), I turn on the `Rot' prior that wraps in information from the Milky Way rotation curve assuming spherical symmetry. This immediately tightens the error envelope but does not drastically reduce the uncertainties at $z \sim 1$\,kpc. In panel c), I turn on the `Scale' prior that puts constraints on the mass and scale length of the visible disc. This prior is quite reasonable in that such information are available for the real Milky Way (\S\ref{sec:massmodel}). The errors are now significantly reduced at $z\simlt 0.5$, but the errors grow significantly at $z \sim 1$\,kpc -- the region where we become sensitive to $\rhodm$. Finally in d), I add both the Rot prior that constraints $\rhodm$ and the Scale prior that constrains the visible disc. Now I obtain rather tight constraints that are closer to previously reported errors in the literature \citep[e.g.][]{1991ApJ...367L...9K,2004MNRAS.352..440H}.

It is clear from Figure \ref{fig:testmock}a-d that with tracer numbers of $n_* \sim 1000$, we are rather sensitive to priors on the mass model. Once the prior from the rotation curve is taken away, the resultant errors on $\rhodm$ are large \citep{bahcall_local_1992,2012MNRAS.425.1445G}.

Figure \ref{fig:testmock}e shows what happens as we raise the sampling to 10,000 stars -- about the number currently available from the SDSS survey data \citep{2013ApJ...772..108Z}. Now, even without any prior constraints, the error envelope is rather tight -- similar to that quoted recently by \citet{2013ApJ...772..108Z}.

\subsubsection{Data high above the disc plane}

Figure \ref{fig:testmock}g and h explore the effect of using tracers high above the disc plane. \citet{2012ApJ...751...30M} recently used a sample of $\sim 500$ stars over heights $\sim 2-4$\,kpc to claim very tight constraints on $\rhodm$, finding -- at odds with previous studies and the Galactic rotation curve -- a dearth of dark matter near the Sun ($\rhodm = 0 \pm 0.001 \Msunpctw$; see the point marked `MB12' in Figure \ref{fig:rhodmhistory}, and Table \ref{tab:measurements}). Their formal uncertainties were also surprisingly small -- much smaller than those in Figure \ref{fig:testmock}g and h. There are likely several reasons for this. Firstly, \citet{2012ApJ...756...89B} showed that the \citet{2012ApJ...751...30M} measurement hinged on an erroneous assumption that the mean azimuthal velocity of stellar tracers $v_\phi(R,z)$ is constant. Assuming instead that the Milky Way {\it rotation curve} is constant in the plane:

\begin{equation} 
v_c^2(R,0) = \left. R\frac{\partial \Phi}{\partial R}\right|_{z=0} = {\rm const.} 
\end{equation} 
which is a statement about the gravitational potential in the plane $\Phi(R,0)$ rather than the stellar kinematics, they derive a value consistent with other measures in the literature (see the point marked `BT12' on Figure \ref{fig:rhodmhistory}). Secondly, it is likely that the observational uncertainties in the MB12 data are underestimated \citep{2012arXiv1205.5397S}. Finally, with just 412 stars, they rely on knowing very well from which photometric sample these stars are drawn (in order to determine the density fall-off with height). Systematic errors could easily creep in here. If the data become inconsistent such that the density fall-off $\nu(z)$ is no longer consistent with $\sigma_z(z)$, then attempts to fit models to these data could push model parameters into corners of parameter space, leading to erroneously small errors. 

\subsubsection{Multiple tracer populations} 

Figure \ref{fig:testmock}f considers modelling different stellar tracers simultaneously in the same potential. I consider two sample populations -- Simple and Simple2 (Table \ref{tab:mockdata}) with 5000 stars in each. We can think of these as being stars that are split, for example, by metallicity or abundance or stellar type. Since each population has a different scale length -- $z_0 = 0.4$\,kpc (Simple); and $z_0 = 0.9$\,kpc (Simple2) --  but they both live in the same potential, we should obtain tighter constraints on $\Sigma_z(z)$ than we would do if modelling the same number of stars with a single population \citep[this trick was recently employed in the context of measuring $\rhodm$ by][for the first time]{2013ApJ...772..108Z}. As can be seen, splitting the populations in this way does not yield significantly improved constraints. As compared to the Simple population with $10^4$ stars, the errors are somewhat larger at high $z$ and smaller at low $z$. This is perhaps surprising given claims from the spherical Jeans modelling community of the power of population splitting \citep[e.g.][]{2008ApJ...681L..13B,2011ApJ...742...20W}. However, the reason for this is that in the spherical Jeans equations there is an unknown cross term that must be marginalised out: the velocity anisotropy $\beta(r)$. In general, the poorly measured $\beta(r)$ can take any value in the range $-\infty < \beta < 1$. By contrast, in the 1D approximation that we employ here, the equivalent cross term is the tilt $\mathcal{T}$ that we can safely assume is small. Thus, population splitting when modelling, for example, dwarf spheroidal galaxies of the Milky Way is invaluable in helping to break a degeneracy between the enclosed mass $M(r)$ and $\beta(r)$. In our 1D disc modelling, here, no such degeneracy exists and population splitting is correspondingly less powerful.

There are, however, two good reasons to still consider population splitting despite the above. Firstly, once we move high up above the disc plane, $\mathcal{T}$ is no longer small and population splitting will likely prove to be a powerful additional constraint. Secondly, population splitting in stellar abundance or age appears to produce stellar tracers that have a remarkably simple distribution function. More on this in \S\ref{sec:beyond1D}. 

\subsection{$N$-body mocks}\label{sec:nbodymock}

\begin{figure}
\begin{center}
\includegraphics[width = 0.99\textwidth]{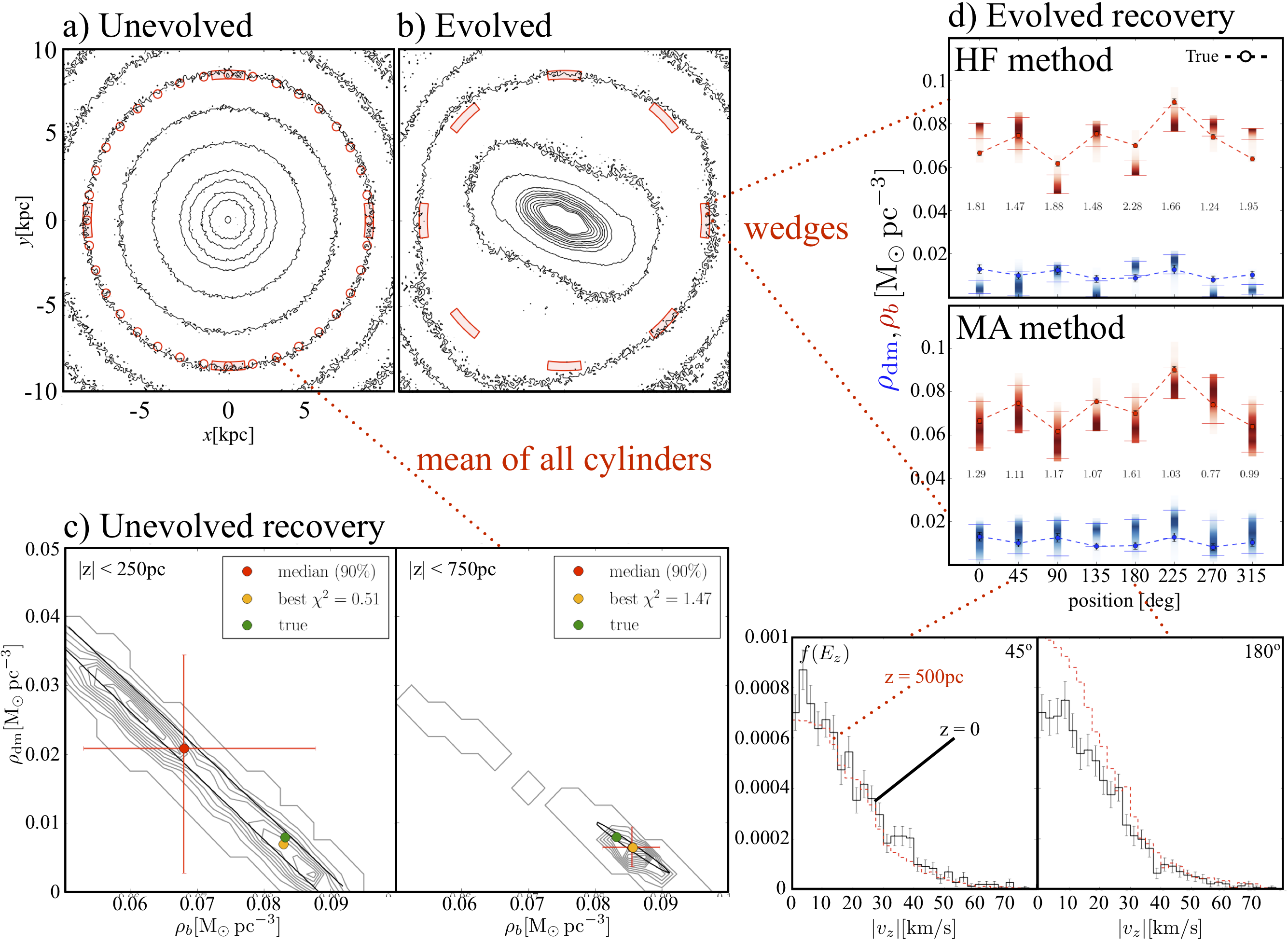}
\end{center}
\vspace{-6mm}
\caption{\small Testing mass modelling methods using dynamically realistic $N$-body mock data; Figures reproduced from \citet{2011MNRAS.416.2318G}. {\bf a)} The $N$-body mock Milky Way disc viewed in stellar density contours from above. The `cylinders' and `wedges' used to represent Solar Neighbourhood volumes are marked in red. {\bf b)} The same simulation evolved for $\sim 4$\,Gyr to form a bar and spiral arms. Notice that the disc is no longer axisymmetric. {\bf c)} Recovery of $\rhodm$ and the in-plane visible mass density $\rho_b$ using the `MA' method on the unevolved disc (see text for details), and conisdering tracers up to $|z|<250$\,pc (left) and $|z|<750$\,pc (right). {\bf d)} Recovery of $\rhodm$ (blue contours) and $\rho_b$ (red contours) for the evolved (non-axisymmetric) disc, using the `HF' method (top) and the `MA' method (bottom). The true values are marked by the dashed lines and solid circles; the horizontal lines on each contour bar mark the 90\% confidence intervals. Notice that the HF method performs well for the wedge at $\theta = 45^\circ$, but poorly at $\theta = 180^\circ$. The bottom two panels plot the distribution function as a function of vertical energy $f(E_z)$ in the plane (black) and at $z = 500$\,pc (red). Notice that these agree for  $\theta = 45^\circ$, but depart strongly at $\theta = 180^\circ$. The $\theta = 180^\circ$ wedge does not satisfy the assumption $f \equiv f(E_z)$ and so the HF method produces a biased result for $\rhodm$.}
\label{fig:silviadisc} 
\end{figure}

The above simple 1D models are instructive in that they already give us a feel for the expected error given perfect data. However, the real Milky Way is dynamically more complex that our simple mock. Apart from observational uncertainties, we have uncertain tracer membership (\S\ref{sec:tracers}), disequilibria (\S\ref{sec:disequilibria}), and potentially significant contributions from the tilt $\mathcal{T}$ and the rotation curve $\mathcal{R}$ terms (\S\ref{sec:theory}). 

One way to test the above is to apply mass modelling methods to dynamically realistic $N$-body mock data. The first to attempt this was \citet{2011MNRAS.416.2318G}; I briefly review the results of that work in this subsection. \citet{2011MNRAS.416.2318G} set up a mock Milky Way using a \citet{2005ApJ...631..838W} model, with $30 \times 10^6$ star `super-particles' (see \S\ref{sec:cosmotheory} for a definition of `super-particles'), and $15 \times 10^6$ and $0.5 \times 10^6$ super-particles for the dark matter halo and stellar bulge, respectively. A contour plot of the stellar distribution viewed from above is shown in Figure \ref{fig:silviadisc} for an `unevolved' disc (a) that was run for $t \sim 50$\,Myrs to ensure equilibrium had been reached; and an `evolved' disc (b) that was run for $t \sim 4$\,Gyr such that a bar and spiral arms similar to those seen in the real Milky Way formed. The unevolved disc satisfies by construction all of the assumptions in the 1D distribution function method outlined in \S\ref{sec:1Ddist}: $\mathcal{T} = 0$ exactly, and $\mathcal{R} \sim 0$. By contrast the evolved disc does not, showing asymmetric variations as a function of angle around the disc. 

\citet{2011MNRAS.416.2318G} test two different mass modelling methods: a generalised 1D moment method (\S\ref{sec:onedimapprox}) that they call the `Minimal Assumption' or (MA) method; and a 1D distribution function method -- the HF method (\S\ref{sec:1Ddist}). The key difference between these two is that the MA method assumes only that $\mathcal{T}$ is {\it small}, while the HF method (like the KG method in \S\ref{sec:1Ddist}) assumes that the distribution function is exactly separable -- i.e. that $\sigma_{Rz} = 0$ and therefore $\mathcal{T} = 0$ exactly. The key results are shown in Figure \ref{fig:silviadisc}. Firstly, \citet{2011MNRAS.416.2318G} apply the MA method to the unevolved disc (Figure \ref{fig:silviadisc}c). Notice that there is a strong degeneracy between $\rhodm$ and the visible in-plane matter density $\rho_b$ if the tracers do not sample high up above the disc plane (compare the left and right panels). This was stressed also by \citet{bahcall_self-consistent_1984}. We must sample several disc scale heights above the disc ($\simgt 750$\,pc) to be able to `see' the dark matter -- at least given current errors on the visible mass density (\S\ref{sec:massmodel}). Secondly, consider the recovery of the evolved disc. Now the disc is axisymmetric and we must consider different `wedges' as a function of angle around the disc, as marked in red on Figure \ref{fig:silviadisc}b. Figure \ref{fig:silviadisc}d shows the recovery of $\rhodm$ (blue contours) and $\rho_b$ (red contours) as a function of wedge angle; the true answers for each wedge are marked by the solid circles and dashed lines. The top plot shows the recovery for the HF method; the bottom for the MA method. Notice that the HF method is biased in several wedges, giving a systematically wrong recovery of $\rhodm$ within the quoted uncertainties (the 90\% confidence intervals are marked by horizontal lines). The reason for this is that the distribution function in most wedges is not a function of vertical energy, whereas the HF method assumes exactly this: $f \equiv f(E_z)$. This is shown in the bottom two panels of Figure \ref{fig:silviadisc}d. Notice that for the wedge at $\theta = 45^\circ$, $f(E_z)$ measured at $z=0$\,pc (black) is in excellent agreement with the same measured at $z = 500$\,pc (red). By contrast, for the wedge at $\theta = 180^\circ$, the distribution function clearly changes as we move from $z=0$\,pc to $z=500$\,pc. This is why the HF method recovers a systematically biased $\rhodm$ and $\rho_b$ for this wedge. 
 
The above demonstrates the importance of testing our methodology on dynamically realistic mock data. At first sight, the HF and MA methods make seemingly identical assumptions. But the subtle difference that the HF method assumes an exactly separable distribution function, while the MA method assumes only approximate separability (via an assumed small tilt term) leads to a potentially strong bias on $\rhodm$ for the HF method. Modern analyses use more sophisticated distribution functions \citep[e.g.][and see \S\ref{sec:torus}]{2012MNRAS.426.1328B,2013arXiv1309.0809B}. However, we must continue to test and hone such parameterised distribution function forms on simulations of ever increasing realism. This is a key goal of the Gaia Challenge project\footnote{\href{http://astrowiki.ph.surrey.ac.uk/dokuwiki/}{http://astrowiki.ph.surrey.ac.uk/dokuwiki/}.}.

\section{Measurements of $\rhodm$ and $\rhodmext$}\label{sec:measurements}

In this section, I summarise historic and recent measurements of $\rhodm$ derived from local tracers in the disc, and $\rhodmext$ derived from the rotation curve.

\subsection{Nearly a century of measurements of $\rhodm$}

A summary of measurements of $\rhodm$ from Kapteyn through to the present day is given in Figure \ref{fig:rhodmhistory}, where I mark also the latest limits on $\rhodmext$ from the rotation curve assuming a spherically symmetric dark matter halo (grey band). In compiling this list, I used the density (black) or surface density (blue) in each study, assuming a baryonic contribution $\rho_b = 0.0914\Msunpcth$ or $\Sigma_b = 55\Msunpctw$, respectively (\S\ref{sec:massmodel}). Note that this can, however, produce a different answer as compared to fitting the raw data using these values as a prior on $\rho_b$ and/or $\Sigma_b$. I illustrate this point using the \citet{2012MNRAS.425.1445G} study in \S\ref{sec:latestlocal}. 

There are a number of fascinating results that crop up from examining Figure \ref{fig:rhodmhistory}. Firstly, notice that right up to the mid and late 1980s, there was an enormous scatter in the results between different groups. \citet{oort_note_1960}, \citet{bahcall_k_1984} and \citet{bahcall_local_1992} claimed evidence for significant dark matter in the disc, while \citet{1987AA...180...94B} and \citet{1989MNRAS.239..651K} found none. Post Hipparcos, there has been a dramatic convergence between groups towards values consistent with spherical extrapolations from the rotation curve (grey band). However, the observant reader will notice that there is quite some difference between the quoted errors. Part of this is explained by volume density (black/red) versus surface density (blue) estimates. The latter average over a region higher up above the disc plane that breaks degeneracies between the visible and dark matter mass \citep{bahcall_self-consistent_1984,2011MNRAS.416.2318G}, leading to greater accuracy. But even accounting for this, the error bars appear to remain static with time or grow despite the arrival of data from SDSS. This is because modern analyses now make many fewer assumptions than previously \citep[][and see \S\ref{sec:theory}]{2012MNRAS.425.1445G,2013ApJ...772..108Z}. As these data have improved, we have begun to address more refined questions about the dynamical state of our Galaxy. Secondly, notice that there are three post-Hipparcos measurements that appear discrepant at greater than $1\sigma$. \citet{creze_distribution_1998} are on the low-side. This is likely because they average over the smallest height of any of the studies: $z_\mathrm{max} = 125$\,pc, which is less than the scale height of the Milky Way thin disc \citep[e.g.][]{BinneyTremaine2008}. At this height, we become very sensitive to errors in $\rho_b$. By contrast, \citet[][G12]{2012MNRAS.425.1445G} is on the high-side, though it does agree with the other measures within 2$\sigma$. I discuss this further in \S\ref{sec:latestlocal}. Finally, there is a third discrepant point. \citet[][MB12]{2012ApJ...751...30M} have recently claimed to find no dark matter near the Sun at very high confidence. As discussed in \S\ref{sec:tests}, this discrepancy results, at least in part, from a poor modelling approximation. \citet[][BT12]{2012ApJ...756...89B} re-analyse their data using more realistic model assumptions, finding a value consistent with the other measures (the BT12 data point is also marked on Figure \ref{fig:rhodmhistory}). 

\begin{wrapfigure}{R}{0.5\textwidth}
\vspace{-8mm}
\begin{center}
\includegraphics[width=0.48\textwidth]{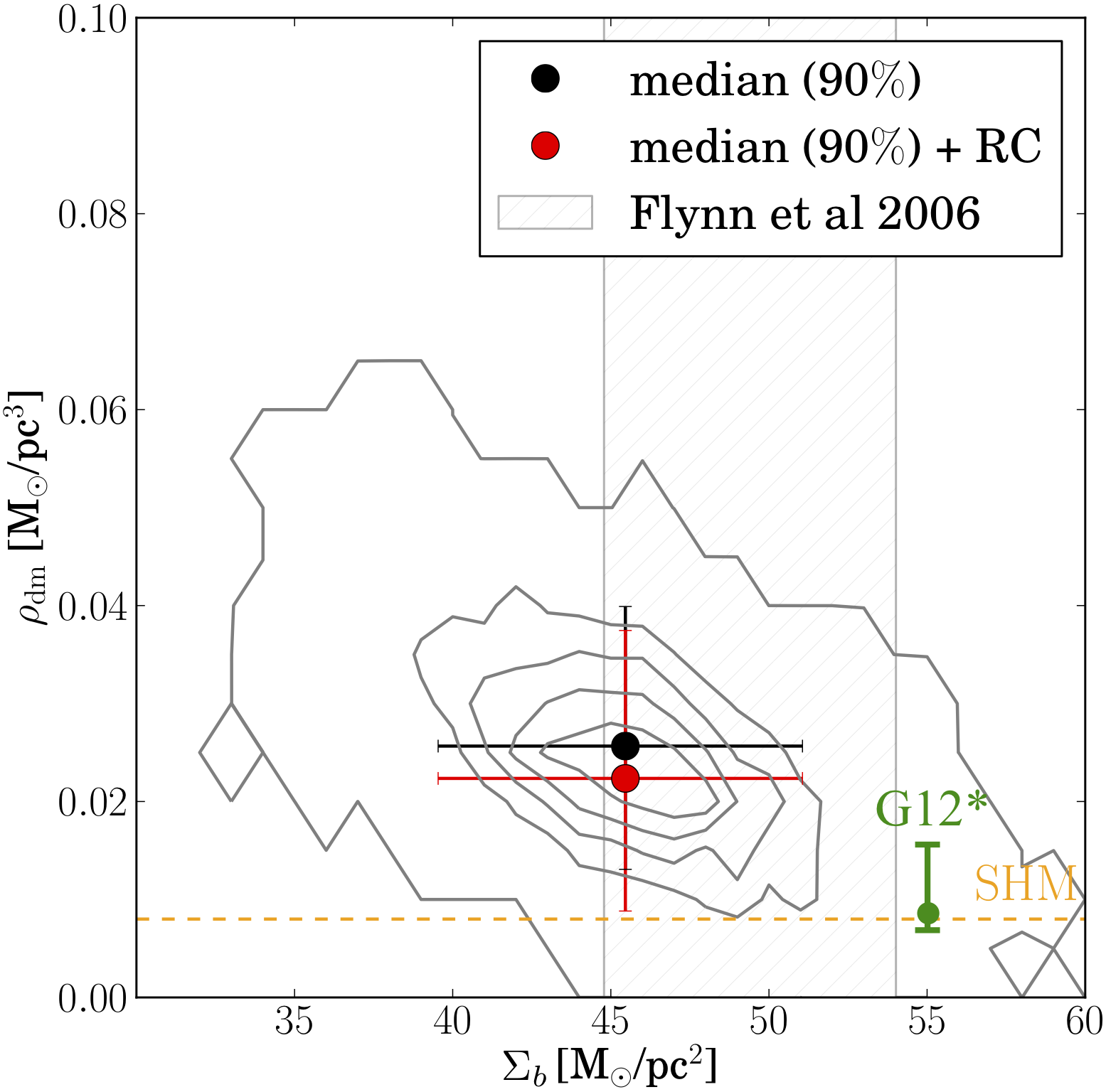}
\vspace{-2mm}
\caption{\small The weakly-broken degeneracy between $\Sigma_b$ and $\rhodm$ in the G12 analysis.}
\label{fig:silvia_explain} 
\end{center}
\vspace{-7mm}
\end{wrapfigure}

\subsection{The latest local measures}\label{sec:latestlocal}
Several groups have recently revisited local measurements of $\rhodm$, as summarised in Figure \ref{fig:rhodmhistory} and Table \ref{tab:measurements}a. All of these measures of $\rhodm$ are complementary in the sense that: i) they rely on different prior assumptions, some stronger than others; and ii) while S12, Z13 and BR13 have $\sim 10,000$ stars within $\sim 2$\,kpc, the $\sim 2000$ K-dwarf stars in the G12 sample (re-calibrated from \citealt{1989MNRAS.239..571K}) have a much simpler, volume complete, selection function.

The first thing to note is that all of the above studies agree within $2\sigma$, while only G12 is discordant at 1$\sigma$. This is already remarkable given the different data sets and methodologies employed. However, it is interesting to understand why G12 is different. The reason can be seen in  Figure \ref{fig:silvia_explain} that plots the derived $\rhodm$ from G12 against $\Sigma_b$ (that is simultaneously fit for in their analysis). The 90\% confidence intervals are marked both with (red) and without (black) correcting for the non-flatness of the local rotation curve. Notice that there is a degeneracy between $\Sigma_b$ and $\rhodm$ that is weakly broken (the contours close), favouring $\Sigma_b = 45.5^{+5.6}_{-5.9}\Msunpctw$. This is systematically lower than Z13 who favour $\Sigma_b = 55 \pm 5 \Msunpctw$. If we invoke a stronger prior on the G12 analysis of $\Sigma_b = 55 \pm 1$\,M$_\odot$\,pc$^{-2}$, in-line with Z13 and with the updated baryonic mass model compiled here (\S\ref{sec:massmodel}), we obtain the green point labelled G12*. This is in much better accord with the other recent measures (see also Figure \ref{fig:rhodmhistory} and Table \ref{tab:measurements}). (Note that using $\Sigma_b = 55 \Msunpctw$ as a prior on the G12 analysis produces a very different result to simply subtracting $\Sigma_b = 55 \Msunpctw$ from the G12 total surface mass density. The former uses knowledge of the baryonic mass distribution in the model fitting, the latter does not.)

The studies S12, Z13 and BR13 all use SDSS survey data that have a Complex Selection Function (CSF). For this reason, S12 choose not to quote uncertainties owing to the difficulty of estimating systematic errors. By contrast, Z13 build on earlier work from \citet{2012ApJ...751..131B} to characterise the survey systematics, computing a final error on their derived $\rhodm$. Ideally, to explore potential systematics in such an analysis we should build sophisticated mock data that are both chemically and dynamically realistic. This is a key goal of the Gaia Challenge initiative\footnote{\href{http://astrowiki.ph.surrey.ac.uk/dokuwiki/}{http://astrowiki.ph.surrey.ac.uk/dokuwiki/}. All mock data tests presented in this paper are available to download from the Gaia Challenge website.}.

\begin{table}
\begin{center}
\resizebox{0.99\textwidth}{!}{%
\begin{tabular}{lllccc}
\thickhline
{\bf Label} & {\bf Reference} & {\bf Description} & {\bf Sampling} & $\rhodm$ [M$_\odot$\,pc$^{-3}$] & $\rhodm$ [GeV\,cm$^{-3}$] \\
\thickhline
\multicolumn{6}{l}{{\bf a)} Local measures ($\rhodm$)}\\
\hline
Kapteyn & \citet{1922ApJ....55..302K} & -- & -- & 0.0076 & 0.285 \\
Jeans & \citet{1922MNRAS..82..122J} & -- & -- & 0.051 & 1.935 \\
Oort & \citet{oort_force_1932} & -- & -- & $0.0006 \pm 0.0184$ & $0.0225 \pm 0.69$ \\
Hill & \citet{1960BAN....15....1H} & -- & -- & $-0.0054$ & $-0.202$ \\
Oort & \citet{oort_note_1960} & -- & -- & $0.0586 \pm 0.015$ & $2.2 \pm 0.56$ \\
Bahcall & \citet{bahcall_k_1984} & -- & -- & $0.033 \pm 0.025$ & $1.24 \pm 0.94$ \\
Bienayme$^\dagger$ & \citet{1987AA...180...94B} & -- & -- & $0.006 \pm 0.005$ & $0.22 \pm 0.187$ \\
\textcolor{blue}{KG}$^\dagger$ & \citet{1991ApJ...367L...9K} & -- & -- & $0.0072 \pm 0.0027$ & $0.27 \pm 0.102$ \\ 
\textcolor{blue}{Bahcall} & \citet{bahcall_local_1992} & -- & -- & $0.033 \pm 0.025$ & $1.24 \pm 0.94$ \\ 
Creze & \citet{creze_distribution_1998} & -- & -- & $-0.015 \pm 0.015$ & $-0.58 \pm 0.56$ \\ 
HF$^\dagger$ & \citet{2000MNRAS.313..209H} & -- & -- & $0.011 \pm 0.01$ & $0.4 \pm 0.375$ \\
\textcolor{blue}{HF}$^\dagger$ & \citet{2004MNRAS.352..440H} & -- & -- & $0.0086 \pm 0.0027$ & $0.324 \pm 0.1$ \\
\textcolor{blue}{Bienayme} & \citet{2006AA...446..933B} & -- & -- & $0.0059 \pm 0.005$ & $0.51 \pm 0.56$ \\ 
\hline
\multicolumn{6}{c}{\it Latest measurements}\\
\hline 
\textcolor{blue}{MB12} & \citet{2012ApJ...751...30M} & CSF & $412$ & $0.00062 \pm 0.001$ & $0.023 \pm 0.042$ \\
& & & & $[0 \pm 0.001]$ & $[0 \pm 0.042]$ \\ 
\textcolor{blue}{BT12} & \citet{2012ApJ...756...89B} & CSF & $412$ & $0.008\pm 0.003$ & $0.3 \pm 0.11$ \\
\textcolor{blue}{G12} & \citet{2012MNRAS.425.1445G} & VC & $2\times10^3$ & 0.022$^{+0.015}_{-0.013}$ & 0.85$^{+0.57}_{-0.5}$ \\
\textcolor{blue}{G12*} & \citet{2012MNRAS.425.1445G} & VC + $\Sigma_b$ & $2\times10^3$ & 0.0087$^{+0.007}_{-0.002}$ & 0.33$^{+0.26}_{-0.075}$\\
\textcolor{blue}{S12} & \citet{2012ApJ...746..181S} & CSF & $10^4$ & $0.005$ [no error] & $0.19$ \\
& & & & $[0.015]$ & $[0.57]$ \\
\textcolor{blue}{Z13} & \citet{2013ApJ...772..108Z} & CSF & $10^4$ & $0.0065 \pm 0.0023$ & $0.25 \pm 0.09$ \\
\textcolor{blue}{BR13} & \citet{2013arXiv1309.0809B} & CSF + MAP & $10^4$ & $0.006 \pm 0.0018$ & $0.22 \pm 0.07$ \\
& & & & $[0.008 \pm 0.0025]$ & [$0.3 \pm 0.094$] \\
\hline
\multicolumn{6}{l}{{\bf b)} Global measures assuming spherical symmetry ($\rhodmext$)}\\
\hline 
S10 & \citet{2010AA...523A..83S} & NP & -- & $0.011 \pm 0.004$ & $0.43 \pm  0.15$ \\
CU10 & \citet{2010JCAP...08..004C} & NFW; SP & -- & $0.0103 \pm 0.00072$ & $0.385 \pm  0.027$ \\
WB10 & \citet{weber_determination_2010} & NFW/ISO; WP & -- & 0.005 - 0.01 & 0.2 - 0.4 \\
I11 & \citet{2011JCAP...11..029I} & gNFW; WP; ML & -- & 0.005 - 0.015 & 0.2 - 0.56 \\
M11 & \citet{2011MNRAS.414.2446M} & NFW; SP & -- & $0.011 \pm 0.0011$ & $0.4 \pm$ 0.04 \\
\thickhline
\end{tabular}
}
\end{center}
\vspace{-5mm}
\caption{\small Measurements of $\rhodm$ (top) and $\rhodmext$ (bottom). The columns show: label; reference; description of the study (for latest measurements only); order of magnitude tracer sample size (for latest measurements of $\rhodm$ only, calculated up to $\sim 1-2$\,kpc); and $\rhodm$ or $\rhodmext$ in $\Msunpcth$ and GeV\,cm$^{-3}$. {\bf Notes: a) ${\bm \rhodm}$:} All values have been calculated from the quoted {\it total density (black) or surface density (blue)} in each study, assuming a baryonic contribution $\rho_b = 0.0914\Msunpcth$ or $\Sigma_b = 55\Msunpctw$, respectively (\S\ref{sec:massmodel}). All error bars represent either $1\sigma$ uncertainties or 68\% confidence intervals. For the latest measurements, if the studies' determination of $\rhodm$ differs from that quoted here, the original determination (using the studies' favoured baryonic contribution) is also marked in square brackets; see text for further details. Studies that use a `rotation curve' prior (see \S\ref{sec:rotprior}), are marked with a dagger $\dagger$. G12 and G12* use re-calibrated volume complete (VC) data from \citet{1989MNRAS.239..571K}; G12* invokes a stronger baryon prior: $\Sigma_b = 55 \pm 1$\,M$_\odot$\,pc$^{-2}$ (see text for details). S12, Z13 and BR13 all use SDSS survey data that have a Complex Selection Function (CSF). S12 choose not to quote uncertainties owing to the difficulty of estimating systematic errors. BR13 slice the data into Mono Abundance Populations (MAPs), assuming that each of these can be fit by a quasi-isothermal distribution function (see \S\ref{sec:torus}). All data points are plotted graphically in Figure \ref{fig:rhodmhistory}. {\bf b) ${\bm \rhodmext}$:} S10 use a non-parametric (NP) method with some of the weakest priors of all of the methods. CU10 use the most restrictive priors, assuming an NFW profile. WB10 explore different halo models (NFW and ISOthermal, amongst others) with weaker priors (WP). I11 wrap in microlensing (ML) data assuming weak priors and a gNFW profile (equation \ref{eqn:nfwprofile} with the power law indices allowed to vary).
}
\label{tab:measurements}
\end{table}

\subsection{The latest global measures}

In addition to recent work on measurements of $\rhodm$, there have been several new measurements of $\rhodmext$. These combine data from a wide range of tracers -- stars and gas -- in the Milky Way, fitting a global model for the Galaxy. Typically, it is assumed that the dark halo is spherical and in the data compilation in Table \ref{tab:measurements}b, I include values only obtained under this assumption. (Note that CU10, WB10 and M11 additionally use the local surface density of matter as a constraint on their models, taking the value from \citet{1991ApJ...367L...9K}. However, since \citet{1991ApJ...367L...9K} use a prior from the rotation curve that assumes a spherical halo (see \S\ref{sec:theory}, \S\ref{sec:tests} and \S\ref{sec:measurements}), I still consider these to be global models that constrain $\rhodmext$ rather than $\rhodm$.)

From Table \ref{tab:measurements}, it is clear that the different studies agree within their quoted uncertainties, but also that a few studies have significantly smaller uncertainties than the others. The reason for this simply comes down to the strength of the assumed priors in each case. S10 and I11 use the weakest priors of all of these studies, the former employing a non-parametric method; the latter using a parametric method but with quite some freedom in the dark matter density distribution (they also include microlensing data constraints). Neither of these studies uses any prior on $\rhodm$ from local measures. This makes their measurements truly independent of local measures of $\rhodm$. Since their results are similar, I use I11 to over-plot results for $\rhodmext$ on Figure \ref{fig:rhodmhistory}.

\subsection{Constraints on the local Milky Way halo shape and an accreted dark disc}\label{sec:shapeddcomp}

\begin{wrapfigure}{R}{0.47\textwidth}
\vspace{-10mm}
\begin{center}
\includegraphics[width=0.47\textwidth]{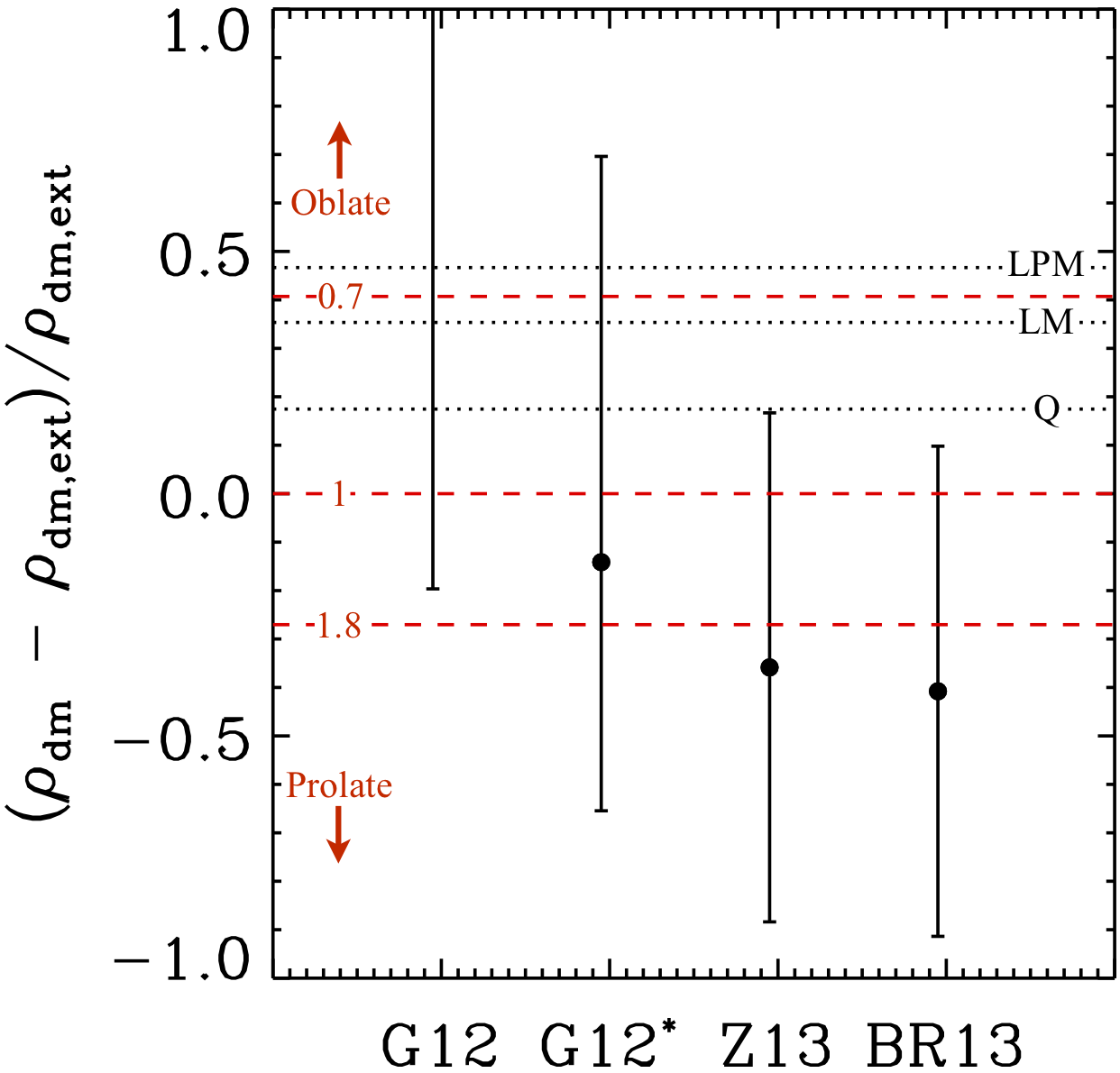}
\vspace{-2mm}
\caption{\small Constraints on the Milky Way halo shape at $R_0$ and/or an accreted dark disc from four recent measurements of $\zeta = (\rhodm - \rhodmext)/\rhodmext$ (data taken from Table \ref{tab:measurements}). The black dotted lines mark $\zeta$ values taken from three cosmological simulations of Milky Way mass halos (Table \ref{tab:darkdisc}); the red dashed lines show $\zeta$ for a  simple flattened Logarithmic halo model (eqauation \ref{eqn:simpleloghalo}).}
\label{fig:darkdisc}
\end{center}
\vspace{-9mm}
\end{wrapfigure}

Comparing the global ($\rhodmext$) and local ($\rhodm$) measures, we can look for evidence for a flattened or prolate dark halo for our Galaxy at $R_0$, or the presence of a dark disc (see Figure \ref{fig:combinedprobes}, and \S\ref{sec:baryons}). I quantify this in Figure \ref{fig:darkdisc}, where I plot $\zeta = (\rhodm - \rhodmext)/\rhodmext$ for four recent measurements of $\rhodm$ from Table \ref{tab:measurements}: G12, G12*, Z13 and BR13. I assume $\rhodmext = 0.38\pm 0.18$\,GeV\,cm$^{-3}$ taken from I11 (see Table \ref{tab:measurements}). Over-plotted are the three $\zeta$ values reported in Table \ref{tab:darkdisc} (dotted lines) for the case of a Quiescent Milky Way (Q), a Milky Way with significant Late Mergers (LM), and a Milky Way with a massive Late Planar Merger (LPM), as marked. I also over-plot the effect of global halo flattening (red dashed lines). To derive these, I assume a Logarithmic halo model, for which the density at the Solar position $[R_0,0]$ is given by \citep[e.g.][]{BinneyTremaine2008}:

\begin{equation} 
\rho_L = \left(\frac{v_0^2}{4\pi G q^2}\right) \frac{(2q^2 + 1) R_c^2 + R_0^2}{(R_c^2 + R_0^2)^2}
\label{eqn:simpleloghalo}
\end{equation} 
where $q$ is the potential flattening in the $z$ direction; $R_c = 15$\,kpc is a halo scale length, and $v_0 = 220$\,km/s sets the halo mass. 

Using the above form for the dark matter halo, we can calculate the increase/decrease in the local dark matter density with respect to the spherical case for different values of $q$: 

\begin{equation} 
\zeta = (\rho_L - \rho_L(q = 1)) / \rho_L(q = 1)
\label{eqn:qrat}
\end{equation}
This is overplotted on Figure \ref{fig:darkdisc} for $q = 0.7, 1$ and $1.8$, as marked (red dashed lines). The small $q$ values correspond to an {\it oblate} (flattened) halo; the large $q$ values to a {\it prolate} (stretched) halo. 

As already reported in G12, notice that G12 favour significant flattening in the plane, suggesting an oblate halo and/or a significant dark disc. By contrast, G12* -- that uses a stronger baryonic surface density prior of $\Sigma_b = 55\pm1 \Msunpctw$ -- is perfectly consistent with a spherical halo at $R_0$. The errors are large, however, permitting both prolate and oblate halos within $1\sigma$,  consistent with all three `dark disc' simulations: Q, LM and LPM. Only the latest SDSS constraints appear constraining at $1\sigma$ (Z13 and BR13). These favour {\it prolate} halos at $R_0$ with negative $\zeta$. If correct, such a local prolate halo would be theoretically rather surprising (see \S\ref{sec:baryons}), and certainly bad news for `alternative gravity' explanations of dark matter \citep[e.g.][]{2005MNRAS.361..971R}. However, the statistical significance for this is low. More interesting is the {\it upper bound} on these data points. Notice that they are only marginally consistent (at 1$\sigma$) with a Quiescent Milky Way with a spherical halo at $R_0$. If the latest measurements from SDSS are correct, the implication is that the Milky Way has a near-spherical or even prolate dark matter halo at $R_0$, no significant dark disc and -- correspondingly -- a rather quiescent merger history. I caution, however, that this result hinges on there being no large systematics that remain to be uncovered in the SDSS data, and on the local baryonic surface density being $\Sigma_b \sim 55 \Msunpctw$.

\subsection{Independent measures of the Milky Way halo shape}\label{sec:streams}

Apart from the $\rhodm/\rhodmext$ comparison, the strongest constraints on the Milky Way halo shape at the moment come from tidal streams. The archetype is the Sagittarius stream, an enormous structure that stretches across the Northern and Southern hemispheres, giving constraints on the halo shape at $\sim 15-50$\,kpc from the Galactic centre \citep{2001ApJ...551..294I}. Early work on the stream suggested a near-spherical halo for the Milky Way \citep{2001ApJ...551..294I}, but more recent data combined with more sophisticated modelling appear to favour a triaxial halo \citep{2010ApJ...714..229L}. The trouble with the latter is that the best fitting model is unstable \citep{1979ApJ...233..872H,1981MNRAS.196..455B,2013MNRAS.434.2971D}. \citet{2013ApJ...773L...4V} have suggested that this problem could be solved if the triaxiality is allowed to vary with ellipsoidal radius, similarly to what is expected from cosmological simulations (\S\ref{sec:baryons}). However, halos that have an axisymmetric inner region that aligns with an outer intermediate axis may also be unstable, once a fully self-consistent `live' halo is taken into account \citep{2013MNRAS.434.2971D}. 

Even if the stability issues for the triaxial solution can be resolved, a key problem remains. None of the current models fit the very latest data that favour a trailing arm with much larger apocentre than the leading arm \citep{2013arXiv1301.7069B}. This puzzling result could imply that the orbit of Sagittarius has evolved \citep{2004MNRAS.351..891Z,2008MNRAS.389.1041R,2013arXiv1308.2235L}, or that the Sagittarius progenitor had a more complex internal structure than is typically assumed \citep{2010MNRAS.408L..26P}. 

The complexity of the Sagittarius stream has driven an increased focus on thinner colder streams that are much simpler to model (though stream-orbit offsets must still be accounted for: \citealt{2011MNRAS.417..198V,2011MNRAS.413.1852E,2013MNRAS.433.1813S,2013arXiv1308.2235L}). \citet{2012MNRAS.424L..16L} have recently argued that a tentative turnaround in the NGC 5466 globular cluster stream at its western edge implies an oblate or triaxial Galactic halo, while \citet{2013arXiv1308.2235L} have shown that with just radial velocity data, the Pal 5 globular cluster stream will determine the flattening of the Milky Way halo -- something that is within reach of current instrumentation. With full proper motion and distance data along these two streams, a triaxial halo could be confirmed or ruled out.

For the time-being, there is sufficient room for uncertainty in all of the above data that a spherical Milky Way halo remains a plausible fit. This could change rapidly, however, as models of already existent data for the Sagittarius stream improve, and as new data for the Pal 5 and NGC 5466 streams become available.

Finally, I stress that all of these stream data give shape constraints at radii significantly larger than that  discussed in \S\ref{sec:shapeddcomp}. Combing local constraints at $R_0$ with stream data over a wide range of Galactocentric radii $R$ holds the exciting promise of constraining radial variations in the shape of our dark matter halo, for the first time.

\subsection{Constraints from \HI\ gas}

As discussed in \S\ref{sec:gas}, the Milky Way \HI\ gas disc also provides information about the Galactic gravitational potential both in and out of the disc plane. \citet{2007A&A...469..511K} have recently fit the Jeans-Poisson equations to new \HI\ data from the Leiden-Argentina-Bonn (LAB) \HI\ survey, finding some rather surprising results. Their favoured mass model has a significant dark disc and a dark `ring' (see also similar results from \citealt{2011JCAP...04..002D}) that appears to align with the known Monoceros stellar over-density \citep{2003MNRAS.340L..21I,2007MNRAS.376..939C,2012ApJ...754..101C}. However, as discussed in \S\ref{sec:gas}, using gas as a tracer presents a range of new complications. Magnetic fields, turbulence driven by gravity or supernovae, radiation pressure, and/or dis-equilibria can all play a role in determining the final distribution of \HI\ gas in the Milky Way -- particularly perpendicular to the Galactic plane. It is not clear whether these are a significant source of uncertainty in deriving the Galactic potential from the \HI\ field, but \citet{2008ApJ...679.1288L} have recently pointed out that the vertical gradient in the \HI\ rotation curve does seem to be too large to be explained by gravity alone.

\subsection{Beyond the 1D approximation}\label{sec:beyond1D}

The most recent measurement of $\rhodm$ from BR13 moves beyond the 1D assumption for the first time to constrain the disc surface density over a range of radii $4\,{\rm kpc}\simlt R \simlt 9\,{\rm kpc}$. Similarly to an earlier study of Geneva Copenhagen Survey (GCS) and RAVE data by \citep{2012MNRAS.426.1328B}, they use the quasi-isothermal distribution function and torus machinery described in \S\ref{sec:torus}. An interesting difference is that \citet{2012MNRAS.426.1328B} slices the data into different stellar populations of a given {\it age}, each of which is assumed to be a quasi-isothermal, whereas BR13 slice on chemical {\it abundance} (Mono Abundance Populations or MAPs). To the extent that abundance is an indicator of age, these two choices are rather similar \citep[e.g.][]{2013A&A...560A.109H}.

Both \citet{2012MNRAS.426.1328B} and BR13 find that the Milky Way disc is close to maximal (where its contribution to the rotation curve is as large as could be allowed by the rotation curve data). Given that these two studies use rather different data sets, this does lend support to their model fits. If the Milky Way disc can really be sliced into quasi-isothermal MAPs as advocated by BR13, or narrow quasi-isothermal age intervals as advocated by \citet{2012MNRAS.426.1328B}, then this is a truly remarkable result. How can our Galactic disc conspire after a cosmic time of violent mergers and gas accretion to have such a simple dynamical structure? This is particularly puzzling given that there is a known and relatively recent merger in the form of the Sagittarius dwarf \citep{1994Natur.370..194I}. More on this, next. 

\subsection{Disequilibria}\label{sec:disequilibria}

\begin{figure}[t]
\begin{center}
\includegraphics[width = 0.9\textwidth]{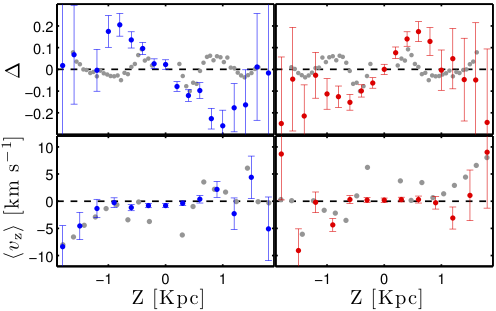}
\end{center}
\vspace{-8mm}
\caption{\small Evidence for disequilibria in the Milky Way, and a possible culprit (Figure reproduced from \citealt{2013MNRAS.429..159G}). The top panels show residual vertical star counts: $\Delta = (n(z) - \langle {n} \rangle (z))/\langle {n} \rangle (z)$ for SDSS data (grey dots taken from \citealt{2012ApJ...750L..41W}); and for a simulated Milky Way disc, recently bombarded by the Sagittarius dwarf (blue/red data points with Poisson errors marked). The bottom plots show the similar results for the vertical velocity averaged in small bins in $z$. The left plots show the raw simulation data; the right, the same phase shifted to better match the observations. Notice that the simulations and observations show qualitatively similar wave-like structures in both density and velocity.}
\label{fig:diseq} 
\end{figure}

I have assumed so far throughout this review that the Milky Way disc is in dynamic equilibrium such that the partial time derivative of the distribution function can be neglected. The very presence of a bar and spiral arms in the Milky Way, along with evidence for `moving groups' in the Solar neighbourhood all point towards disequilibria \citep[e.g.][]{1991ApJ...379..631B,Dehnen1998,2002MNRAS.330..591B,2011MNRAS.418.1423A}. As we have seen in \S\ref{sec:nbodymock}, however, this sort of disequilibria is not a major source of systematic error, at least for current data quality. Interestingly, however, \citet{2012ApJ...750L..41W} have recently reported evidence for a different type of disequilibria that might be more problematic. Using SDSS star counts and kinematics, they find evidence for vertical waves in the disc at $R_0$. This has been largely confirmed -- at least in the stellar kinematics -- by the RAVE survey \citep{2013arXiv1302.2468W}. I show the key plots that define the asymmetry in Figure \ref{fig:diseq}. Such density waves could be illusory, resulting from complex survey selection functions -- after all, the agreement between SDSS and RAVE is only qualitative rather than quantitative \citep{2013arXiv1302.2468W}. However, if such waves are real, it raises some interesting questions. I discuss these, next. 

\paragraph{What could have caused the perturbation?} 
\citet{2011ApJ...731L..35S} recently considered the longevity of perturbations to a simple 1D model of the Milky Way disc \citep[see also a similar analysis in][]{2012ApJ...750L..41W}. They showed that any excited modes decay rapidly on the order of $\sim 10$ vertical crossing times for the disc. For the Milky Way thin disc this is $\sim 20 h / \sigma_z \sim 200$\,Myrs which is extremely short in astronomical terms. However, \citet{2011Natur.477..301P} present a numerical model of the Sagittarius merger where it has undergone three close pericentric passages, the latest being close to the present time. Such continued and current interaction with the disc could excite modes that are still present today. Indeed, \citet{2013MNRAS.429..159G} show that this same simulation leads to vertical modes in the disc that are similar to those found by \citet{2012ApJ...750L..41W} and \citet{2013arXiv1302.2468W} (see Figure \ref{fig:diseq}). This is compelling but not necessarily conclusive. There are a number of unknowns in the Sagittarius modelling, not least the mass and properties of the progenitor system (see \S\ref{sec:streams}). These exquisite stream data do, however, hold out some hope that we can quantitatively predict the effects of such a merger on the Milky Way disc in the not too distant future. 

Alternatively, we need not necessarily appeal to Sagittarius to perform the perturbations. Our current $\Lambda$CDM cosmological model (\S\ref{sec:cosmotheory}) has long predicted the presence of thousands of massive satellites orbiting the Milky Way that are not observed as visible galaxies \citep{1999ApJ...524L..19M,1999ApJ...522...82K}. A fascinating possibility is that such satellites really are there, orbiting as ghostly dark halos that constantly perturb the Milky Way disc.

Finally, it is possible that the perturbations are induced -- at least in part -- by the Milky Way spiral arms. \citet{2014arXiv1403.0587F} have recently used 3D test particle simulations to show that these can also induce vertical modes in the disc similar to those reported by the SDSS and RAVE surveys, though it is unclear if such a mechanism can explain the asymmetry in stellar density at large heights above the disc plane reported by \citet{2012ApJ...750L..41W}.

\paragraph{How do disequilibria bias $\rhodm$ measurements?}
Ideally, we should address this question by performing the sort of mock data tests outlined in \ref{sec:nbodymock} but applied to discs that have been recently perturbed. This exercise remains to be performed and is certainly beyond the scope of this present review. However, \citet{2012ApJ...750L..41W} do perform some simple 1D experiments of a perturbed disc. Like \citealt{2011ApJ...731L..35S}, they find that oscillations damp after $\sim 200$\,Myrs, but they also consider the associated oscillations excited in the vertical force $K_z$. For oscillations that match the amplitude of perturbations in the SDSS data (grey dots, Figure \ref{fig:diseq}), \citet{2012ApJ...750L..41W} find a vertical force oscillation of $\delta |K_z| / 2\pi G \sim 2 \Msunpctw$ at $z \sim 1$\,kpc. Since this is about 10\% of the expected dark matter contribution at this height, such disequilibria are unlikely to have a major impact on measures of $\rhodm$. 

Note that such a small effect should not be surprising. Consider some wave-like perturbation to the disc density $\rho \rightarrow \rho(1+\delta)$, with: 

\begin{equation}
\delta \sim \delta_0 \sin(2\pi z/\lambda)
\end{equation}
Assuming an exponential disc $\rho = \rho_0 \exp(-|z|/z_0)$, this gives a perturbation to the vertical force at $z \gg z_0$:

\begin{equation} 
\Delta_K = \frac{\delta |K_z|}{|K_z|} = \frac{\delta \Sigma_z}{\Sigma_b} = \delta_0 \int_{0}^{\infty} e^{-x} \sin\left(2\pi \frac{z_0}{\lambda} x\right)dx
\end{equation} 
Assuming $\delta_0 \sim 0.1$, $z_0 \sim 0.3$\,kpc and $\lambda \sim 1$\,kpc \citep[][and see Figure \ref{fig:diseq}]{2012ApJ...750L..41W}, this gives $\Delta_K \sim 0.04$. For a disc surface density of $\Sigma_b = 55\Msunpctw$, this gives $\delta|K_z| / 2\pi G = \delta \Sigma_z \sim 2.2 \Msunpctw$, which is in excellent agreement with the number reported in \citet{2012ApJ...750L..41W}.

\subsection{Gaia: precision measurements of $\rhodm$}\label{sec:gaia}
\begin{figure}[t]
\begin{center}
\includegraphics[width = 0.99\textwidth]{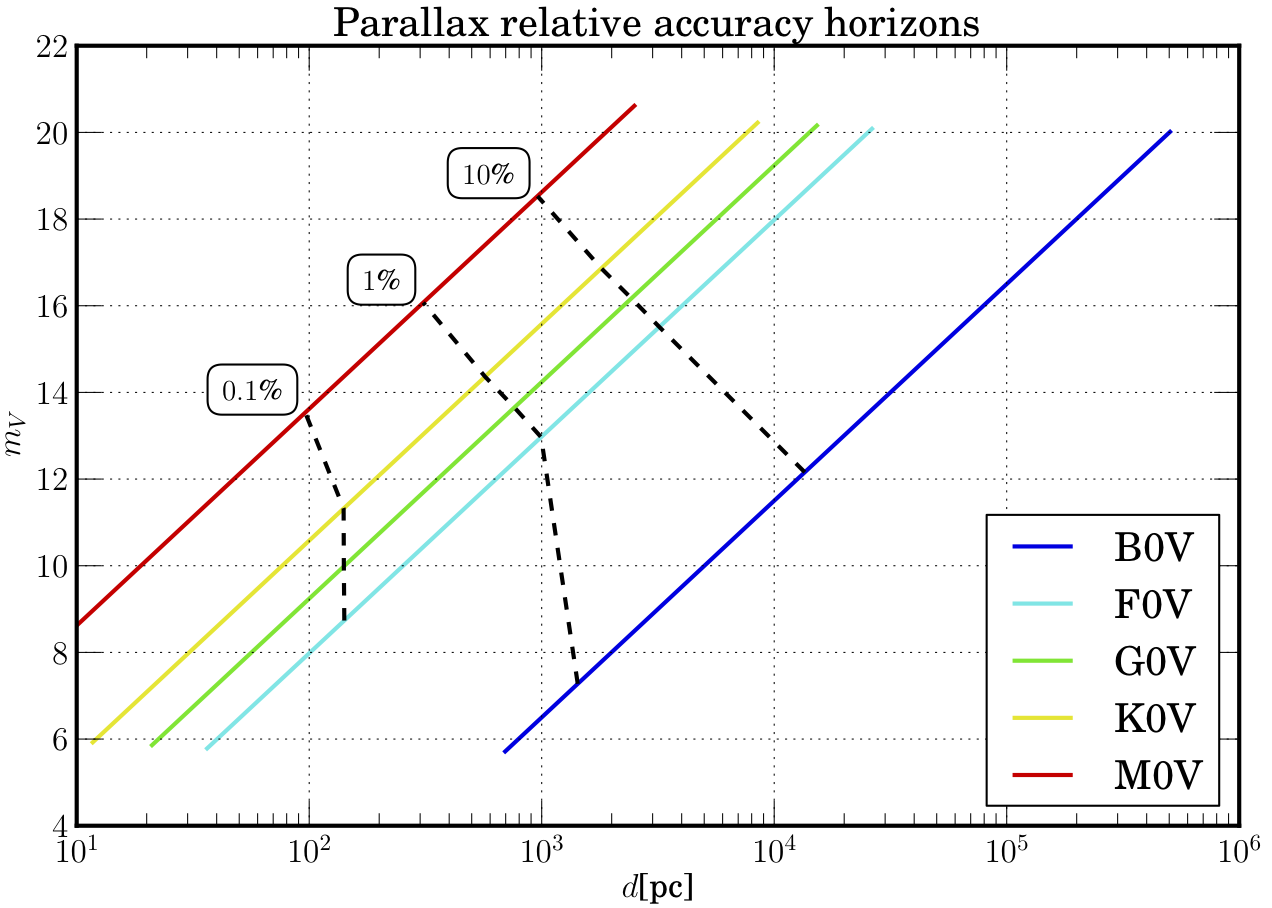}
\end{center}
\vspace{-8mm}
\caption{\small The distance accuracy of the Gaia mission at completion. The coloured lines show the apparent magnitude $m_V$ of stars of different spectral type (see Figure \ref{fig:photodist} for a definition of spectral type) as a function of distance $d$. Distance accuracy horizons of 0.1, 1 and 10\% are marked by the dashed lines. This Figure was produced using \href{https://pypi.python.org/pypi/PyGaia/}{https://pypi.python.org/pypi/PyGaia/} and a template from Anthony Brown.}
\label{fig:gaia_accuracy} 
\end{figure}

Over a mission lifetime of $\sim 9$\,years, the Gaia satellite will catalogue the positions and velocities of $\sim$ a billion stars in our Galaxy \citep{2001A&A...369..339P}. Such a dataset will be transformative for measures of $\rhodm$. Figure \ref{fig:gaia_accuracy} shows the expected distance error for stars of different spectral type as a function of distance $d$ and apparent magnitude $m_V$ (equation \ref{eqn:apparentmag}) at the end of the Gaia mission. I have plotted the spectral types: B0V, F0V, G0V, K0V and M0V, as in Figure \ref{fig:photodist} (for a definition of spectral type and apparent magnitude see footnote 11). For K-dwarf stars, Gaia will measure distance to better than 10\% accuracy out to $\sim 1$\,kpc even for the very faintest stars. Just considering K stars over $5.5 < M_V < 7.5$, this amounts to some $\sim 18 \times 10^6$ stars\footnote{I estimate this number using the Besan\c{c}on Galactic model: \href{http://model.obs-besancon.fr/}{http://model.obs-besancon.fr/}.}. The position and proper motions for these stars will be similarly accurate out to this distance \citep[e.g.][]{2013arXiv1310.3485B}.

With such data -- available for bright stars over a large volume around the Sun -- we will be pushed beyond the simple 1D models typically employed to date. Including all of these stars in the analysis and splitting by spectral type and/or abundance \citep[e.g.][]{2013arXiv1309.0809B}, we will have an enormously valuable dataset for measuring $\rhodm$, largely free from complex survey selection functions. Some problems will remain, however. An accurate 3D dust model for the Galaxy will be necessary to ensure that stars are not mis-classified \citep[e.g.][]{2010HiA....15..782M}. Ideally, we should fit such a dust model simultaneously alongside the dynamical model fit. Furthermore, it may be preferable to fit a full {\it chemo-dynamic} model to the Gaia data, rather than splitting on spectral type or abundance. This has several advantages: i) all data may be used simultaneously; ii) prior information about the inter-relationship between different stellar types could help to break model degeneracies \citep[e.g.][]{2013arXiv1309.2794B}; and iii) the result of such a fit would give us much more information than just the Galactic potential or $\rhodm$ -- it would simultaneously constrain the formation history of these stars within our Galaxy. In the end, however, all of these different approaches are likely to be complementary. For the question of interest here -- the measurement of $\rhodm$ -- the cleanest approach of fitting volume complete stellar tracers seems like a good place to start. 

\section{Conclusions}\label{sec:conclusions} 

I have presented a review of nearly a century of measurements of the mean density of dark matter near the Sun: $\rhodm$. We are about to enter a golden age where such measurements become truly precise. Such accurate measures encode valuable dynamical information about our Galaxy, and are also of great importance for `direct detection' dark matter experiments. I have reviewed theoretical expectations for $\rhodm$, its laboratory extrapolation $\rhodmlab$ and the local velocity distribution function of dark matter $f({\bf v})$ (that is important for direct detection experiments). I presented the key theory behind measurements of $\rhodm$ in the Milky Way, and I collated both historical and modern measures. Finally, I looked ahead to what will soon be possible with the Gaia satellite. My key conclusions are as follows: 

\paragraph{Numerical simulations of $\rhodm$}
\begin{itemize} 
\item State of the art Dark Matter Only (DMO) cosmological simulations make accurate predictions for the local phase space distribution function of dark matter (its mean density $\rhodm$ and velocity distribution function $f({\bf v})$) on scales of $\simgt 20$\,pc. Unresolved structure on smaller scales is unlikely to affect the conclusions of these simulations. 

\item Although unresolved structures in the simulations are not likely important, baryonic processes are. Gas cooling, star formation and stellar feedback during galaxy formation likely rearrange the dark matter distribution in galaxies, even if the dark matter and baryons interact only via gravity. Baryons act to make halos oblate and aligned with the central disc, at least out to $\sim 10$ disc scale lengths; to transform central dense dark matter cusps into cores (if stellar/black hole feedback is strong enough); and -- through biased accretion -- to lead to the formation of an accreted dark matter disc. Each of these processes affects the expectation values of $\rhodm$ and $f({\bf v})$ near the Sun.
\end{itemize} 

\paragraph{Measurements of $\rhodm$}
\begin{itemize} 

\item A key source of uncertainty on $\rhodm$ is the baryonic contribution to the local dynamical mass: $\Sigma_b$. I have compiled a new measurement from literature data: $\Sigma_b = 54.2 \pm 4.9 \Msunpctw$, where the dominant source of uncertainty is in the \HI\ gas contribution. Improving our determination of $\Sigma_b$ warrants renewed attention. 

\item Homogenising $\Sigma_b$ across different studies (using the above value), I find excellent agreement between different groups. In Table \ref{tab:measurements}, I have compiled a list of recent measures of both $\rhodm$ (calculated locally) and $\rhodmext$ (extrapolated from the rotation curve assuming spherical symmetry). Each of these studies is complementary. One -- G12 -- uses a very clean dataset with a simple selection function, but poorer sampling ($\sim 2000$ stars). Three -- S12, Z13, and BR13 -- use SDSS data with significantly improved sampling ($\sim 10,000$ stars), but with a significantly more complex data selection function. The latter studies have smaller formal errors, but present a greater challenge when estimating systematic errors. 

\item Comparing the above measures of $\rhodm$ with spherical extrapolations from the Milky Way's rotation curve ($\rhodmext = 0.005 - 0.015 \Msunpctw$; $0.2 - 0.56$\,GeV\,cm$^{-3}$), the Milky Way is consistent with having a spherical dark matter halo at $R_0 \sim 8$\,kpc. The latest measurements of $\rhodm$ from SDSS appear to favour little halo flattening in the disc plane, suggesting that the Galaxy has little or no accreted dark matter disc and a correspondingly quiescent merger history (see Figure \ref{fig:darkdisc}). I caution, however, that this result hinges on there being no large systematics that remain to be uncovered in the SDSS data, and on the local baryonic surface density being $\Sigma_b \sim 55 \Msunpctw$. 

\item There is a continuing need for detailed tests of our methodologies on dynamically realistic mock data. I illustrated this using both simple 1D tests and full 6D mock data based on an $N$-body simulation of the Milky Way. This latter reveals the surprising result that seemingly sensible assumptions about the distribution function of tracer stars in the disc can lead to significant systematic biases on $\rhodm$. Such model systematics will likely become a dominant source of uncertainty on $\rhodm$ in the Gaia era.

\item Two groups have recently found evidence for disequilibria in the Milky Way in the form of vertical density/velocity waves in the Milky Way disc stars. I showed that, at the currently quoted wave amplitudes, these contribute a systematic error on $\rhodm$ of order $\sim 10$\%. This is not likely to be the dominant source of uncertainty on $\rhodm$ even with Gaia quality data. However, if such oscillatory modes persist as the data continue to improve, they will provide us with a brand new probe of Galactic structure.
\end{itemize} 

\section{Acknowledgements}
This work has made use of the {\it IAC-STAR} synthetic CMD computation code. {\it IAC-STAR} is supported and maintained by the computer division of the Instituto de Astrof\'isica de Canarias. I would like to thank the Gaia Project Scientist Support Team and the Gaia Data Processing and Analysis Consortium (DPAC) for providing the {\tt PyGaia} package that was used to make Figure \ref{fig:gaia_accuracy}. I would like to acknowledge support from SNF grant PP00P2\_128540/1 and ESF funding for the {\it Gaia Challenge} conference where much of of the work in \S\ref{sec:1dmock} was conceived and undertaken. I would like to thank the Oxford University Press publication Monthly Notices of the Royal Astronomical Society and each of the individual authors concerned for the permission to reproduce material that contributed to Figures \ref{fig:cdmsims}, \ref{fig:baryonsims}, \ref{fig:photodist}, \ref{fig:silviadisc}, \ref{fig:silvia_explain} and \ref{fig:diseq}. Finally, I would like to thank Chris Flynn, Silvia Garbari, Paul McMillan, Andrew Pontzen, Jo Bovy, George Lake and the anonymous referees for reading through early drafts of this review and for very helpful feedback. 

\bibliographystyle{mn2e}
\bibliography{../../BibTeX/refs}

\end{document}